\documentclass[12pt,preprint]{aastex}
\usepackage{epstopdf}
\usepackage{graphicx,rotating}
\usepackage{amssymb}
\usepackage{amsmath}
\usepackage{natbib}
\usepackage{lscape}
\usepackage{adjustbox,lipsum}
\newcommand{\beq}{\begin{equation}}
\newcommand{\eeq}{\end{equation}}
\newcommand{\etal}{{\sl et~al.~}}
\newcommand{\kms}{km s$^{-1}$}
\newcommand{\msy}{mas y$^{-1}$}
\newcommand{\HST}{{\it HST~}}
\newcommand{\HSTns}{{\it HST}}
\newcommand{\HIPs}{{\sl Hipparcos~}}
\newcommand{\HIP}{{\sl Hipparcos}}
\newcommand{\Gtf}{{G 250-029\,AB}}
\newcommand{\Get}{{GJ\,831\,AB}}
\newcommand{\Gsn}{{GJ\,791.2\,AB}}
\newcommand{\Gfs}{{GJ\,469\,AB}}
\newcommand{\Gst}{{GJ\,623\,AB}}
\newcommand{\Gsf}{{GJ\,748\,AB}}
\newcommand{\Gtt}{{GJ\,22\,AC}}
\newcommand{\Gttf}{{GJ\,234\,AB}}
\newcommand{\Got}{{GJ\,1245\,AC}}
\newcommand{\Goz}{{GJ\,1081\,AB}}
\newcommand{\Gozz}{{GJ\,1005\,AB}}
\newcommand{\Lst}{{GJ\,1005\,AB}}
\newcommand{\Lsts}{{GJ\,65\,AB}}
\newcommand{\Gsfa}{{GJ\,65\,AB}}
\newcommand{\Gff}{{GJ\,54\,AB}}
\newcommand{\Gon}{{G 193-027\,AB}}
\newcommand{\Gfst}{{GJ\,473\,AB}}

\def\fdg{\hbox{$.\!\!^\circ$}}

\def\Ha{H$\alpha$}

\begin{document}

\received{}
\revised{}
\accepted{}

\shorttitle{M Dwarf Masses}
\shortauthors{Benedict}

\bibliographystyle{/Active/my2}

\title{The Solar Neighborhood XXXVII: The Mass-Luminosity
  Relation for Main Sequence M Dwarfs \footnote{Based on observations
    made with the NASA/ESA Hubble Space Telescope, obtained at the
    Space Telescope Science Institute, which is operated by the
    Association of Universities for Research in Astronomy, Inc., under
    NASA contract NAS5-26555.} }


\author{ G.\ F. Benedict\altaffilmark{2}, T. J. Henry\altaffilmark{3,11},
  O. G. Franz\altaffilmark{4}, B. E.  McArthur\altaffilmark{2},
  L.\ H. Wasserman\altaffilmark{4}, Wei-Chun Jao\altaffilmark{5,11},
  P. A. Cargile\altaffilmark{6}, S. B. Dieterich \altaffilmark{7,11},
  A. J. Bradley\altaffilmark{8}, E. P. Nelan\altaffilmark{9}, and
  A.\ L.\ Whipple\altaffilmark{10}}
	
\altaffiltext{2}{McDonald Observatory, University of Texas, Austin, TX 78712}
\altaffiltext{3}{RECONS Institute, Chambersburg, PA 17201}
\altaffiltext{4}{Lowell Observatory, 1400 West Mars Hill Rd., Flagstaff, AZ 86001}
\altaffiltext{5}{Dept.~of Physics \& Astronomy, Georgia State University, Atlanta GA  30302}
\altaffiltext{6}{Harvard-Smithsonian Center for Astrophysics, 60 Garden Street, Cambridge, MA 02138 USA}
\altaffiltext{7}{Carnegie Institiution of Washington, Washington, DC, 20005}
\altaffiltext{8}{Spacecraft System Engineering Services, PO Box 91, Annapolis Junction, MD 20701}
\altaffiltext{9}{Space Telescope Science Institute, 3700 San Martin Dr., Baltimore, MD 21218}
\altaffiltext{10}{Conceptual Analytics, LLC, Greenbelt, MD 20771}

\altaffiltext{11}{Visiting Astronomer, Cerro Tololo Inter-American
  Observatory. CTIO is operated by the Association of Universities for
  Research in Astronomy, Inc., under contract to the National Science
  Foundation.}



\begin{abstract}

We present a Mass-Luminosity Relation (MLR) for red dwarfs spanning a
range of masses from 0.62 ${\cal M}_{\sun}$ to the end of the stellar
main sequence at 0.08 ${\cal M}_{\sun}$.  The relation is based on 47
stars for which dynamical masses have been determined, primarily 
using astrometric data from Fine
Guidance Sensors (FGS) 3 and 1r, white-light interferometers on the
{\it Hubble Space Telescope (HST)}, and radial velocity data from
McDonald Observatory.  For our \HSTns/FGS sample of 15 binaries,
component mass errors range from 0.4\% to 4.0\% with a median error of
1.8\%.  With these and masses from other
sources, we construct a $V$-band MLR for the lower main sequence with 47
stars, and a $K$-band MLR with 45 stars with fit
residuals half of those of the $V$-band.

We use GJ\,831~AB as an example, obtaining an
absolute trigonometric parallax, $\pi_{abs} = 125.3 \pm 0.3$
milliseconds of arc, with orbital elements yielding ${\cal M}_{\rm A}
= 0.270 \pm 0.004{\cal M}_{\sun}$ and ${\cal M}_{\rm B} = 0.145 \pm
0.002 {\cal M}_{\sun}$.  The mass precision rivals that derived for
eclipsing binaries.

A remaining major task is the interpretation
of the intrinsic cosmic scatter in the observed MLR for low mass stars
in terms of physical effects.  In the meantime, useful mass values can be
estimated from the MLR for the ubiquitous red dwarfs that account for
75\% of all stars, with applications ranging from the
characterization of exoplanet host stars to the contribution of red
dwarfs to the mass of the Universe.

\end{abstract}




%

\section{Introduction}

With the exception of the Hertzsprung-Russell (HR) diagram, the
main sequence Mass-Luminosity Relation (MLR) is perhaps the single most important
``map" of stellar astronomy because the entire evolution of an
individual star depends on its mass.  The MLR allows astronomers to
convert a $relatively$ easily observed quantity, luminosity, to a more
revealing characteristic, mass, providing a better understanding of
the object's nature. When considering populations of stars, an
accurate MLR permits luminosity functions to be converted into mass
functions, and drives estimates of the stellar contribution to the
mass of the Galaxy.

Basic observations of stars include measurements of apparent
magnitudes, colors, surface temperatures ($T_{\rm eff}$), and
metallicities.  From these attributes we can derive luminosities and
radii, given combinations of distance measurements and/or directly
measured angular sizes from long-baseline interferometry.  Yet, the
crucial masses remain elusive, typically only measured dynamically in
binary systems.  Thus, when stellar modelers compare their theoretical
results with real stars to test the accuracy of their efforts, they
often assume masses for their test stars.  For example, at a mass of
0.3${\cal M}_{\sun}$, models from \cite{All00},
\cite{Bar15}, and \cite{Dot16} predict $M_V$ values of 11.42, 11.61, and 11.25 respectively.
An accurate MLR allows choices to be made between various modeling
approaches, such as the treatment of stellar ages, convection,
turbulent mixing, rotation, magnetic activity, and mass loss (Andersen
1991)\nocite{And91}.  Furthermore, the need for accurate mass
estimates for the smallest stars has come to the forefront with the
discovery of exoplanets associated with M dwarf stars (see Lurie \etal
2014 for a recent list of the nearest M dwarfs with
planets)\nocite{Lur14}.  More accurate estimates of host star masses
translate directly into more reliable companion planet masses, and
consequently, more accurate bulk compositions for planets that
transit.
 
The first robust MLR for M dwarf stars was reported in \cite{Hen93},
who provided relations at $VJHK$ for 37 stars, of which 26 had masses
less than $0.6{\cal M}_{\sun}$.  Early results from the project
described here can be found in \cite{Hen99}, where an improved MLR at
$V$ was pivotal.  A modest update was presented in \cite{Del00}, which
included many of the same stars.  \cite{Tor10a} summarized the entire
stellar MLR using eclipsing systems as of a few years ago, outlining
how few M dwarf stars have accurate masses because these tiny stars
rarely eclipse.

Here we present astrometry from the {\it Hubble Space Telescope (HST)}
and other sources, in combination with ground-based radial velocities
(RVs), to improve the low-mass region of the MLR that is complicated by
age, metallicity, and magnetic effects.  This is the realm of the red
dwarfs, typically of spectral type M, that slowly descend to the Main
Sequence along a nearly vertical Hyashi track in the HR diagram
\citep{Pal12}.  This results in an M dwarf population that in the
solar neighborhood spans a wide range of ages \nocite{Rie14} (Riedel
et al. 2014).  In addition, M dwarf luminosities are affected by
composition (metallicity), which is a function of the mixing of the
interstellar medium (with sources internal and external to the Galaxy), 
star formation rate, and birth date within the Galaxy (e.g., Sneden,
Lawler, \& Cowan 2002). \nocite{Sne02} To complicate matters even
further, a more poorly understood factor affecting M dwarf luminosity
is magnetism, which induces short and long-term photometric
variability \citep{Hos15}.

In this paper we model \HST Fine Guidance Sensor (FGS) astrometric,
and, if available, radial velocity (RV) measurements simultaneously to
obtain parallaxes, proper motions, orbits, and component masses for 15
systems containing 30 M dwarfs (Table~\ref{tbl-ID}).  Our resulting
masses have a median precision of 1.8\%, which constitutes a
significant advance from previous efforts.  The derived parallaxes,
which take into account the orbital motions of the binaries, are
usually marked improvements --- typically by factors of 5--10 over
available parallaxes in the Yale Parallax Catalog (van Altena, Lee \&
Hoffleit 1995, hereafter YPC) and {\it Hipparcos} \citep{Lee07b} ---
and are key in deriving the high-accuracy masses.  We derive the first
masses for stars in six systems, improve masses over previous efforts
by others for seven systems, and improve upon our own \HST efforts for
two systems: \Gsf~in \citep{Ben01} and \Gsn~in \citep{Ben00b}.  Our
reductions and analyses for the latter two systems include a few new
\HSTns/FGS measurements and have improved radial velocity
measurements, yielding more precise masses.

In Sections~\ref{ast} and \ref{rv} we describe the various sources of
astrometric and RV data that were used in our modeling.  In
Section~\ref{GET} we discuss in detail the data and modeling process
for \Get, as an example of our technique to obtain masses.
Section~\ref{ALL} summarizes modeling notes and resulting masses for
the other 14 systems.  In Section~\ref{OGC} we describe the
measurements for the $M_V$ and $M_K$ values used to create the MLRs.
Section~\ref{othersystems} describes additional systems in which red
dwarfs with high-quality masses are available to be used in the MLRs.
In Section~\ref{mlr} we construct a $V$-band MLR from the set of 47
stars and a $K$-band MLR using 45 stars.  We discuss our results in
Section~\ref{bd}, and conclude with a summary in Section~\ref{summ}.

\section{Astrometric Data} \label{ast}
\subsection{\HST Fine Guidance Sensors}

An \HST Fine Guidance Sensor (FGS) provides two types of astrometric
data.  Like all interferometers, an FGS produces a fringe that we
interrogate either by tracking the fringe position (POS mode) or by
scanning to build up a fringe image (TRANS mode).  TRANS mode yields
the structure of the fringe from which it is possible to derive the
separation and position angle of a secondary relative to a primary
in a binary system, essential in establishing the relative orbit.  POS
mode permits the measurement of the position of the primary star
(brighter component) relative to a local frame of reference stars,
essential for determining proper motion, parallax, and mass fraction.
\cite{Fra98} and \cite{Ben11} contain details of \HST FGS's TRANS and
POS mode reduction and analysis, respectively.  Detectability and
measurement precisions for a given binary star's components critically
depend on star brightnesses, component brightness differences, and the
binary angular separation.  All \HSTns/FGS astrometry also depends on
the Optical Field Angle Distortion (OFAD) calibration \citep{McA02}.
This calibration reduces as-built {\it HST} telescope and FGS
distortions with magnitude $\sim1\arcsec$ to below 2 milliseconds of
arc (mas) over much of the FGS field of regard.  The present work
benefits from a POS mode OFAD revision more recent (2012) than
reported in \cite{McA06}.  The 15 red dwarf systems targeted in this
program and discussed in the following sections rely on \HST TRANS
and/or POS mode measurements, and are listed in Tables~\ref{tbl-ID}
and~\ref{tbl-ST}.

For binary stars the FGS transforms images of the two components into
two overlapping fringes, sometimes called the ``S-curves''.  For
perfect fringes and component separations greater than the resolution
of \HST at $\lambda =$ 580 nm, about 40 mas, the presence of a
companion should have little effect on the component A position
obtained from the fringe-zero crossing.  For separations smaller than
40 mas the measurement of the position of the brighter component of a
blended image will be biased toward the fainter component
\citep{vdK67}, possibly requiring a photocenter correction.  Because
an FGS has two orthogonal axes, the separation along each
interferometer axis, not the total separation on the sky, determines
the photocenter correction.  For any arbitrary \HST orientation,
components A and B can have separations along the FGS axes from 0.0
mas to the actual full separation.

During this project we observed with two FGS units: FGS\,3 from
1992--2000, and FGS\,1r from 2000--2009. FGS\,1r replaced the original
FGS\,1 during \HST Servicing Mission 3A in late 1999. Our original
analysis of \Gsf~\citep{Ben01} incorporated POS mode photocenter
corrections, as does the present re-analysis. These corrections,
ranging from $-$3.2 to $+$2.3 mas, are for the FGS\,3 X-axis a
complicated function of separation because FGS\,3 delivered far from
perfect fringes with substantial instrumental structure \citep{Fra91}.
FGS\,1r produces higher quality fringes \citep{Nel12}, and corrections
are similar along both axes and similar in size to the Y-axis
correction seen in \cite{Ben01} for \Gsf. These corrections (typically
a maximum of 2 mas) are negligible for systems with component $\Delta
V \geq 2.3$ ($\Delta V$ values listed in Table~\ref{tbl-MMVMK}). The
systems with $\Delta V \leq 2.3$ and typically smaller separations are
\Gfs, \Gsf, \Get, and \Goz, and \Gtf.  Generally, only a few
measurements require photocenter corrections.

\subsection{Additional Astrometry Measurements}
\subsubsection{Traditional Ground-Based Methods}

We included visual, photographic, and CCD observations of separations
and position angles from \citet{Gey88} for our analysis of
\Gsfa~(Section~\ref{G65}). These (relative to \HSTns) low-precision
measurements are primarily from the USNO 61in reflector and stretch to
nearly twice the full  period of \Gsfa, proving 
extremely useful for this long orbital period system.
  
\subsubsection{Adaptive Optics}

We include a single observation of G\,193-027\,AB from \cite{Beuz04},
who used the Adaptive Optics Bonnette system on the
Canada-France-Hawaii Telescope. For GJ\,65\,AB we include five VLT/NACO
measures of position angle and separation \citep{Ker16}.

\subsubsection{Aperture Masking Interferometry}

For our analysis of \Gst~(Section~\ref{G62}) we included astrometric
observations~\citep{Mar07} performed with the PHARO instrument on the
Palomar 200in (5m) telescope and with the NIRC2 instrument on the
Keck II telescope. Separations have typical errors of 2 mas. Position
angle errors average 0\fdg5.

\subsubsection{Speckle Interferometry}

Measurements are included for \Gtt~from \cite{McC91} and for
\Gfst~from \cite{Hen92} and Torres et al.~(1999), who used a
two-dimensional infrared speckle camera containing a 58$\times$62
pixel InSb array on the Steward Observatory 90in telescope.  We also
include infrared speckle observations by \cite{Woi03}, who obtained
fourteen separation and position angle measurements for \Gtt~with the
near-infrared cameras MAGIC and OMEGA Cass at the 3.5m telescope on
Calar Alto.  These instruments are capable of taking fast sequences of
short time exposures (t$_{exp} \sim0.1$ sec) that enable speckle
interferometry observations at the infrared wavelengths where red
dwarfs emit significant flux, and where atmospheric turbulence is
reduced compared to visual wavelengths.  We also include a few speckle
observations at optical wavelengths from the Special Astrophysical Observatory 6m BTA and 1m Zeiss \citep{Bal94}, from the CFHT \citep{Bla87},
and from the DSSI on the WIYN 3.5 m \citep{Hor12}.

\subsubsection{Other \HST Astrometry}

Where available, we use astrometric observations from \HST instruments
other than the FGSs, including the Faint Object Camera (FOC)
\citep{Bar96}, the Faint Object Spectrograph (FOS) \citep{Sch98}, the
Near-Infrared Camera and Multi-Object Spectrometer (NICMOS)
\citep{Gol04}, and the Wide-Field Planetary Camera 2 (WFPC2)
\citep{Sch00, Die12}.

\section{Radial Velocities} \label{rv}
Adding radial velocities (RVs) to the astrometry allows us to
completely sample the motion of a binary system; in tandem, the two
data sets improve the accuracies of the final component masses.  RVs
have particular value for the longer-period systems lacking consistent
astrometric coverage of the entire orbit (\Gtt, \Gttf, \Gfs, and
\Gst). We have RVs for 11 system components out of the 30 for which we
present new masses.  The value of the RV data sets depends the
brightnesses and magnitude differences of components in each binary,
given in Table~\ref{tbl-ST} and Table~\ref{tbl-MMVMK}.  Our RV
measurements, listed in Table 3, are from two sources, described next.
RV orbit semi-amplitudes and systemic velocities for the 11 stars in 7
systems are given in Table~\ref{tbl-RV}.

\subsection{McDonald Cassegrain Echelle Spectrograph}
 
We obtained most RV data with the McDonald 2.1m Struve telescope and
Sandiford Cassegrain Echelle spectrograph \citep{McC93}, hereafter CE.
The CE delivers a dispersion equivalent to 2.5 \kms/pix ($R =
\lambda/\Delta\lambda = $60,000) with a wavelength range of 5500
$\leq$ $\lambda$ $\leq$ 6700 \AA~spread across 26 orders (apertures).
The McDonald data were collected during thirty-three observing runs
from 1995 to 2009 and reduced using the standard IRAF \citep{Tod93}
{\tt echelle} package tools, including the cross-correlation tool {\tt
  fxcorr}.  We visually inspected the resulting cross correlation
functions (CCF) to select the better of the 26 apertures, typically
using half for each binary.  Four systems (\Gfs, \Gsf, \Gsn, \Get) had
detectable double peaks in their CCF; these have $\Delta V$ = 1.59 to
3.27.  For these we used an IRAF script to average the lower-noise
CCF.  We then used a Gaussian multi-peak fitting routine in the
GUI-based commercial package IGOR\footnote{https://www.wavemetrics.com} to derive component $\Delta$RVs.  We
obtained the RVs for the three single peak CCF systems (\Gtt, \Gttf,
and \Gst) through a weighted average of the {\tt fxcorr} output
velocities.  For all systems, we averaged the velocities derived from
observations taken during the span of a few days (our typical
observing run lasted four nights) to form average absolute heliocentric velocities at
the epochs plotted in Sections~\ref{plot831} and \ref{ALL}.

\subsection{HET and Other Sources}

Some \Gst~velocities came from the Hobby-Eberly Telescope using the
Tull Spectrograph, as did our high-resolution, high signal-to-noise
spectrum of \Gst~used as a template for all cross correlations. We
identify a few other sources of radial velocities in the individual
object modeling notes below (Section~\ref{ALL}).  The HET became fully
functional about halfway through our CE campaign, so we decided to
remain with the McDonald 2.1m CE system for the entire datasets for
the six other systems with RVs.

\section{An Analysis of GJ\,831\,AB} \label{GET}

As a guideline for all 15 systems for which we determine masses, the
following sections provide details of the analysis path leading to the
determination of component masses for \Get. Section~\ref{ALL} provides
summary information for the other 14 systems, indicating where the
mass derivations differ from \Get.

\cite{Seg00} summarizes previous knowledge of the orbit and mass of this system.

\subsection{The \Get~Astrometric Reference Frame}  \label{AstRefs}

\begin{figure}
\includegraphics[width=6.5in]{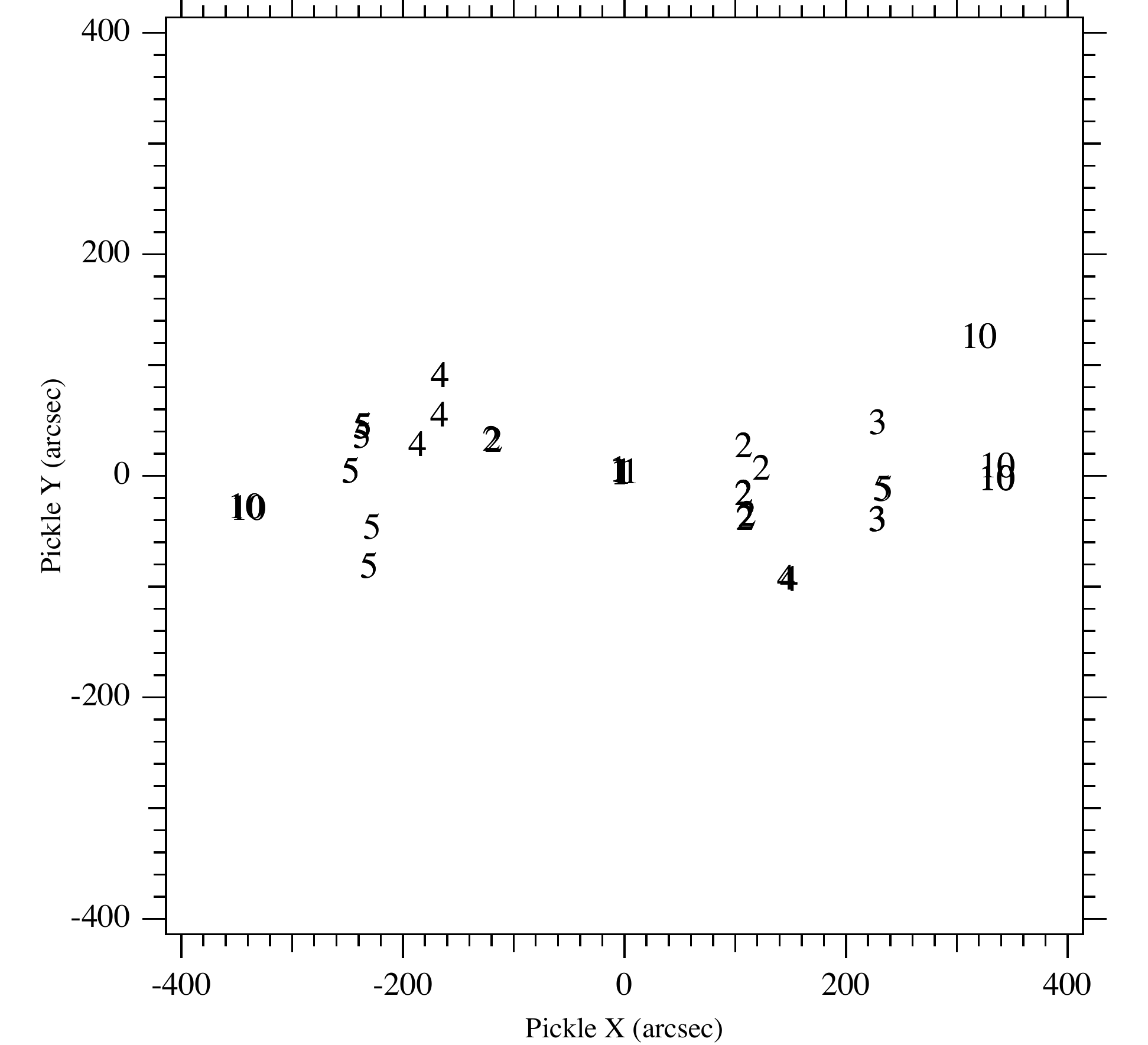}
\caption{Positions within FGS\,3 of GJ\,831\,AB and reference stars
  (ID numbers from Table~\ref{tbl-pis}) for all ten observation
  sets. The ``1" indicates the location of the GJ\,831\,AB
  photocenter. Reference stars 6--9, identified in the original \HST
  Phase 2 submission, were never observed, either due to faintness or
  placement within the FGS FOV.}
\label{fxy}
\end{figure}

Figure \ref{fxy} shows the distribution in FGS  coordinates of
the 10 sets of five reference star (numbers 2, 3, 4, 5, and 10) POS
mode measurements for the \Get ~reference frame, where 1 indicates
\Get.  The elongated pattern is impressed by the pickle-like shape of the FGS field of regard and
the requirement that {\it HST} roll to keep its solar panels fully
illuminated throughout the year.  Not all reference stars were
measured at each epoch, nor were all stars measured by the same FGS.
We acquired the first six POS data sets with FGS\,3 and the last four
sets with FGS\,1r.  We obtained a total of 71 reference star
observations and 29 observations of \Get.  We note in
Table~\ref{tbl-DSR} that we have more position angle and separation
TRANS mode measurements (18) over 6.61 yr than we have for component A
POS mode measurements (10) over 4.07 yr.

\subsection{\Get~Radial Velocities}

We have a single source of RV data for this system, including 13
measurement epochs spanning 11.9 years (Table~\ref{tbl-RVs}) obtained with
the McDonald 2.1m telescope and Sandiford Cassegrain Echelle
spectrograph \citep{McC93} (hereafter, CE). 
Treating \Get~as a double lined
spectroscopic binary, we obtained velocities for both components at
orbital phases $0.3\leq \phi \leq 0.98$ (Figure~\ref{G831},
right). Unfortunately, our monitoring missed a critical phase,
periastron.  \

\subsection{Prior Knowledge and Modeling Constraints} \label{MODCON}
\subsubsection{Reference Stars}

As for of our previous parallax projects, e.g., \cite{Ben07,Ben09,
  Ben11, McA11} we include as much prior information as possible for
our modeling.  Quantities that inform the modeling include estimates
of reference star absolute parallaxes, various color indices for the
reference stars, proper motions from the PPMXL \citep{Roe10}, and FGS
lateral color calibrations (e.g., Benedict \etal 1999, section
3.4)\nocite{Ben99}.  In contrast to much of our previous parallax work
(e.g., Harrison \etal 1999)\nocite{Har99}, for this project we
estimate our reference star absolute parallaxes using only color and
proper motion information.

Because the parallaxes determined for the binary systems will be
measured with respect to reference frame stars that have their own
parallaxes, we must either apply a statistically derived correction
from relative to absolute parallax (van Altena, Lee, \& Hoffleit 1995)
\nocite{WvA95} or estimate the absolute parallaxes of the reference
frame stars.  We choose the latter methodology for the FGS reference
fields discussed here.  The colors and luminosity class of a star can
be used to estimate the absolute magnitude, $M_V$, and $V$-band
absorption, $A_V$.  The absolute parallax is then simply,

\beq
\pi_{abs} = 10^{-{(V-M_V+5-A_V)}\over5}
\eeq

\begin{figure}
\includegraphics[width=6.5in]{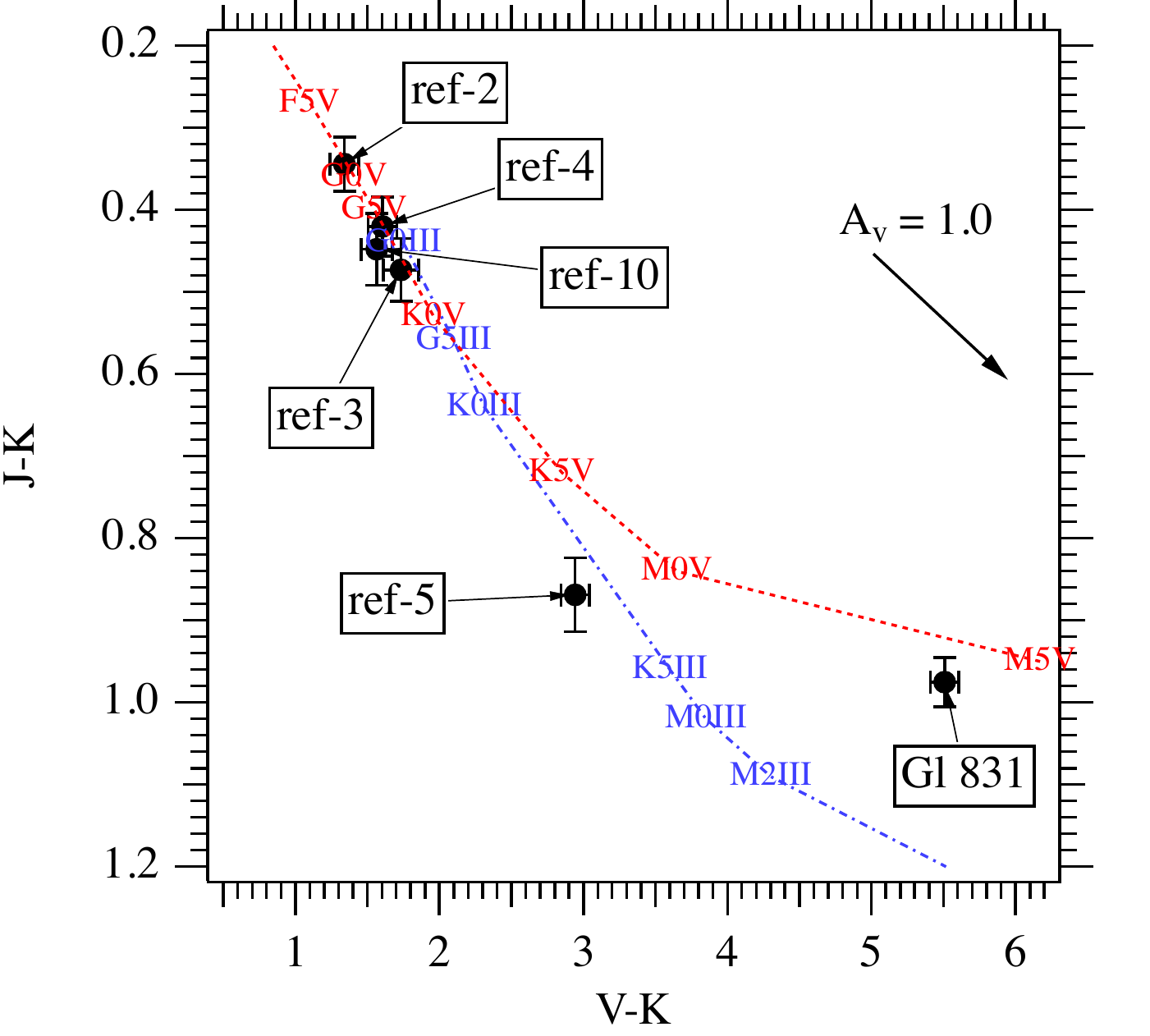}
\caption{$J-K$ vs. $V-K$ color-color diagram for \Get~and reference stars
  described in Table~\ref{tbl-pis}. The dashed line is the locus of
  dwarf (luminosity class V) stars of various spectral types; the
  dot-dashed line is for giants (luminosity class III) from
  \cite{Cox00}. The reddening vector indicates A$_V$=1.0 for the
  plotted color systems. For this field at Galactic latitude
  $\ell^{II}=-40\arcdeg$ we estimate $\langle A_V\rangle$ = 0.00 $\pm$
  0.06 magnitude with a maximum of 0.11 \citep{Sch11}.}
\label{CCD}
\end{figure}

Given the galactic latitude of \Get~($\ell^{II}=-40\arcdeg$) and
external estimates of low reddening \citep{Sch11}, a $J-K$ vs.~$V-K$
color-color diagram (Figure~\ref{CCD}) supports an initial assignment
of spectral type and dwarf luminosity class to all but reference star
ref-5 (with an initial estimate of K2III).

\begin{figure}
\includegraphics[width=5in]{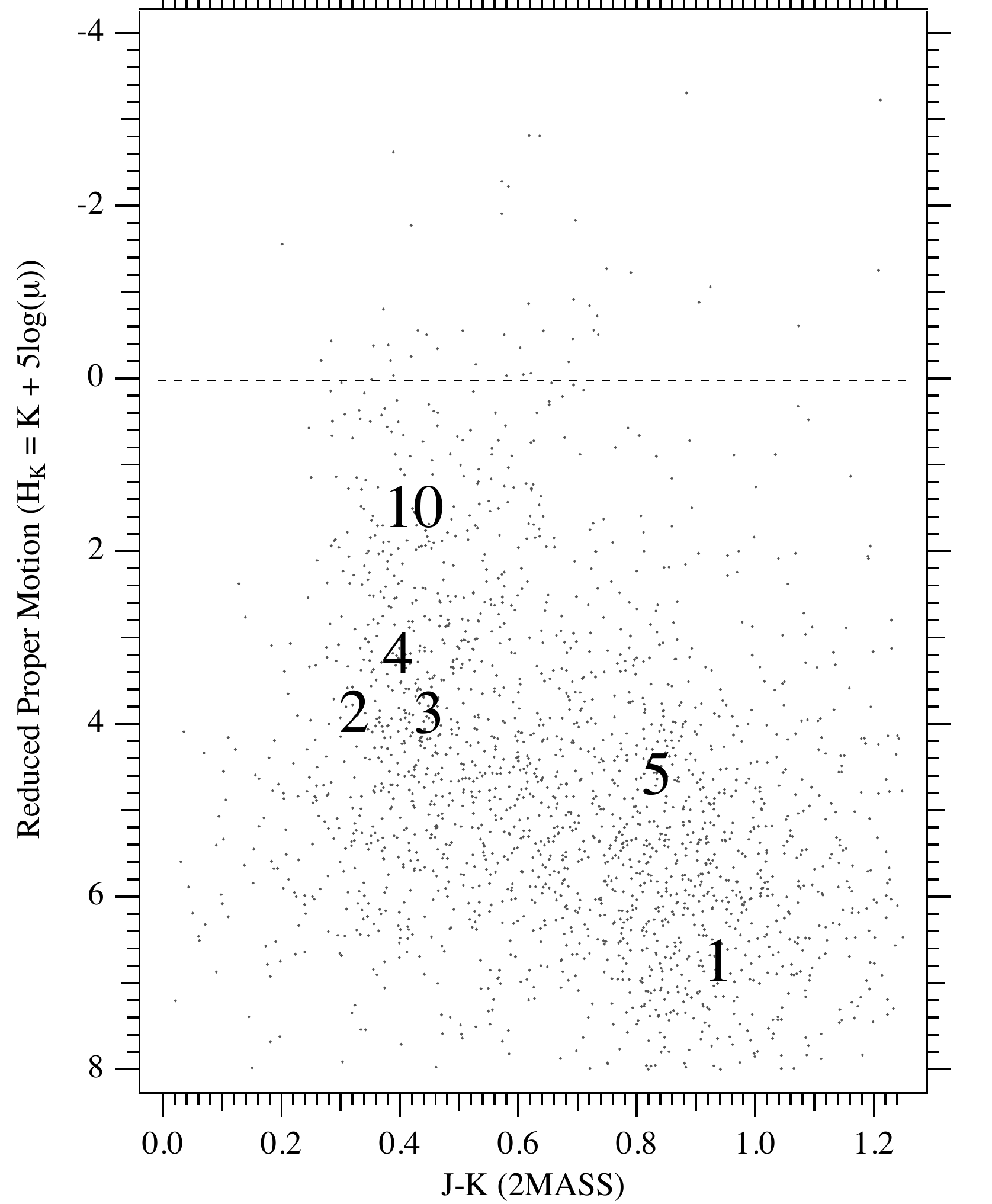}
\caption{Reduced proper motion diagram for 2378 stars taken from a
  ${2}$\arcdeg $\times$ ${2}$\arcdeg~field centered on \Get. Star
  identifications are shown for astrometric reference stars
  2,3,4,5, and 10 in Table~\ref{tbl-pis}.  Star ``1" is \Get. H$_K$
  for all numbered stars is calculated using proper motions from the
  PPMXL and $K$ magnitude from 2MASS. See Section~\ref{MODCON} for a
  discussion of ref-5 and ref-10.} \label{Hk}
\end{figure}

To confirm the reference star spectral type and luminosity classes estimated  
from all available photometry we employ
the technique of reduced proper motions \citep{Str39,Gou03,Gou04}.  We
obtain proper motions from PPMXL and $J$ and $K$ photometry from 2MASS
for sources in a ${2}$\arcdeg $\times$ ${2}$\arcdeg~field centered on
\Get.  Figure~\ref{Hk} shows a reduced proper motion diagram (RPM),
$H_K$ = $K$ + 5log($\mu$) plotted against $J-K$ color index for that
field, where $\mu$ is the proper motion vector absolute value in arcsec yr$^{-1}$. 
 If all stars had the same transverse velocities,
Figure~\ref{Hk} would be equivalent to an HR diagram.  \Get ~and
associated reference stars are plotted with ID numbers from
Table~\ref{tbl-pis}.  Comparing with our past reduced proper motion
diagrams \citep{Ben11,Ben14X,McA11}, all but ref-5 and ref-10 lie on
or near the main sequence.

  
In a quasi-Bayesian approach we input all priors as observations with
associated errors, not as hardwired quantities known to infinite
precision.  Input proper motion values have typical errors of 4--6 mas
yr$^{-1}$ for each coordinate.  The lateral color and cross-filter
calibrations and the $B-V$ color indices are also treated as
observations with error.  Where there is tension between the
color-color and RPM diagrams, in this case for ref-5 and -10, we run the
models with two sets of inputs and adopt the classification that
produces the smaller $\chi^2$.  The best model results included ref-5
as an M dwarf and ref-10 as a G subgiant.  Table~\ref{tbl-pis}
contains $V$ magnitudes, colors, estimated spectral types, $M_V$,
input prior parallaxes (estimated from photometry and proper motion)
and final parallaxes (as final astrometric modeling results) for the
five reference stars used in the \Get~field.

\subsubsection{The M Dwarf Binaries} \label{PJC}

Parallax was not the primary goal of the FGS effort.  Consequently,
for the more reference star-poor fields and those systems with very
poor sampling of the parallactic ellipse we introduce previously
determined science target parallaxes as priors.  While not included as
a prior in the \Get~modeling, we will flag parallax prior inclusion in
the modeling notes for other systems in Section~\ref{ALL}.
Ultimately, the \HST/FGS do provide significantly improved parallaxes
in most cases.

The derived orbital solutions benefit from a relationship between the
two astrometric modes (POS and TRANS) and the RV measurements, which
together are enforced by the constraint \citep{Pou00}

\beq
\displaystyle{{\alpha_{\rm A}\,sin\,i \over \pi_{\rm abs}} = {P K_{\rm A} \sqrt{(1 -e^2)}\over2\pi\times4.7405}} 
\label{PJeq}
\eeq

\noindent Quantities derived only from astrometry (parallax,
$\pi_{abs}$, primary perturbation orbit size, $\alpha_{\rm A}$, and
inclination, $i$) are on the left, and quantities derivable from both
radial velocities and astrometry (the period, $P$ in years and eccentricity,
$e$), or radial velocities only (the RV amplitude for the primary,
$K_{\rm A}$ in \kms), are on the right.  An object traveling 4.7405 \kms~will
move 1 AU in 1 year. When radial velocities for both
components exist, we employ an additional constraint, substituting
$K_{\rm B}$ for $K_{\rm A}$ and the component B orbit size,
$\alpha_{\rm B}$={\it a}-$\alpha_{\rm A}$, for $\alpha_{\rm A}$, where
$a$ is the orbital semimajor axis.  Finally, given the simple orbital
mechanics of these binary systems, we constrain the longitudes of
periastron passage, $\omega_{\rm A}$ and $\omega_{\rm B}$ to differ by
180\arcdeg.

\subsection{The Astrometric Model}

From the astrometric data we determine the scale, rotation, and offset
``plate constants" relative to an arbitrarily adopted constraint epoch
(the so-called ``master plate") for each observation set. The
\Get~reference frame contains five stars, but only four were observed
at each epoch. Hence, we constrain the scales along X and Y to
equality and the two axes to orthogonality. The consequences of this
choice are minimal. For example, imposing these constraints on the
Barnard's Star astrometry discussed in \cite{Ben99} results in an
unchanged parallax and increases the error by 0.1 mas, compared to a
full 6 parameter model (substituting $D$ for $-B$ and $E$ for $A$ in
Equation 6, below).

Our reference frame model becomes, in terms of standard coordinates
\beq
        x'  =  x + lc_x(\it B-V) 
\eeq
\beq
        y'  =  y + lc_y(\it B-V) 
\eeq
\beq
\xi = Ax' + By' + C  - \mu_\alpha \Delta t  - P_\alpha\pi_\alpha - ORBIT_\alpha
\eeq
\beq
\eta = -Bx' + Ay' + F  - \mu_\delta \Delta t  - P_\delta\pi_\delta - ORBIT_\delta
\eeq

\noindent 
where $\it x$ and $\it y$ are the measured coordinates from {\it HST},
$\it lc_x$ and $\it lc_y$ are lateral color corrections (see section
3.4 of Benedict \etal 1999)\nocite{Ben99}, and $B-V$ represents the
color of each star, either from $SIMBAD$ or estimated from the
spectral types suggested by Figure~\ref{CCD}.  $A$ and $B$ are scale
and rotation plate constants, $C$ and $F$ are offsets, $\mu_\alpha$ and
$\mu_\delta$ are proper motions, $\Delta$$t$ is the epoch difference from
the mean epoch, $P_\alpha$ and $P_\delta$ are parallax factors, and
$\it \pi_\alpha$ and $\it \pi_\delta$ are the parallaxes in RA and DEC.  
$\xi$ and $\eta$ are 
relative positions that (once scale, rotation, parallax, the proper motions 
and the ORBIT are determined) should not change with time. All Equation 
5 and 6 subscripts are in RA and DEC because the master constraint plate was rolled
into the RA DEC coordinate system before the analysis. We obtain
the parallax factors from a JPL Earth orbit predictor (Standish
1990\nocite{Sta90}), upgraded to version DE405.  Orientation to the
sky for the master plate is obtained from ground-based astrometry from
the PPMXL \citep{Roe10} with uncertainties in the field orientation of
$\pm 0\fdg1$.  This orientation also enters the modeling as an
observation with error.  Note that because we switched FGS units on
\HST~during the sequence of \Get~observations, two different OFAD and lateral
color calibrations entered the modeling.  Finally, $ORBIT$ is an
offset term that is a function of the traditional astrometric and RV
orbital elements listed in Table~\ref{tbl-OE}.

For the astrometric solution we solve simultaneously for a position
within our reference frame, a parallax, and a proper motion for the
target \Get~and all reference stars.  In addition, for \Get~the
orbital period (P), the epoch of passage through periastron in
modified Julian days (T$_0$), the eccentricity (e), and the position
angle of the line of nodes ($\Omega$), are constrained to be equal in
the RV and two modes of astrometry. We also constrain the angle
($\omega$) in the plane of the true orbit between the line of nodes
and the major axis to differ by 180\arcdeg~for the component A and B
orbits. Only RV provides information with which to determine the
velocity half-amplitudes ($K_{\rm A}$, $K_{\rm B}$) and $\gamma$, the
systemic velocity. For systems with both radial velocities and
astrometry, the constraint described in Section~\ref{PJC} ties POS,
TRANS, and radial velocities together, yielding a single
self-consistent description of the binary system.

\subsection{Assessing Reference Frame Residuals}
\begin{figure}
\includegraphics[width=4in]{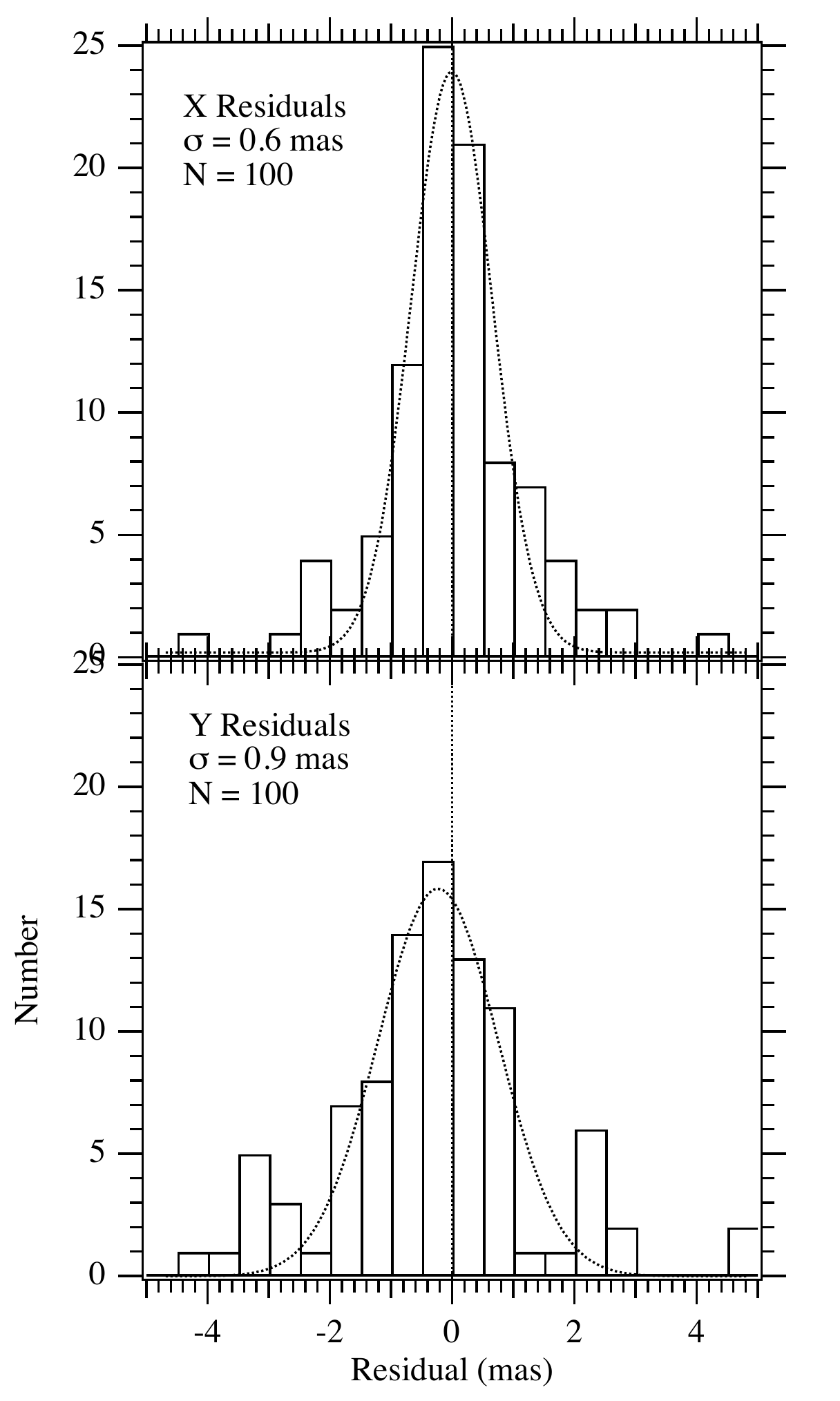}
\caption{Histograms of GJ\,831\,AB and reference star residuals
  resulting from the application of the model (Equations 5 and 6)
  incorporating both astrometry and radial velocities.}
\label{his}
\end{figure}

 From histograms of the astrometric residuals for 71 reference star and
29 \Get~position measurements (Figure~\ref{his}), we conclude that we
have obtained corrections at the $\sim 1$ mas level in the region
available at all {\it HST} roll angles (an inscribed circle centered
on the pickle-shaped FGS field of regard).  The resulting reference
frame ``catalog'' in $\xi$ and $\eta$ standard coordinates (Equations
5 and 6) was determined with median absolute errors of $\sigma_\xi=
0.6$ and $\sigma_\eta = 0.9$ mas in X and Y, respectively.

To determine if there might remain unmodeled --- but possibly
correctable --- systematic effects at the 1 mas level, we plotted the
\Get~reference frame X and Y residuals against a number of spacecraft,
instrumental, and astronomical parameters.  These included X, Y
positions within the pickle, radial distances from the pickle center,
reference star $V$ magnitudes and $B-V$ colors, \HST spacecraft roll
angles, and epochs of observation.  We saw no obvious trends, other
than an expected increase in positional uncertainty with reference
star magnitude.

\subsection{Results of \Get~Simultaneous Modeling} \label{plot831}

The results of this simultaneous solution are as follows.  Average
absolute value residuals for the RV and POS/TRANS observations of
\Get~are given in Tables \ref{tbl-RV} and \ref{tbl-DSR},
respectively.  The absolute parallax and the proper motion are
presented in Table~\ref{tbl-PPM} (errors are $1\sigma$), where
previous parallaxes are also listed.  Compared to {\it Hipparcos}, our
precision has improved knowledge of the parallax by a factor of 20.
Residuals to the orbit fits for each astrometric measurement are
individually listed in Table~\ref{tbl-TR}, while Table~\ref{tbl-OE}
contains the orbital parameters with formal ($1 \sigma$)
uncertainties.

\begin{figure}
\includegraphics[width=7in]{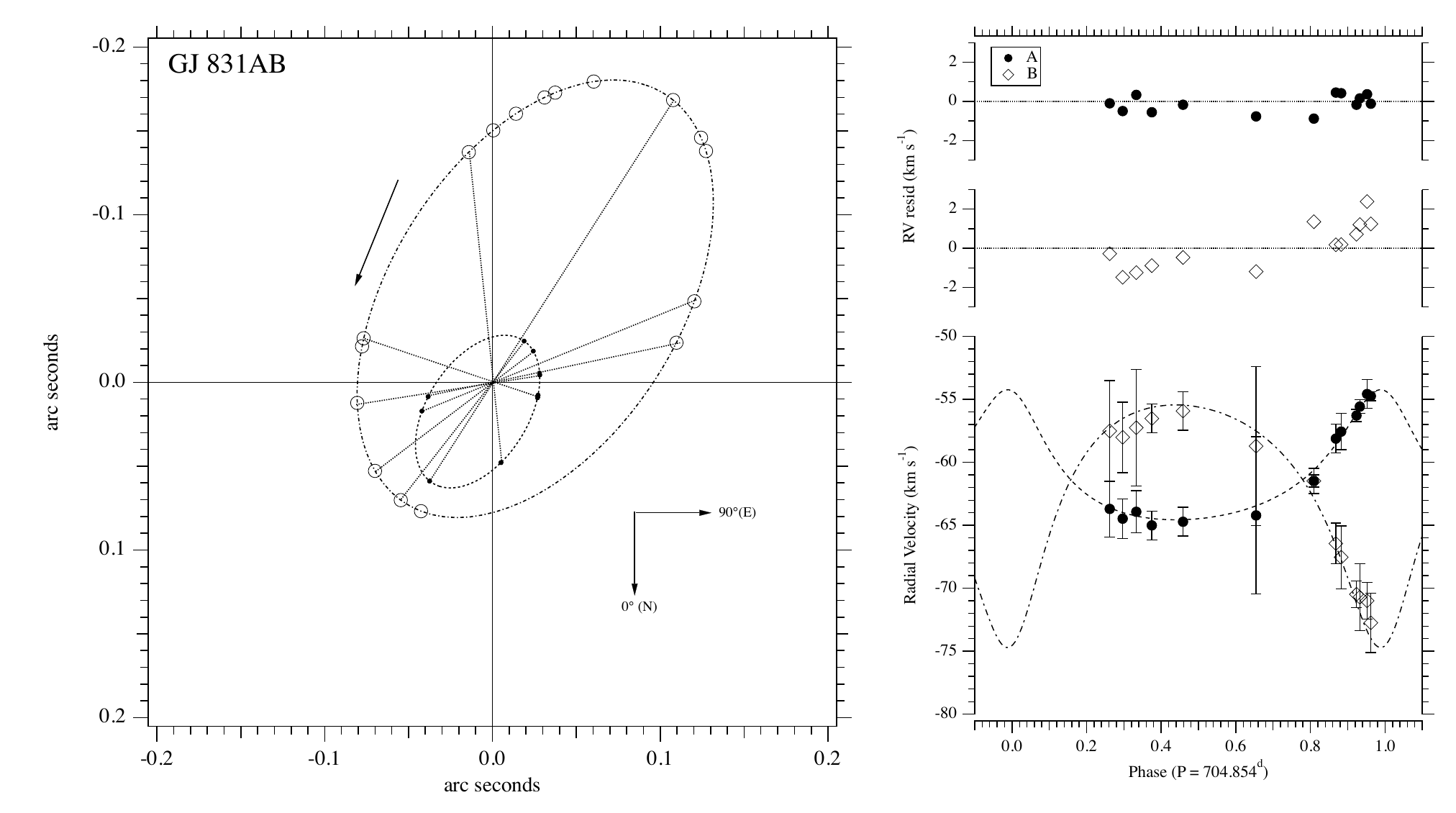}
\caption{Left: GJ\,831\,A (dots, POS orbit predicted positions) and
  component B (open circles, TRANS orbit predicted positions).  All
  observations, POS and TRANS and A and B component radial velocities,
  were used to derive the orbital elements listed in
  Table~\ref{tbl-OE}. POS and TRANS astrometric residuals (average
  absolute value for which listed in Table~\ref{tbl-DSR}) are smaller
  than the symbols. The POS mode points either side of p.a=90\arcdeg~each 
  represent two temporally close epochs. The arrow indicates the direction of orbital
  motion. Right: RV measurements (Table~\ref{tbl-RVs}) using CE on
  the McDonald 2.1m, phased to the orbital period determined from a
  combined solution including astrometry and RVs. The error bars are
  1-$\sigma$. The dashed lines are component velocities predicted from
  the orbital parameters derived in the combined solution. Middle and
  top panels illustrate component B and A RV residuals (average
  absolute value for which listed in Table~\ref{tbl-RV}) from the
  combined solution.  }
\label{G831}
\end{figure}

Figure~\ref{G831}, left, illustrates component A and B astrometric
orbits.  The POS mode component A residuals and the TRANS residuals
(component A--B separations) are all smaller than the dots in
Figure~\ref{G831}.  Figure~\ref{G831}, right, shows all RV
measurements, RV residuals, and the predicted velocity curve from the
simultaneous solution, phased to the derived orbital period.

\subsection{\Get~Component Masses}
Our orbit solution and derived absolute parallax (Equation~\ref{PJeq}) provide an orbital
semimajor axis, $a$ in AU, from which we can determine the system mass
through Kepler's Third Law.  Given P and $a$, we solve the expression

\beq
a^3 / P^2 =({\cal M}_{\rm A}+ {\cal M}_{\rm B})= {\cal M}_{tot}
\eeq

\noindent to find (in solar units) ${\cal M}_{tot} = 0.414 \pm 0.006{\cal M}_{\sun}$.
At each instant in the orbits of the two components around the common
center of mass,

\beq
{\cal M}_{\rm A} / {\cal M}_{\rm B} = \alpha_{\rm B} / \alpha_{\rm A}
\eeq

\noindent a relationship that contains only one observable,
$\alpha_{\rm A}$, the perturbation orbit size. Instead, we calculate
the mass fraction

\beq
f = {\cal M}_{\rm B}/({\cal M}_{\rm A}+{\cal M}_{\rm B}) = \alpha_{\rm A} / (\alpha_{\rm A} + \alpha_{\rm B}) = \alpha_{\rm A} /a,
\eeq

\noindent where $\alpha_{\rm B}$ = $a-\alpha_{\rm A}.$ This parameter,
f (also given in Table~\ref{tbl-OE}), ratios the two quantities
directly obtained from the observations --- the perturbation orbit
size ($\alpha_{\rm A}$ from POS mode) and the relative orbit size ($a$
from TRANS mode), both shown in Figure~\ref{G831} and listed in
Table~\ref{tbl-OE}. From these we derive a mass fraction of 0.3489
$\pm$ 0.0031.  Equations 7, 8, and 9 yield ${\cal M}_{\rm A} = 0.270
\pm 0.004{\cal M}_{\sun}$ and ${\cal M}_{\rm B} = 0.145 \pm 0.002
    {\cal M}_{\sun}$, indicating that both components are low mass red
    dwarfs with mass errors of $\sim$1.5\%, a considerable improvement over the
    \cite{Seg00} 4\% determinations.  We collect these component masses
    and those for the other systems discussed below in
    Table~\ref{tbl-MMVMK}, which also includes $V$, $K$, the magnitude
    differences $\Delta V$, $\Delta K$, and the $M_V$, and $M_K$ values used to create the MLRs.

\section{Modeling Notes and Masses for Other Systems} \label{ALL}

The presentation order in this section for the other 14 systems
observed with \HSTns/FGS\,3 and FGS\,1r is dictated by dwindling
observational resources.  We first discuss component mass results for
the six other systems with astrometric and RV measurements, then
provide notes on the derivation of component masses for the remaining
eight systems with astrometric measurements only. The final two
systems discussed analyze only position angle and separation
measurements (TRANS and/or ground-based).  Among all 15 systems, we
provide the first mass determinations for six binary pairs: \Gff,
\Gfs, \Get~(above), \Goz, \Gon, and \Gtf.  For historical context, we
provide comparisons with previous mass
determinations.

In parallel with our detailed discussion of \Get, RV measurements are
listed in Table~\ref{tbl-RVs} for the first six systems, with RV
results (component semi-amplitudes, system center of mass velocities,
number of observations, and residuals to the orbital fits) in
Table~\ref{tbl-RV}.  Astrometric results for all 15 systems are
collected in Table~\ref{tbl-DSR} (reference star information, study
durations, and average absolute value residuals to the orbital fits
for POS, TRANS), Table~\ref{tbl-PPM} (previous parallaxes and derived
\HST parallaxes and proper motions), and Table~\ref{tbl-TR}
(observation dates, individual position angles, separations, residuals
to the orbital fits, and measurement sources). All analyses yield the
orbital elements in Table~\ref{tbl-OE} and component masses in
Table~\ref{tbl-MMVMK}. All of these Tables list the
systems in Right Ascension order.

\subsection{\Gtt} \label{G22sec}

The \Gtt~system is a triple, with the two close components known as A
and C.  Our modeling of \Gtt~included the four sources of position angle and separation measurements
noted in Table~\ref{tbl-TR}.  \HST TRANS astrometric measurements came
from FGS\,3 and FGS\,1r, while modeling included POS from only FGS\,1r
in a reference frame of four stars.  We also utilized infrared speckle
interferometry measurements from \cite{McC91} and \cite{Woi03}.  All
but one of the POS mode epochs were secured at one year intervals,
very poorly sampling the parallactic ellipse.  In this situation,
simultaneous modeling including RVs and the Equation 2 constraint could
push the parallax to unrealistic values.    Therefore we
modeled radial velocities (only for GJ\,22\,A due to the large $\Delta
V=3.08$ value) in combination with the astrometry, with a parallax
from \HIP~\citep{Lee07b} used as a prior.

Our derived parallax has a larger error (0.6 mas) than typical for an
\HST parallax (0.2--0.3 mas) due to the less than ideal sampling.
Still, even though all epochs sampled nearly the same part of the
parallactic ellipse, they serve to scale the size of that ellipse well
enough to provide a useful parallax. We obtain a parallax similar to
\HIP~\citep{Lee07b}, but the precision has improved by a factor of
four.  The average absolute value of the residuals to the orbit fit
(Table~\ref{tbl-DSR}) for both the TRANS and speckle observations is
$\langle |res| \rangle$ = 10.1 mas.  For TRANS only, the average
absolute value residuals are $\langle |res| \rangle$ = 3.3 mas, and
for speckle-only, $\langle|res| \rangle$ = 13.4 mas.

\begin{figure}
\includegraphics[width=6.5in]{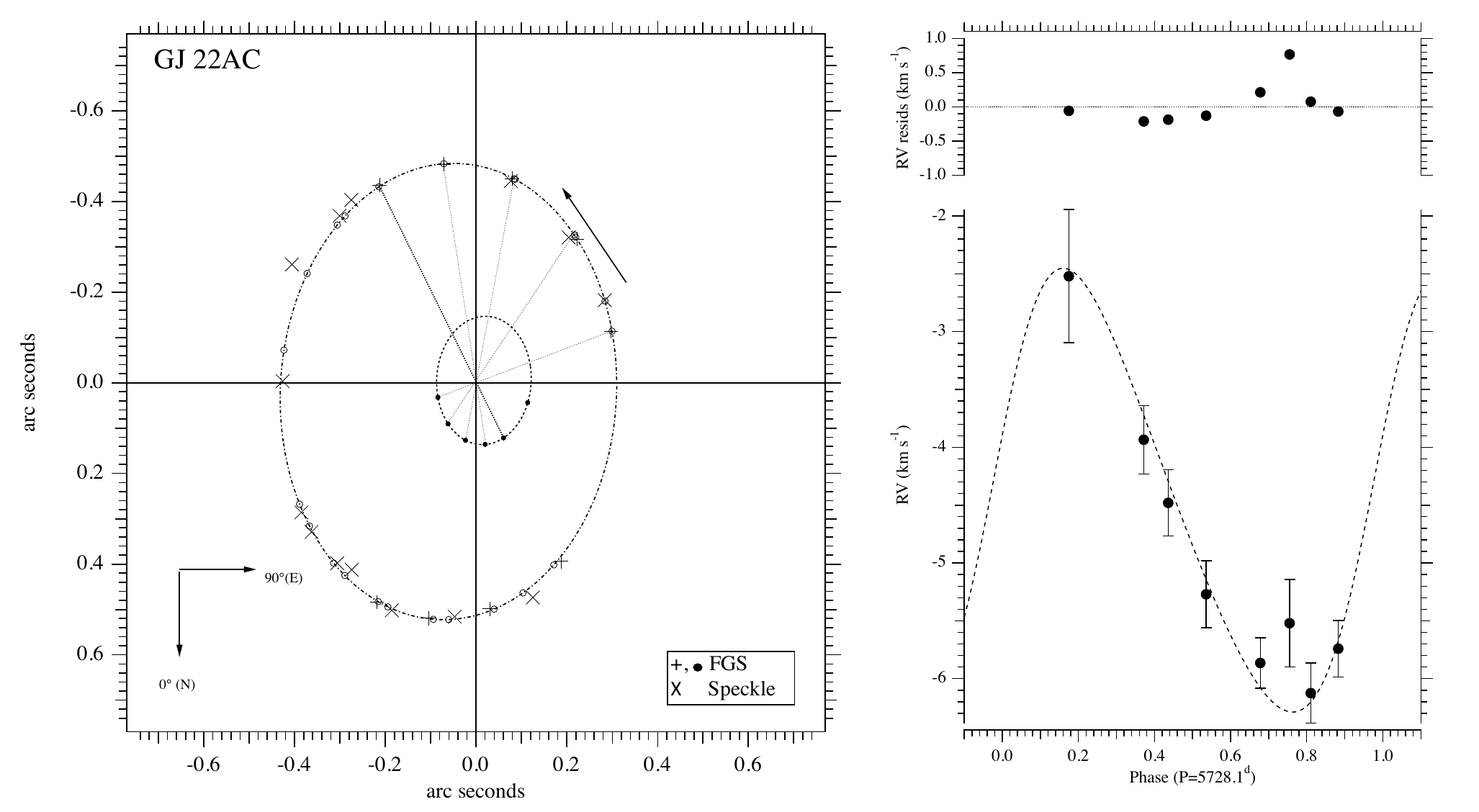}
\caption{Left: GJ\,22\,A (dots, POS orbit predicted positions) and
  component C (open circles) TRANS and speckle \citep{McC91,Woi03}
  orbit predicted positions.  All observations, POS, TRANS ($+$), and
  speckle ($\times$), and component A radial velocities were used to
  derive the orbital elements listed in Table~\ref{tbl-OE}. POS
  astrometric residuals (average absolute value for which listed in
  Table~\ref{tbl-DSR}) are smaller than the dot symbols. Right:
  component A RV measurements from the present study using CE on the
  McDonald 2.1m, phased to the orbital period determined from a
  combined solution including astrometry and RVs. The error bars are
  1-$\sigma$. The dashed line is the velocity predicted from the
  orbital parameters derived in the combined solution. Top panel:
  Component A RV residuals (average absolute values, listed in
  Table~\ref{tbl-RV}) from the combined solution.  }
\label{G22}
\end{figure}

\begin{figure}
\includegraphics[width=5in]{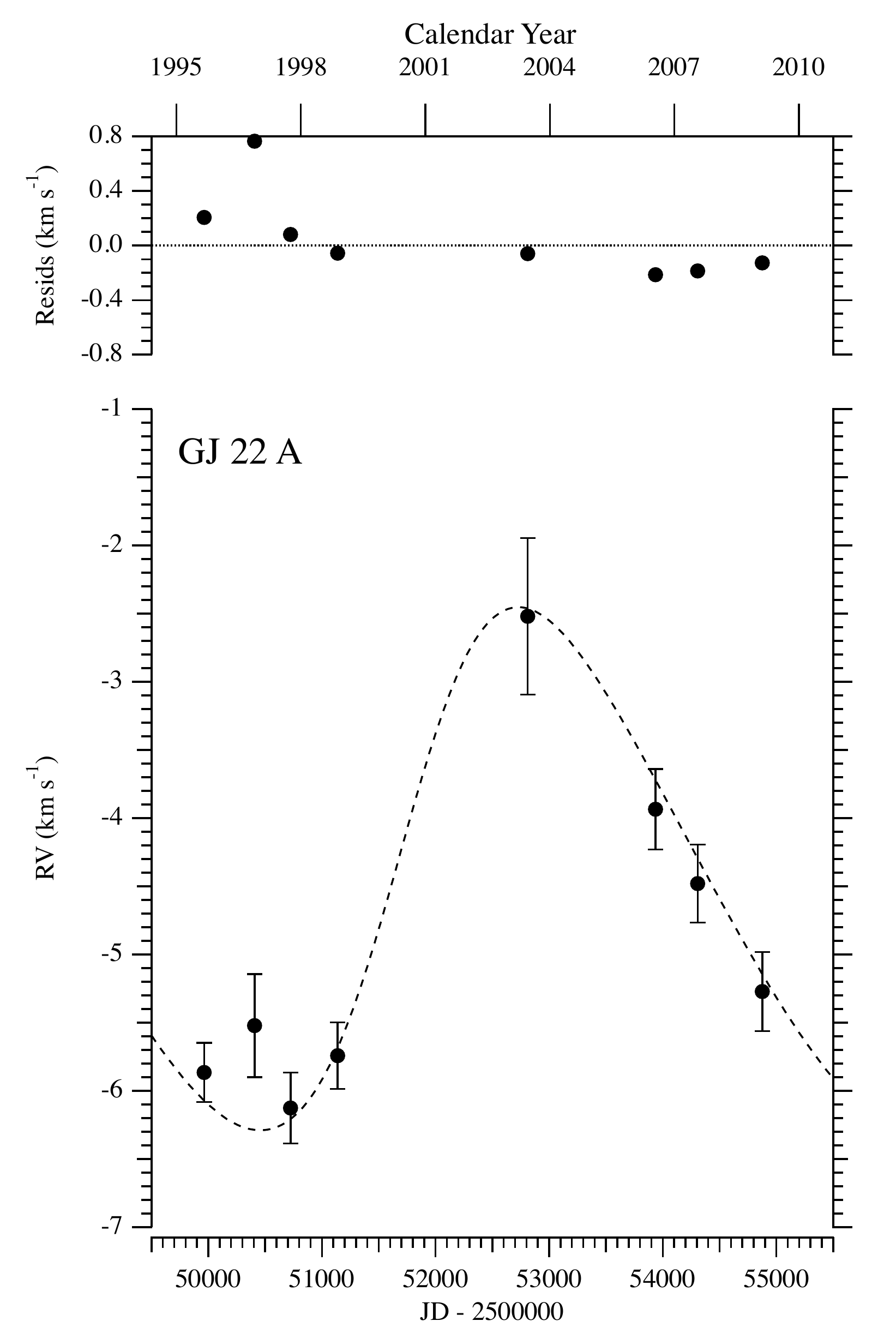}
\caption{GJ\,22\,A radial velocities used to derive the orbital
  elements listed in Table~\ref{tbl-OE} plotted against mJD rather
  than orbital phase.  \Gtt~has a companion, GJ\, 22\,B, separated by
  4", with a period, P $\simeq$ 320 yr. We may be detecting that motion as a
  slope in the RV residuals (top panel).  }
\label{G22y}
\end{figure}

Table~\ref{tbl-OE} contains the \Gtt~orbital parameters with formal
($1 \sigma$) uncertainties.  The orbital semimajor axis is now
extremely well-determined, with an error approaching 0.1\%.  The
orbital period is known to 0.3\%, while the parallax error (0.6\%)
dominates the derived mass errors.  The left panel of Figure~\ref{G22}
provides component A and B orbits and shows the observations that
entered into the modeling.  Component A--B separation residuals from
TRANS mode and speckle observations are illustrated by the offsets (+
and $\times$ next to o).  Regarding the RV measurements
(Figure~\ref{G22}, right), we note that \Gtt~has a companion, GJ\,
22\,B, separated from \Gtt~by 4\arcsec, with an orbital period on the
order of 320 yr \citep{Her73}.  Figure~\ref{G22y} shows the RV orbit
as a function of time, not phase. We may  (at a low level of significance) be detecting the AC--B
orbital motion as a slope in the RV residuals.  We obtain component
masses of ${\cal M}_{\rm A} = 0.405 \pm 0.008{\cal M}_{\sun}$ (2.0\%
error) and ${\cal M}_{\rm C} = 0.157 \pm 0.003 {\cal M}_{\sun}$
(1.9\%).  These masses support the excellent early work on this system
by \cite{Her73}, who found ${\cal M}_{\rm A} = 0.40{\cal M}_{\sun}$
and ${\cal M}_{\rm C} = 0.13{\cal M}_{\sun}$, followed by \cite{McC91}
who found ${\cal M}_{\rm A} = 0.362 \pm 0.053{\cal M}_{\sun}$ (15\%)
and ${\cal M}_{\rm C} = 0.123 \pm 0.018 {\cal M}_{\sun}$ (15\%).


\subsection{\Gttf} 

\begin{figure}
\includegraphics[width=6.5in]{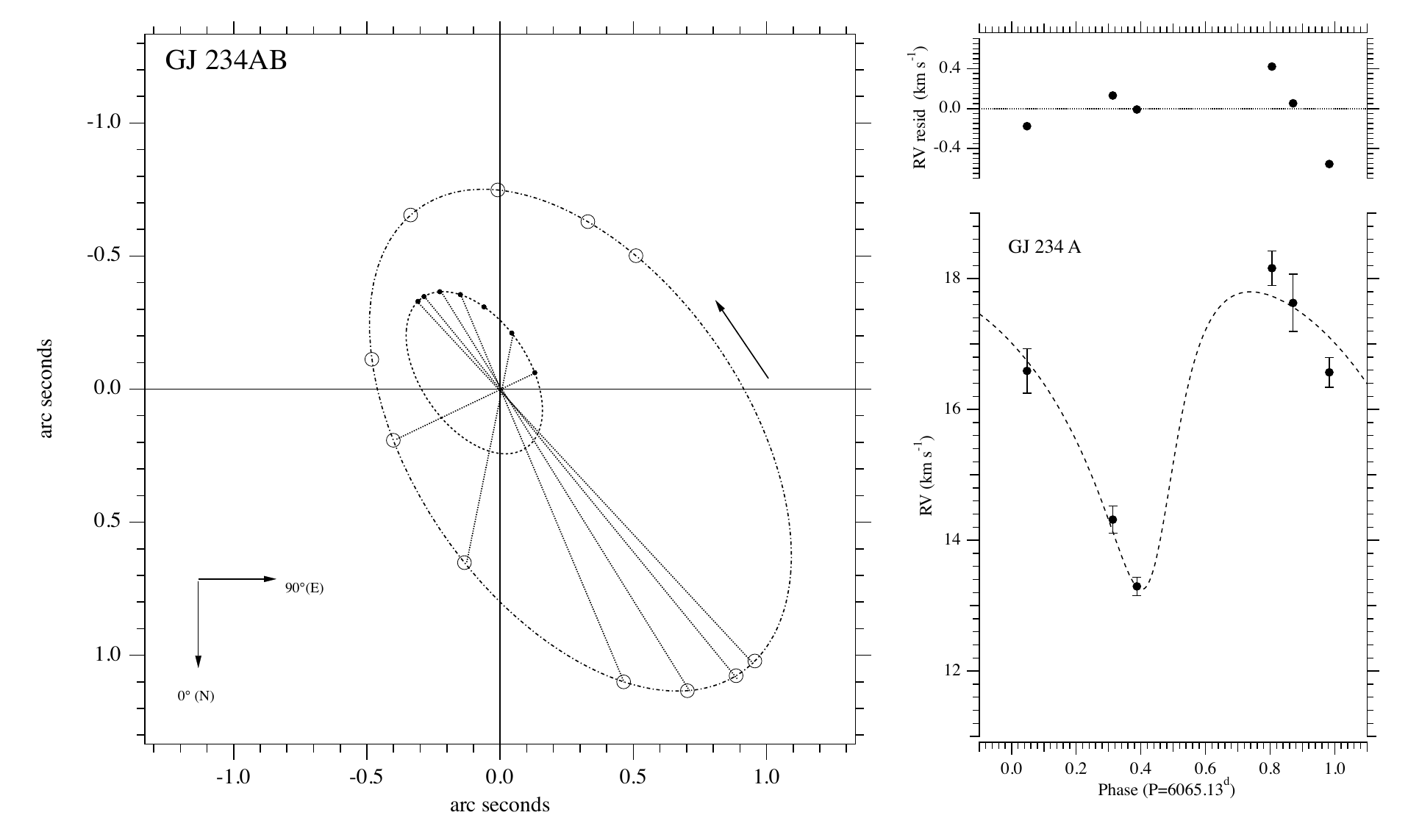}
\caption{Left: GJ\,234\,A (dots, POS orbit predicted positions) and
  component B (open circles) TRANS orbit predicted positions.  All
  observations, POS, TRANS, and component A radial velocities were
  used to derive the orbital elements listed in
  Table~\ref{tbl-OE}. POS, TRANS astrometric residuals are smaller
  than the symbols. Right: Component A RV measurements from the
  present study using CE on the McDonald 2.1m, phased to the orbital
  period determined from a combined solution including astrometry and
  RV. The error bars are 1-$\sigma$. The dashed line is the velocity
  predicted from the orbital parameters derived in the combined
  solution. Top panel: Component A RV residuals (average absolute
  values, listed in Table~\ref{tbl-RV}) from the combined solution.  }
\label{G234}
\end{figure}

Our modeling of \Gttf~included astrometry measurements from both
FGS\,3 and FGS\,1r. Treating \Gttf~as a single-lined spectroscopic
binary (due to the large $\Delta V$ = 3.08 value), we obtained
velocities from the CE for component A at most orbital phases, as seen
in Figure~\ref{G234}, right.  Modeling included RVs for the A component
and astrometry with five reference stars and only four plate
parameters.  Most of the POS mode observations were secured at one
year intervals, very poorly sampling the parallactic ellipse, with
only one POS measurement at the other extreme of the parallactic
ellipse, so we used a \HIPs parallax as a prior.  We obtained 11 TRANS
observations of the separation and position angle for AB and 7 POS
observations of component A and the reference frame.


Our observations have reduced the parallax error by a factor of 7--10
compared to YPC and {\it Hipparcos}.  Table~\ref{tbl-OE} contains the
\Gttf~orbital parameters with formal ($1 \sigma$) uncertainties, with
both $a$ and $P$ now known to better than 0.2\%.  The left panel of
Figure~\ref{G234} illustrates component A and B orbits with observed
positions indicated. The POS mode component A residuals and the TRANS
residuals are all smaller than the dots and circles.  We obtain masses
of ${\cal M}_{\rm A} = 0.223 \pm 0.002{\cal M}_{\sun}$ (0.9\% error)
and ${\cal M}_{\rm B} = 0.109 \pm 0.001 {\cal M}_{\sun}$ (0.9\%).
These are among the best mass measurements known for red dwarfs,
rivaling accuracies for components in eclipsing systems.  These masses
are larger than those reported by \cite{Pro77}, ${\cal M}_{\rm A} =
0.13 \pm 0.04{\cal M}_{\sun}$ (31\%) and ${\cal M}_{\rm B} = 0.07 \pm
0.02{\cal M}_{\sun}$ (29\%), who summarized much of the early work on
this important binary.  Our new masses are larger but consistent with
those of \cite{Cop94}, who found ${\cal M}_{\rm A} = 0.179 \pm
0.047{\cal M}_{\sun}$ (26\%) and ${\cal M}_{\rm B} = 0.083 \pm
0.023{\cal M}_{\sun}$ (28\%).

\subsection{\Gfs} 

\Gfs~astrometric measurements came from FGS\,3 and FGS\,1r and radial
velocities came from the CE.  Modeling was identical to that used for
\Get, although the astrometric reference frame included only three
stars. Hence, we included a \HIPs parallax \citep{Lee07b} as a prior.

\begin{figure}
\includegraphics[width=6.5in]{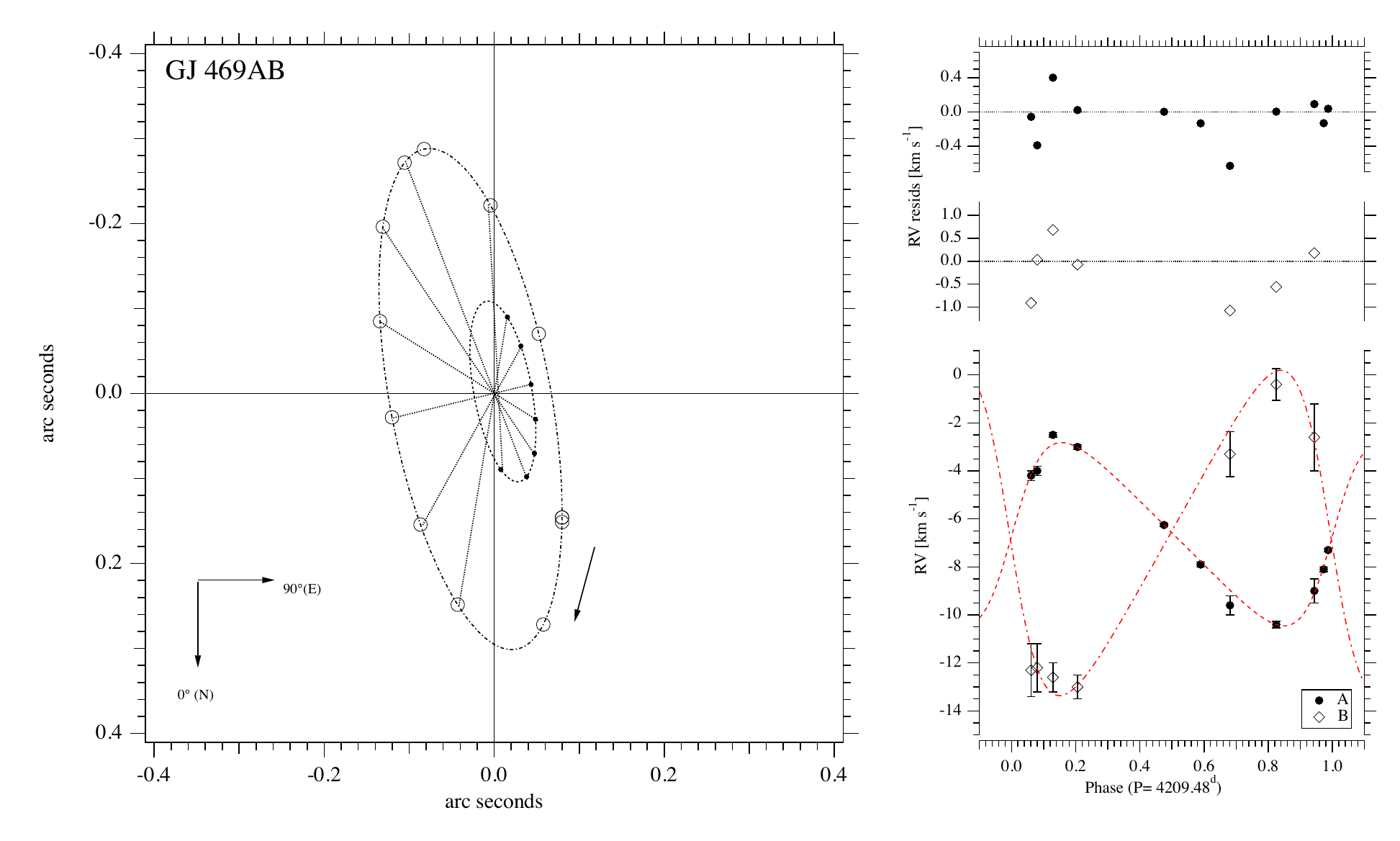}
\caption{Left: GJ\,469\,A (dots, POS orbit predicted positions) and
  component B (open circles, TRANS orbit predicted positions).  All
  observations, POS and TRANS and A and B component radial velocities,
  were used to derive the orbital elements listed in
  Table~\ref{tbl-OE}. POS and TRANS astrometric residuals (average
  absolute values listed in Table~\ref{tbl-DSR}) are smaller
  than the symbols. Right: RV measurements from the present study
  using CE on the McDonald 2.1m, phased to the orbital period
  determined from a combined solution including astrometry and RVs. The
  error bars are 1-$\sigma$. The dashed lines are velocities predicted
  from the orbital parameters derived in the combined solution. Middle
  and top panels: Component B and A RV residuals (average absolute
  values listed in Table~\ref{tbl-RV}) from the combined solution.  }
\label{G469}
\end{figure}

The FGS measurements have improved our knowledge of \Gfs's absolute
parallax of by a factor of 8--13 compared to YPC and {\it Hipparcos}.
Table~\ref{tbl-OE} contains the system's orbital parameters and formal
($1 \sigma$) uncertainties, with the largest error, 0.3\%, in the
semimajor axis.  Figure~\ref{G469}, left, illustrates component A and
B orbits, with effectively complete coverage of the orbit for the
TRANS observations.  The POS mode component A residuals and the TRANS
residuals (component A--B separations) are all smaller than the dots.
Figure~\ref{G469}, right, shows the RV measurements, residuals, and
velocity curves predicted from the simultaneous solution.  We obtain
component masses of ${\cal M}_{\rm A} = 0.332 \pm 0.007{\cal
  M}_{\sun}$ (2.4\% error) and ${\cal M}_{\rm B} = 0.188 \pm 0.004
{\cal M}_{\sun}$ (2.7\%) listed in Table~\ref{tbl-MMVMK}.  These are
the first mass determinations for this system.

\subsection{\Gst} \label{G62} 

\begin{figure}
\includegraphics[width=6.5in]{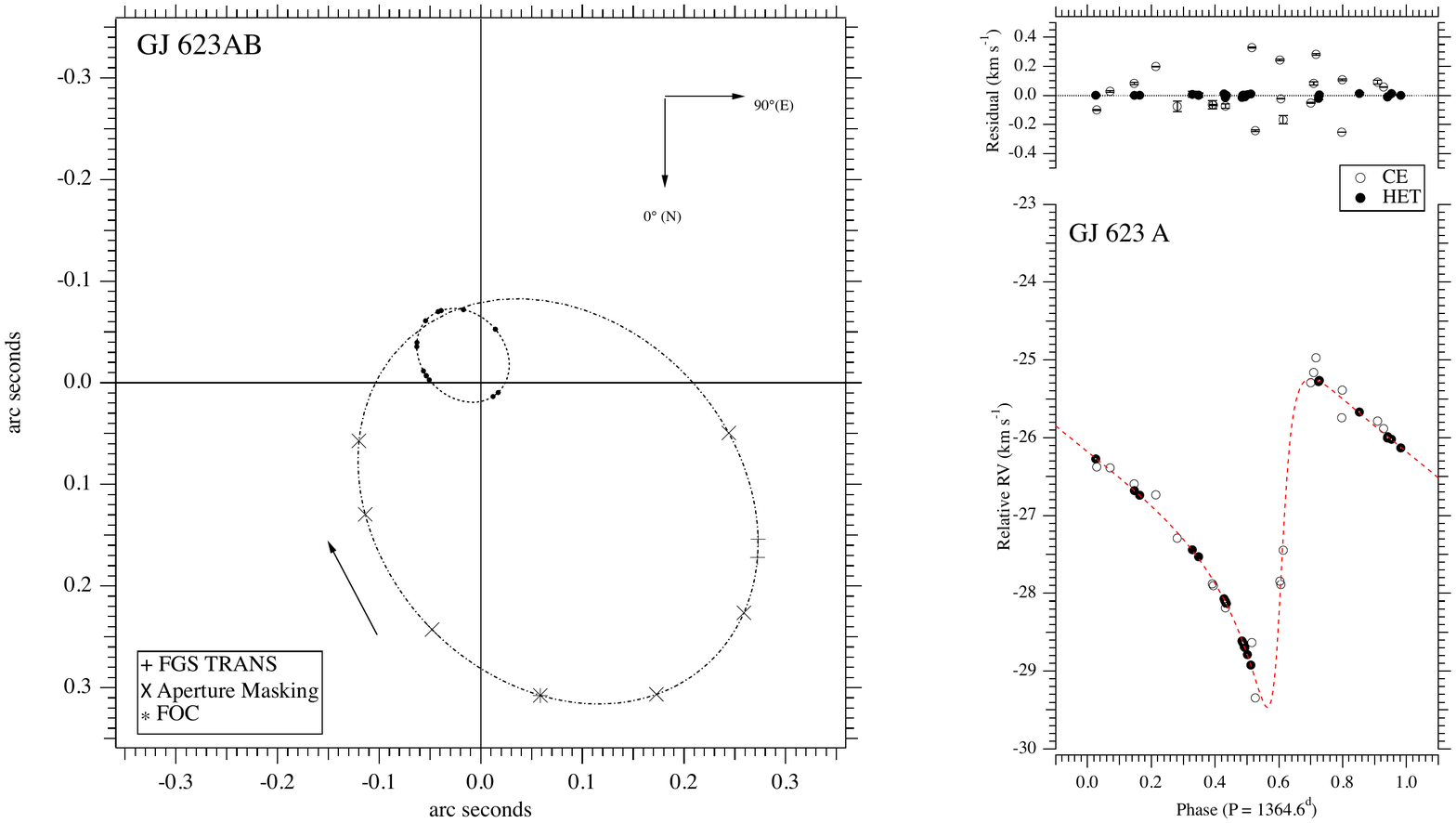}
\caption{Left: GJ\,623\,A (dots, POS orbit predicted positions) and
  component B (+) TRANS, FOC (*) \citep{Bar96}, and ($\times$)
  Aperture Masking \citep{Mar07} orbit predicted positions.  All
  observations, POS, TRANS, FOC, Aperture Masking, and component A
  radial velocities, were used to derive the orbital elements listed
  in Table~\ref{tbl-OE}. POS, TRANS, FOC, and Aperture Masking
  astrometric residuals (RMS for which listed in Table~\ref{tbl-DSR})
  are smaller than the symbols. Right: Component A RV measurements
  from the present study using CE on the
  McDonald 2.1m and the Tull Spectrograph on HET, phased to the orbital period
  determined from a combined solution including astrometry and RVs. The
  dashed lines are velocities predicted from the orbital parameters
  derived in the combined solution. Top panel: Component A RV
  residuals (average absolute values, CE and HET, listed in
  Table~\ref{tbl-RV}) from the combined solution. We plot 1-$\sigma$ error bars in this case on the residuals. They are smaller than the points in the lower panel.)  }
\label{G623}
\end{figure}

Our modeling of the challenging \Gst~system included three sources of
astrometry and two sources of RVs. With $\Delta V = 5.3$, \Gst~is an
extremely difficult system for \HST TRANS mode observations; our
campaign yielded only two usable measurements.  Fortunately, in
addition to the \HST POS and TRANS astrometric observations, there are
six aperture masking observations from \cite{Mar07} and a single
\HSTns/FOS observation \citep{Bar96} available that were included in
our modeling.  Figure~\ref{G623}, left, illustrates the orbital phase
coverage contributions of the three astrometric sources. The inclusion
of ground-based relative orbit observations greatly increased the
number of measurements of position angle and separation and the time
span over which to establish orbital elements.  Treating \Gst~as a
single lined spectroscopic binary, we obtained from the CE and from
the HET \citep{End06} velocities for component A at most orbital
phases, as shown in the right panel of Figure~\ref{G623}.  In addition
to including velocities only for component A, we modeled \Gst~(and 5 reference stars, all
observed at each of the 13 POS mode epochs of observation) with six
plate coefficients, replacing $-B$ with $D$ and $A$ with $E$ in
Equation 6, thereby allowing unequal scales along each axis. The
addition of two additional coefficients reduced the $\chi^2$ to
degrees of freedom ratio by 47\%, to near unity.

We have improved our knowledge of the parallax of \Gst~by a factor of
four compared to {\it Hipparcos}, and by over a factor of ten compared
to the YPC.  Table~\ref{tbl-OE} contains the \Gst~orbital parameters
with formal ($1 \sigma$) uncertainties, with the semimajor axis now
known to 0.6\%.  We obtain component masses of ${\cal M}_{\rm A} =
0.379 \pm 0.007{\cal M}_{\sun}$ (1.8\% error) and ${\cal M}_{\rm B} =
0.114 \pm 0.002 {\cal M}_{\sun}$ (1.8\%).  We have improved upon the
previous results of \cite{Mar07}, who found ${\cal M}_{\rm A} = 0.371
\pm 0.015{\cal M}_{\sun}$ (4.0\%) and ${\cal M}_{\rm B} = 0.115 \pm
0.002 {\cal M}_{\sun}$ (1.7\%), and the first masses measured by
McCarthy and Henry (1986), ${\cal M}_{\rm A} = 0.51 \pm 0.16{\cal
  M}_{\sun}$ (31\%) and ${\cal M}_{\rm B} = 0.11 \pm 0.029 {\cal
  M}_{\sun}$ (26\%), which provided a breakthrough in infrared speckle
imaging sensitivity at the time.

\subsection{\Gsf} 

The binary \Gsf~was the first system with a relative orbit determined
using the \HST FGSs \cite{Fra98}.  The system was subsequently
analyzed for mass determinations in \cite{Ben01}, and we revisit
\Gsf~here for several reasons.  We have an additional TRANS
observation obtained with FGS\,1r to add to the previous observations
secured with FGS\,3, extending our coverage by 12 years.  This reduces
the uncertainty in the orbital period modestly, from 0.5 to 0.3 days.
The OFAD has improved from that applied in 2000, and we now use all
reference star POS measurements; the previous study discarded any
reference star not observed at each epoch, while the present analysis
averages four reference stars per epoch.  The primary RV source
remains \cite{Ben01}, although we add a single new epoch.  Finally, we
now apply our quasi-Bayesian modeling technique to these data.

\begin{figure}
\includegraphics[width=6.5in]{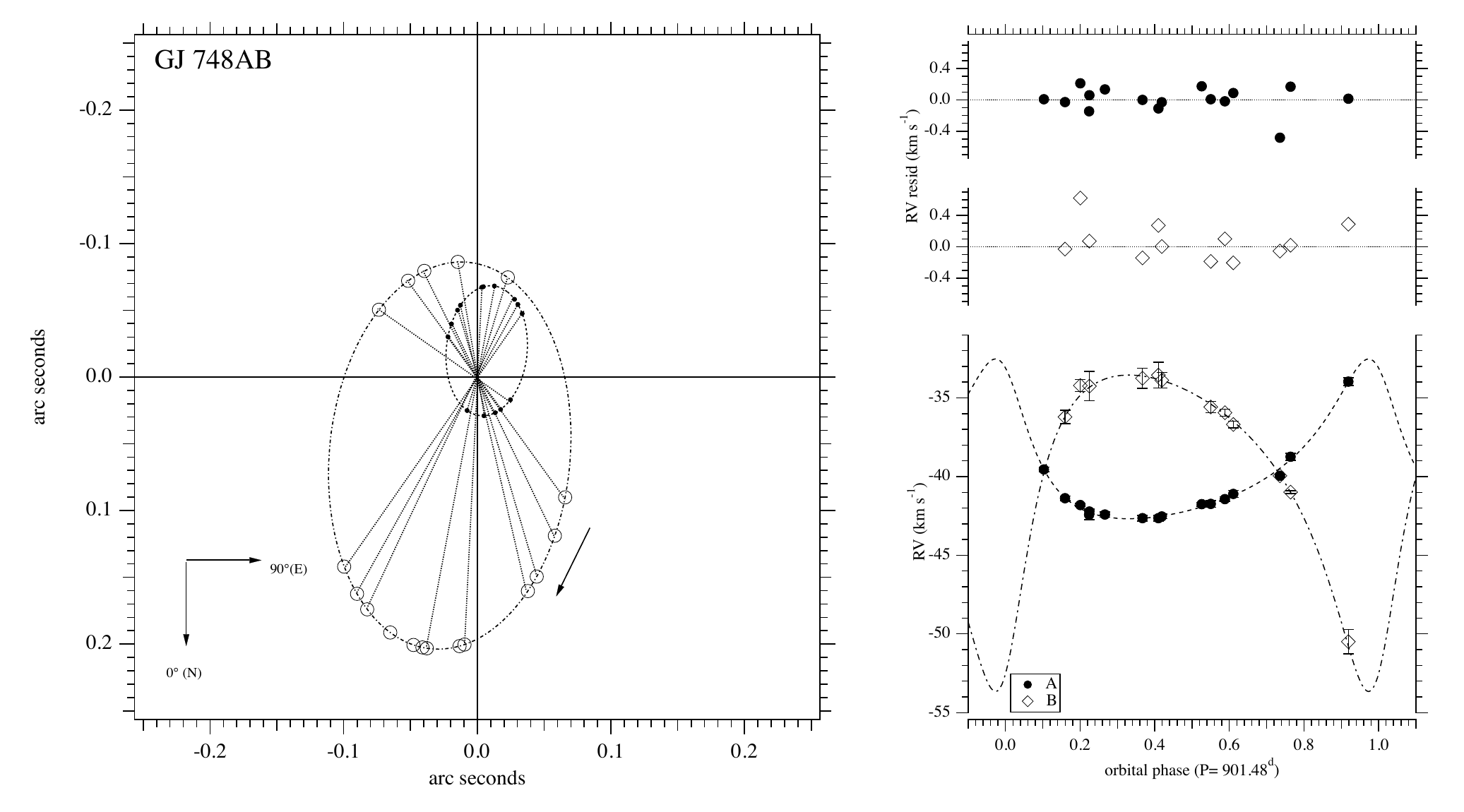}
\caption{Left: GJ\,748\,A (dots, POS orbit predicted positions) and
  component B (open circles, TRANS orbit predicted positions).  All
  observations, POS and TRANS and A and B component radial velocities,
  were used to derive the orbital elements listed in
  Table~\ref{tbl-OE}. POS and TRANS astrometric residuals (average
  absolute value for which listed in Table~\ref{tbl-DSR}) are smaller
  than the symbols. Right: RV measurements from the present study
  using CE on the McDonald 2.1m, phased to the orbital period
  determined from a combined solution including astrometry and RVs. The
  error bars are 1-$\sigma$. The dashed lines are velocities predicted
  from the orbital parameters derived in the combined solution. Middle
  and top panels: Component B and A RV residuals (average absolute
  value for which listed in Table~\ref{tbl-RV}) from the combined
  solution. The four component B velocities with errors larger than 0.5
  \kms~come from \cite{Mar89}.}
\label{G748}
\end{figure}

We have improved our knowledge of the parallax by a factor of 8--10
compared to YPC and \HIP.  We note that the revised proper motion now
agrees more closely with \HIP.  Table~\ref{tbl-OE} contains the
\Gsf~orbital parameters with formal ($1 \sigma$) uncertainties, and we
find that the parameters are consistent with the orbit in
\cite{Ben01}, as expected.  Figure~\ref{G748}, left, provides the
component A and B orbits, in which the POS mode component A residuals
and the TRANS residuals (component A--B separations) are all smaller
than the dots.  Figure~\ref{G748}, right, contains all RV
measurements, RV residuals, and velocity curves predicted from the
simultaneous solution.  We obtain component masses of ${\cal M}_{\rm
  A} = 0.369 \pm 0.005{\cal M}_{\sun}$ (1.3\% error) and ${\cal
  M}_{\rm B} = 0.190 \pm 0.003 {\cal M}_{\sun}$ (1.6\%). The previous
result \citep{Ben01} had slightly higher masses and similar errors:
${\cal M}_{\rm A} = 0.379 \pm 0.005{\cal M}_{\sun}$ and ${\cal M}_{\rm
  B} = 0.192 \pm 0.003 {\cal M}_{\sun}$.

\subsection{\Gsn} 

We also re-analyze \Gsn, previously investigated in \cite{Ben00b}. Our
motivations, similar to those for \Gsf, now include a set of
\Ha~radial velocities for both components.  With system total
magnitude $V=13.06$ and $\Delta V = 3.27$, detecting both components
of \Gsn~in both the RV data and with \HST/FGS is difficult.  In
addition, \Gsn~is a rapid rotator \citep{Del98}, further hindering RV
efforts.  However, both components exhibit strong \Ha~emission,
permitting RV determination for each component, although with less
precision (0.7 \kms) than for our other systems with RVs (typically 0.3
\kms).  We also now have two additional TRANS observations and an
additional epoch of POS measurements.  All but the last set of POS
astrometric measurements came from FGS\,3; the last is from FGS\,1r.
Modeling is exactly as for \Get.

\begin{figure}
\includegraphics[width=6.5in]{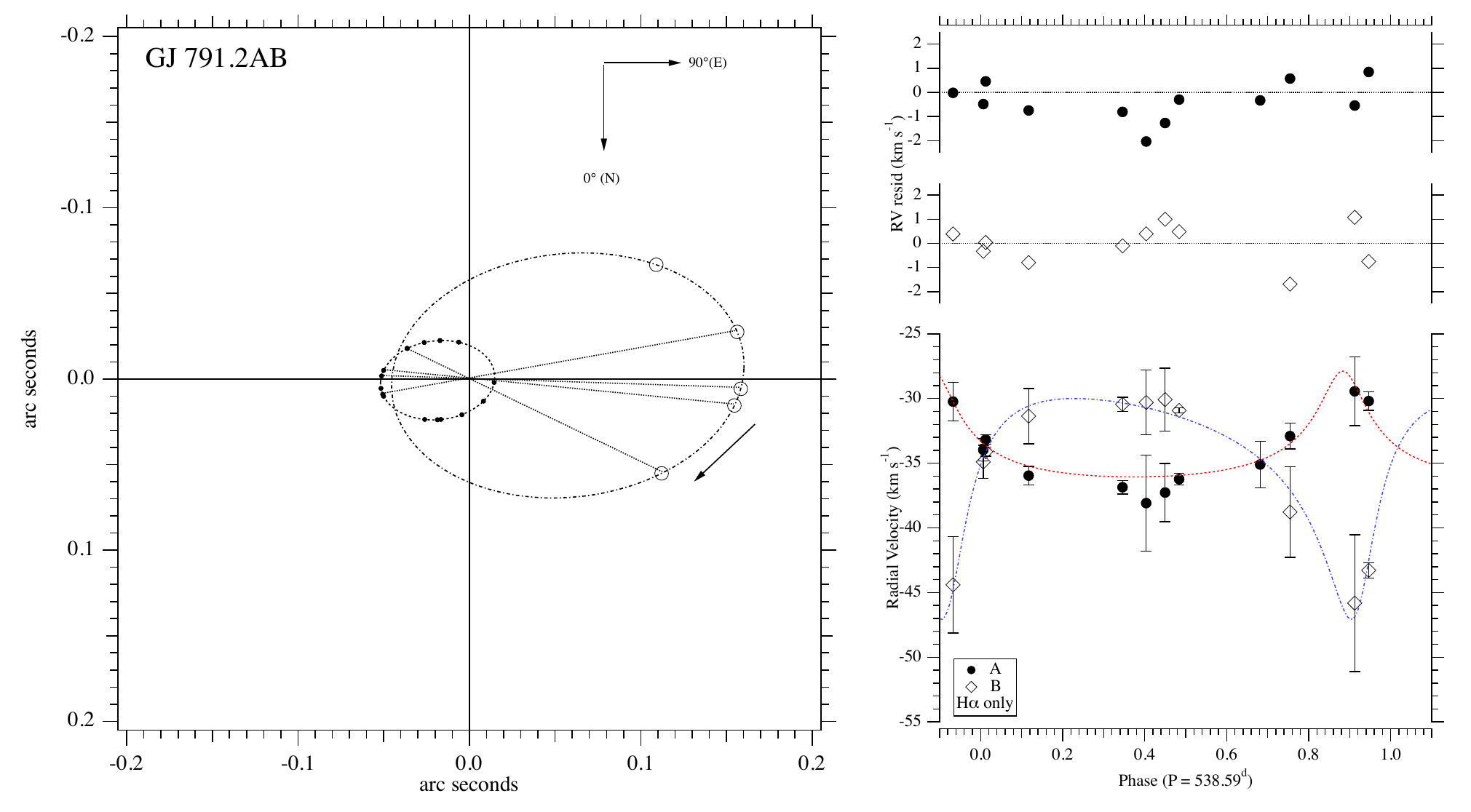}
\caption{Left: GJ\,791.2\,A (dots, POS orbit predicted positions) and
  component B (open circles, TRANS orbit predicted positions).  All
  observations, POS and TRANS and A and B component radial velocities
  (from H$\alpha$ emission lines only), were used to derive the
  orbital elements listed in Table~\ref{tbl-OE}. POS and TRANS
  astrometric residuals (average absolute value for which listed in
  Table~\ref{tbl-DSR}) are smaller than the symbols. Right: RV
  measurements of the \Ha~emission lines from the present study using
  CE on the McDonald 2.1m, phased to the orbital period determined
  from a combined solution including astrometry and RVs. The error bars
  are 1-$\sigma$. The dashed lines are velocities predicted from the
  orbital parameters derived in the combined solution. Middle and top
  panels: Component B and A RV residuals (average absolute value for
  which listed in Table~\ref{tbl-RV}) from the combined solution. Note
  that the less massive component B has stronger \Ha~emission,
  resulting in slightly smaller RV residuals for B (0.66 \kms) than
  for component A (0.70 \kms).}
\label{G791}
\end{figure}

This re-analysis has improved our knowledge of the parallax of \Gsn~by
30\% compared to our previous effort, and by a factor of ten versus
the YPC. The new analysis, incorporating priors rather than relying on
a correction to absolute derived from a Galactic model, has increased
the accuracy of the parallax.  Table~\ref{tbl-OE} contains the
\Gsn~orbital parameters with formal ($1 \sigma$) uncertainties with an
error of 0.5\% in the semimajor axis, which is 5 mas smaller than in
\cite{Ben00b}. The derived orbital period is
unchanged. Figure~\ref{G791}, left, provides the component A and B
orbits, in which the POS mode component A residuals and the TRANS
residuals (component A--B separations) are all smaller than the dots.
Figure~\ref{G791}, right, contains all \Ha~RV measurements, RV
residuals, and velocity curves predicted from the simultaneous
solution.  We obtain component masses of ${\cal M}_{\rm A} = 0.237 \pm
0.004{\cal M}_{\sun}$ (1.7\% error) and ${\cal M}_{\rm B} = 0.114 \pm
0.002 {\cal M}_{\sun}$ (1.8\%). The previous analysis \citep{Ben00b}
yielded significantly higher masses; ${\cal M}_{\rm A} = 0.286 \pm
0.006{\cal M}_{\sun}$ and ${\cal M}_{\rm B} = 0.126 \pm 0.003 {\cal
  M}_{\sun}$, which placed both components far from other stars in the
MLR. The new masses, due to a smaller orbital semimajor axis, bring
the components of GJ\,791.2 closer to the MLR defined by the ensemble
of system components (Section~\ref{mlr}, below).

\subsection{\Lst} 

We have chosen to reanalyze these data, originally collected and
analyzed by \cite{Her98}, with the hope that our Bayesian approach
might offer some small improvement. \cite{Her98} obtained component
masses of ${\cal M}_{\rm A} = 0.179 \pm 0.003{\cal M}_{\sun}$ and
${\cal M}_{\rm B} = 0.112 \pm 0.002 {\cal M}_{\sun}$, precision
difficult to improve upon.

We have no RVs for this system, nor for any of the remaining systems
discussed below.  All \HST astrometric observations, POS and TRANS,
were made using FGS\,3.  The parallactic ellipse is well-sampled, as
is the component A--B separation with 17 TRANS observations. The
component A orbit is somewhat less-well sampled, but better than, for
example, \Gtt.  The major difference in modeling, compared to \Get,
involves the number of plate coefficients.  The reference frame
consist of only two stars.  Consequently, we replace the coefficients
$A$ and $B$ in Equations 5 and 6 with $cos(A)$ and $sin(A)$, where A
is a rotation angle.  In essence we assume the scale given by our
improved OFAD, just as done by \cite{Her98} with an older OFAD.  The
average absolute value residuals for the POS and TRANS observations of
\Lst~given in Table~\ref{tbl-DSR} are $\sim$1 mas, indicating adequate
astrometry even with only three coefficients.

\begin{figure}
\includegraphics[width=6.5in]{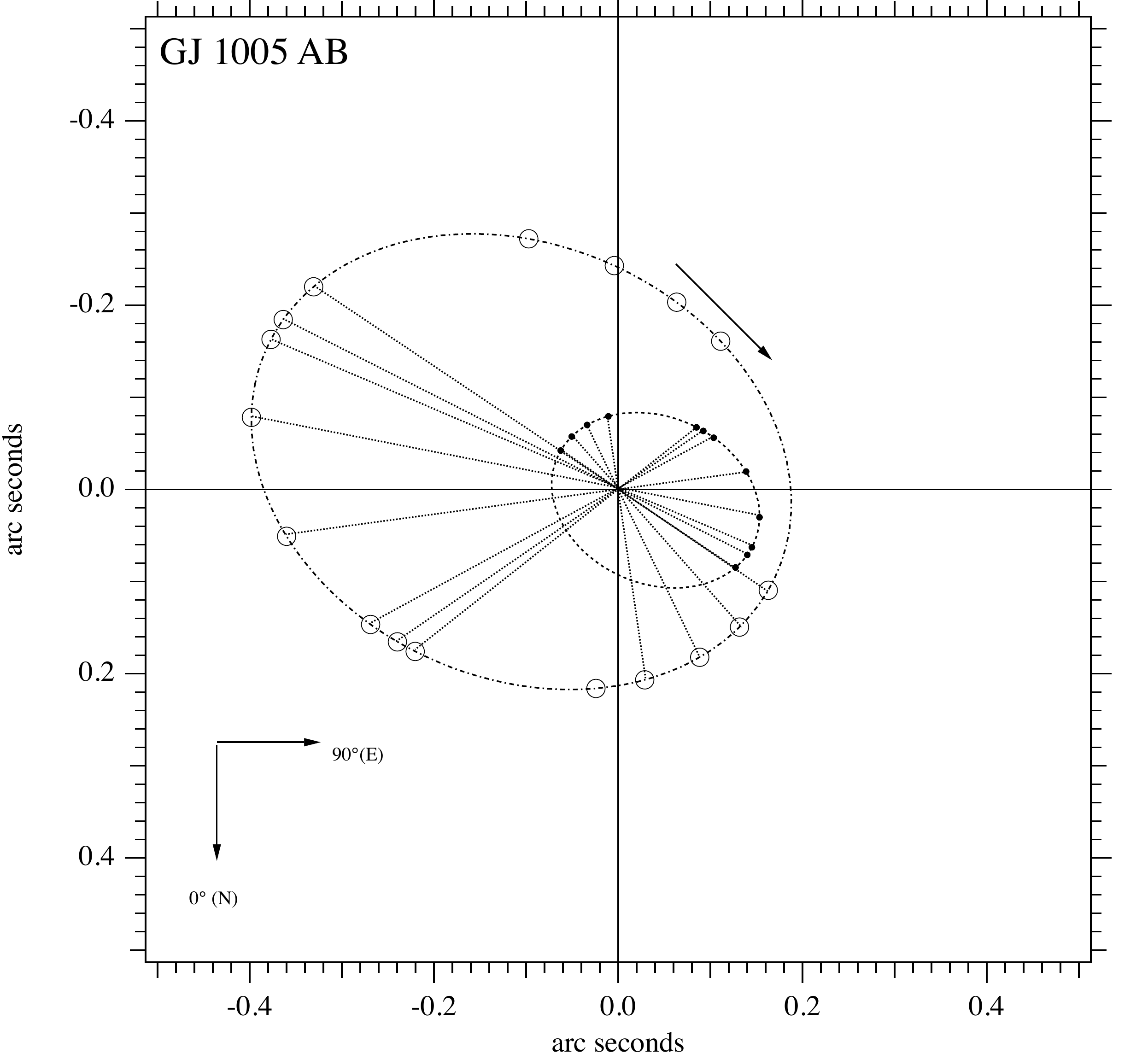}
\caption{GJ\,1005\,A (dots, POS orbit predicted positions) and
  component B (open circles) TRANS orbit predicted positions.  All
  observations were used to derive the orbital elements listed in
  Table~\ref{tbl-OE}. Component B TRANS and component A POS
  astrometric residuals (average absolute value for which listed in
  Table~\ref{tbl-DSR}) are smaller than their symbols.  }
\label{L722}
\end{figure}

Our parallax agrees almost perfectly with the \cite{Her98} value, but
our approach has improved the precision by a factor of four, and
represents a vast improvement by factors of $\sim$20--30 over the YPC
and \HIPs values.  Table~\ref{tbl-OE} contains the \Gozz~orbital
parameters with formal ($1 \sigma$) uncertainties, with both $a$ and
$P$ now known to 0.2\%.  Figure~\ref{L722} provides component A and B
orbits with observed positions indicated. The POS mode component A
residuals and the TRANS residuals (component A--B separations) are all
smaller than the dots and circles in Figure~\ref{L722}.  We obtain
masses of ${\cal M}_{\rm A} = 0.179 \pm 0.002{\cal M}_{\sun}$ (1.1\%
error) and ${\cal M}_{\rm B} = 0.112 \pm 0.001 {\cal M}_{\sun}$
(0.9\%), values that are virtually identical to and with errors
slightly smaller than reported in \cite{Her98}.

\subsection{\Got} 

\begin{figure}
\includegraphics[width=6.5in]{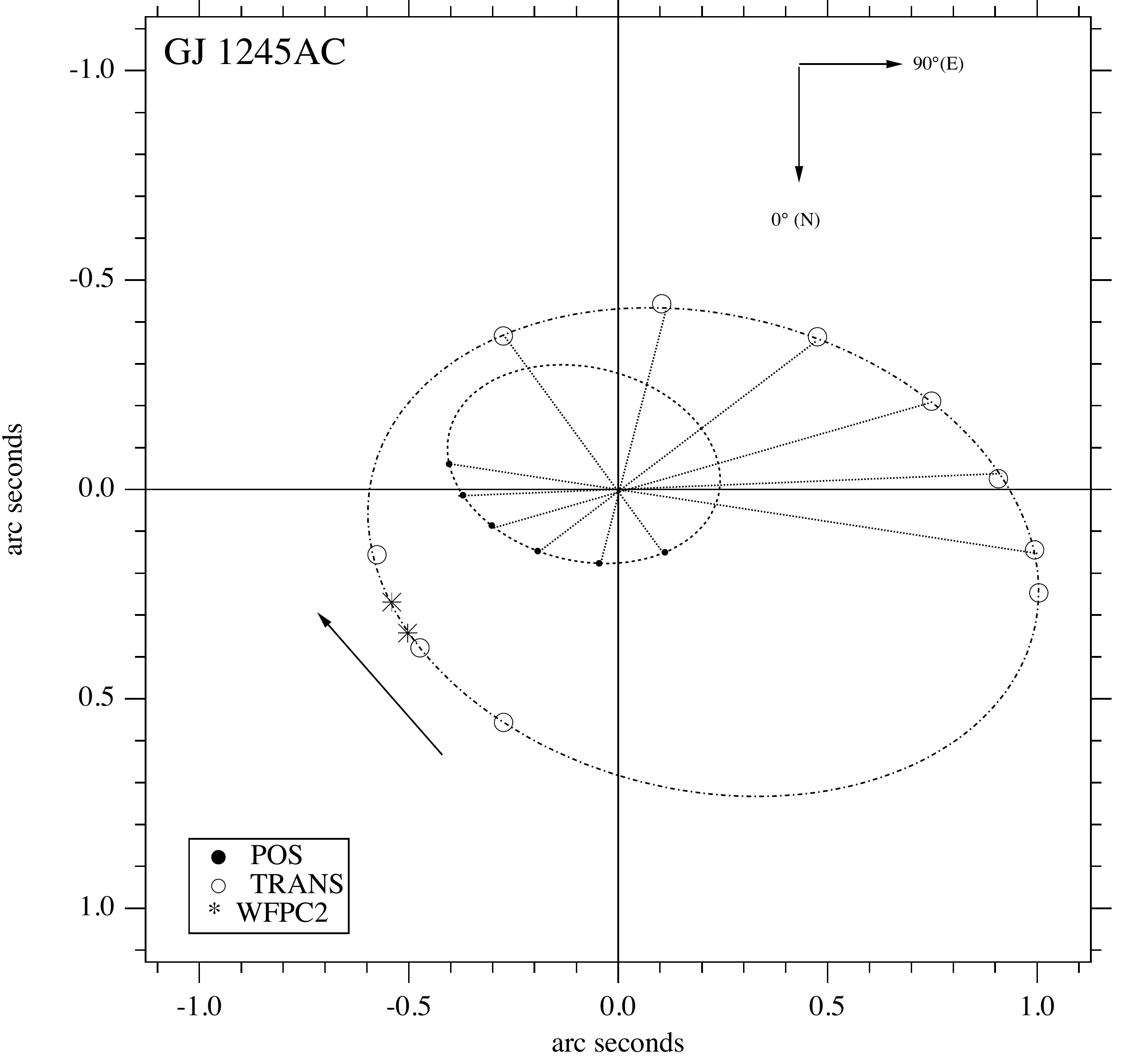}
\caption{GJ\,1245\,A (dots, POS orbit predicted positions); WFPC2 (*) and
  component C (open circles) TRANS orbit predicted positions.  All
  observations were used to derive the orbital elements listed in
  Table~\ref{tbl-OE}. Component C TRANS and component A POS
  astrometric residuals (average absolute value for which listed in
  Table~\ref{tbl-DSR}) are smaller than their symbols.  }
\label{G1245}
\end{figure}

The GJ\,1245 system is a triple, consisting of components A, B, and C,
with three components near the end of the stellar main sequence.  Our
modeling of \Got~included six \HST POS and 10 TRANS astrometric
measurements from FGS\,3 and FGS\,1r, and two observations from WFPC2
\citep{Sch98}. The POS mode epochs of observations included seven
reference stars and were secured at one year intervals, again, very
poorly sampling the parallactic ellipse, and sampling less than half
the component A orbit (Figure~\ref{G1245}). Consequently, we included
a YPC parallax as a prior. Note that the parallax error (0.5 mas,
Table~\ref{tbl-PPM}) is larger than for a typical \HST parallax due to
the less than ideal sampling.  Even though all epochs sampled almost
exactly the same part of the parallactic ellipse, they serve to scale
the size of that ellipse well enough to provide a parallax whose error
does not adversely affect the determination of the final masses.  In
our final solution, we obtain a parallax error that is half as large
at the YPC parallax used as a prior.

The 10 TRANS mode data points sample three-quarters of the relative
orbit and nicely map the component A--C separations and position
angles over time, as shown in Figure~\ref{G1245}.  The POS mode
component A residuals and the TRANS residuals (component A--C
separations) are all smaller than the dots and circles. We obtain
masses of ${\cal M}_{\rm A} = 0.111 \pm 0.001{\cal M}_{\sun}$ (0.9\%
error) and ${\cal M}_{\rm C} = 0.076 \pm 0.001 {\cal M}_{\sun}$
(1.3\%).  These masses are consistent with the first detailed study of
the system by \citet{McC88}, who
found ${\cal M}_{\rm A} = 0.14 \pm 0.03{\cal M}_{\sun}$ (21\%) and
${\cal M}_{\rm C} = 0.10 \pm 0.02 {\cal M}_{\sun}$ (20\%).

These are among the best red dwarf masses yet determined, which is
important because component C has the lowest mass of any object in
this study.  Its mass corresponds to $79.8 \pm 1.0~{\cal M}_{\rm
  Jup}$, placing it at the generally accepted main sequence hydrogen burning 
demarkation boundary of $\sim80~{\cal M}_{\rm Jup}$ (Dieterich et
al. 2014 and references therein)\nocite{Die14}. In addition, {\it
  Kepler} has collected extensive photometric data. Astrometry carried
out with these data \citep{Lur15} are consistent with our \Got~orbit
and may have detected AC--B motion.

\subsection{\Gtf} 

\begin{figure}
\includegraphics[width=6.5in]{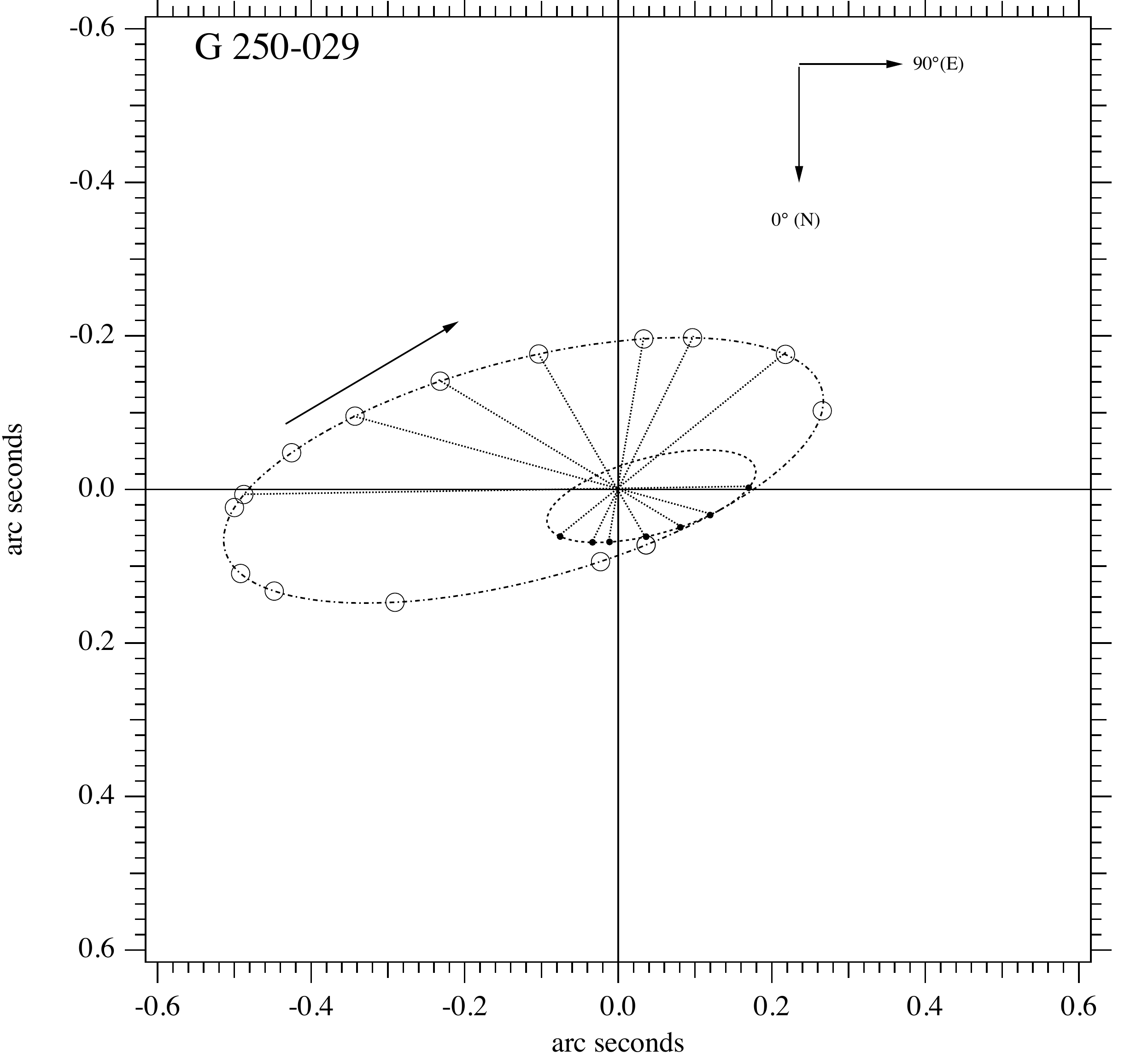}
\caption{G\,250-029\,A (dots, POS orbit predicted positions) and
  component B (open circles) TRANS orbit predicted positions. All
  observations were used to derive the orbital elements listed in
  Table~\ref{tbl-OE}. Component B TRANS and component A POS
  astrometric residuals (average absolute value for which listed in
  Table~\ref{tbl-DSR}) are smaller than their symbols.  }
\label{G250}
\end{figure}

All astrometric measurements came from \HSTns/FGS\,3 and FGS\,1r.  The
eight POS measurements contain only three reference stars, but sample
the parallactic ellipse better than for \Got, although not well-enough
to preclude the introduction of a lower-precision parallax prior from
\HIP~as a prior.  Fourteen TRANS observations of the component A--B
separations sample nearly the entire orbit (Figure~\ref{G250}).


Our parallax agrees with the {\it Hipparcos} value, but has reduced
the error by a factor of eight, and represents a factor of 16
improvement over the value in YPC.  Figure~\ref{G250} provides
component A and B orbits with observed positions indicated. Again, the
POS mode component A residuals and the TRANS residuals (component A--B
separations) are all smaller than the dots and circles in the
Figure. We obtain masses of ${\cal M}_{\rm A} = 0.350 \pm 0.005{\cal
  M}_{\sun}$ (1.7\% error) and ${\cal M}_{\rm B} = 0.187 \pm 0.004
{\cal M}_{\sun}$ (2.2\%).  These are the first mass determinations for
this system.

\subsection{\Goz} 

\begin{figure}
\includegraphics[width=6.5in]{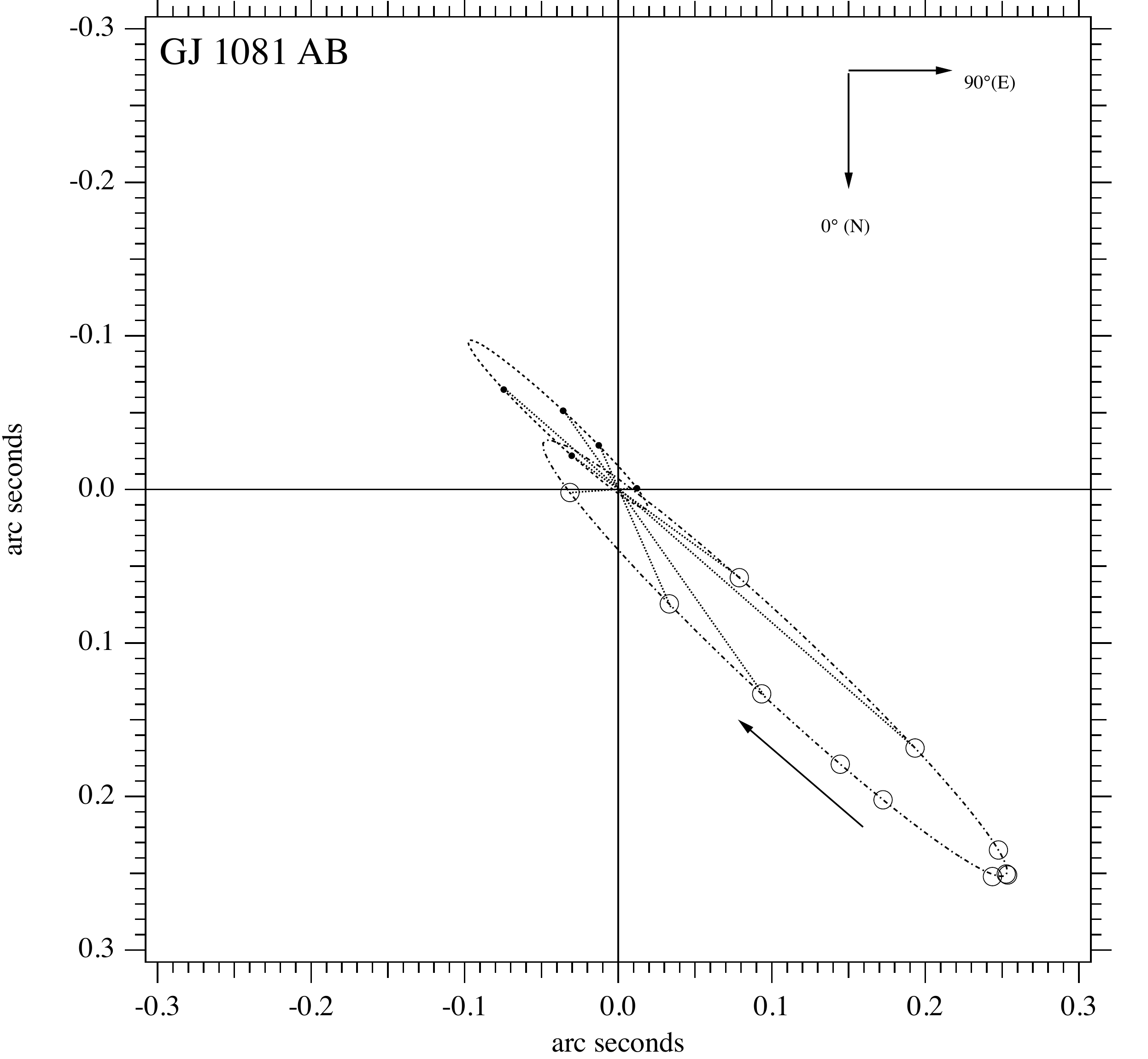}
\caption{GJ\,1081\,A (dots, POS orbit predicted positions) and
  component B (open circles) TRANS orbit predicted positions.  All
  observations were used to derive the orbital elements listed in
  Table~\ref{tbl-OE}. Component B TRANS and component A POS
  astrometric residuals (average absolute value for which listed in
  Table~\ref{tbl-DSR}) are smaller than their symbols.  }
\label{G1081}
\end{figure}

Our modeling of \Goz~included only \HST POS and TRANS astrometric
observations made using FGS\,1r.  The five epochs of POS mode
observations had six reference stars and were secured at one year
intervals, very poorly sampling the parallactic ellipse.  Hence, we
included a parallax from the YPC as a prior. The POS observations
sparsely sample the perturbation of the primary, while the 11 TRANS
observations nicely map the entire relative orbit
(Figure~\ref{G1081}).


Our parallax agrees with the YPC value, but our precision has reduced
the parallax error of \Goz~by a factor of four.  Figure~\ref{G1081}
provides component A and B orbits with observed positions indicated,
where the POS and TRANS residuals are smaller than the dots and
circles in the Figure. Due to the high orbital inclination and paucity
of POS observations, we obtain relatively poor mass precision for this
system. We find component masses of ${\cal M}_{\rm A} = 0.325 \pm
0.010{\cal M}_{\sun}$ (3.2\% error) and ${\cal M}_{\rm B} = 0.205 \pm
0.007 {\cal M}_{\sun}$ (3.4\%).  These are the first mass
determinations for this system.

\subsection{\Gff} 

\begin{figure}
\includegraphics[width=6.5in]{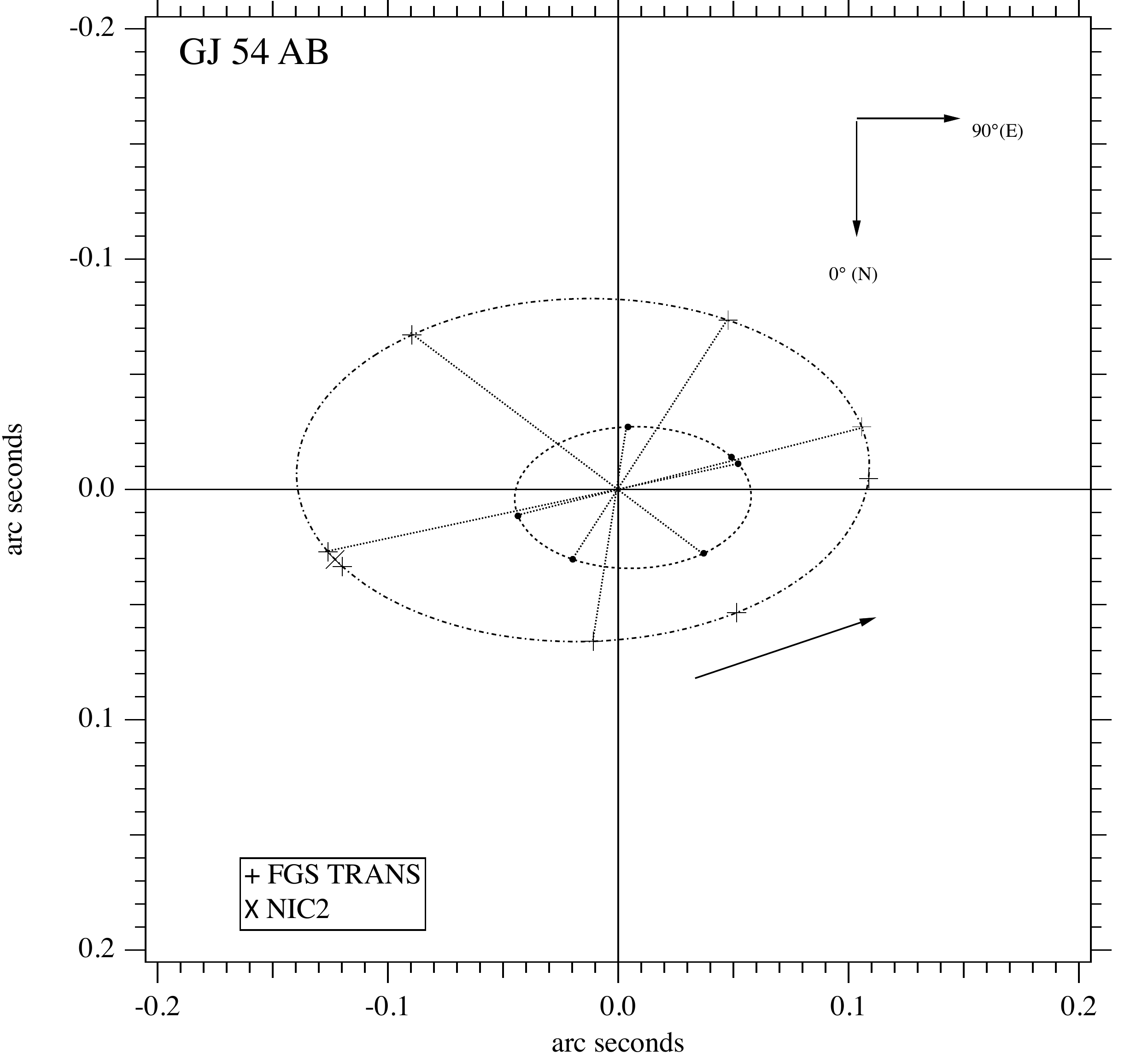}
\caption{GJ\,54\,A (dots, POS orbit predicted positions) and component
  B (+) TRANS orbit predicted positions.  The solution included one
  \HST NICMOS observation ($\times$). All observations were used to
  derive the orbital elements listed in Table~\ref{tbl-OE}. Component
  B (TRANS and NICMOS) and component A POS astrometric residuals
  (average absolute value for which listed in Table~\ref{tbl-DSR}) are
  smaller than their symbols.  }
\label{G54}
\end{figure}

\Gff~has the shortest orbital period in our sample, with a
 period of 1.15 yr.  Our modeling of \Gff~included \HST POS
(with three reference stars) and TRANS from FGS\,1r, and one NICMOS
astrometric measurement \citep{Gol04}.  Again, the timing of the six
epochs of POS mode observations sparsely sampled the parallactic
ellipse, so we included a \HIPs parallax prior.  The eight TRANS
observations sample the entire relative orbit (Figure~\ref{G54}), and
the NICMOS measurement falls squarely between two TRANS observations
on the orbit, although occurs nearly two years before the first
FGS\,1r observation.

Our derived parallax is larger than the \HIPs value and has an error
six times smaller.  \HIPs likely struggled to measure an accurate
value because the orbital period is so close to the 1.00 yr
periodicity of the tracing of the parallax ellipse --- without
knowledge of the resolved positions of the two stars in the binary, an
accurate parallax determination is a challenge.  Figure~\ref{G54}
provides the component A and B orbits with observed positions
indicated.  The model yields component masses of ${\cal M}_{\rm A} =
0.432 \pm 0.008{\cal M}_{\sun}$ (2.0\% error) and ${\cal M}_{\rm B} =
0.301 \pm 0.006 {\cal M}_{\sun}$ (2.0\%). These are the first mass
determinations for this system.

\subsection{\Gon} 

Our modeling of \Gon~included \HST POS and TRANS from FGS\,1r, one AO
observation \citep{Beuz04}, and one NICMOS astrometric measurement
\citep{Gol04}.  The POS mode observations included five reference
stars and most of the six epochs were secured at one year intervals,
very poorly sampling the parallactic ellipse.  Hence, we initially
included a parallax from \cite{Khr10} as a prior, although that
parallax had no correction from relative to absolute parallax.
Ultimately, the $\chi^2$ markedly decreased when we removed that
low-precision and perhaps inaccurate prior.

\begin{figure}
\includegraphics[width=6.5in]{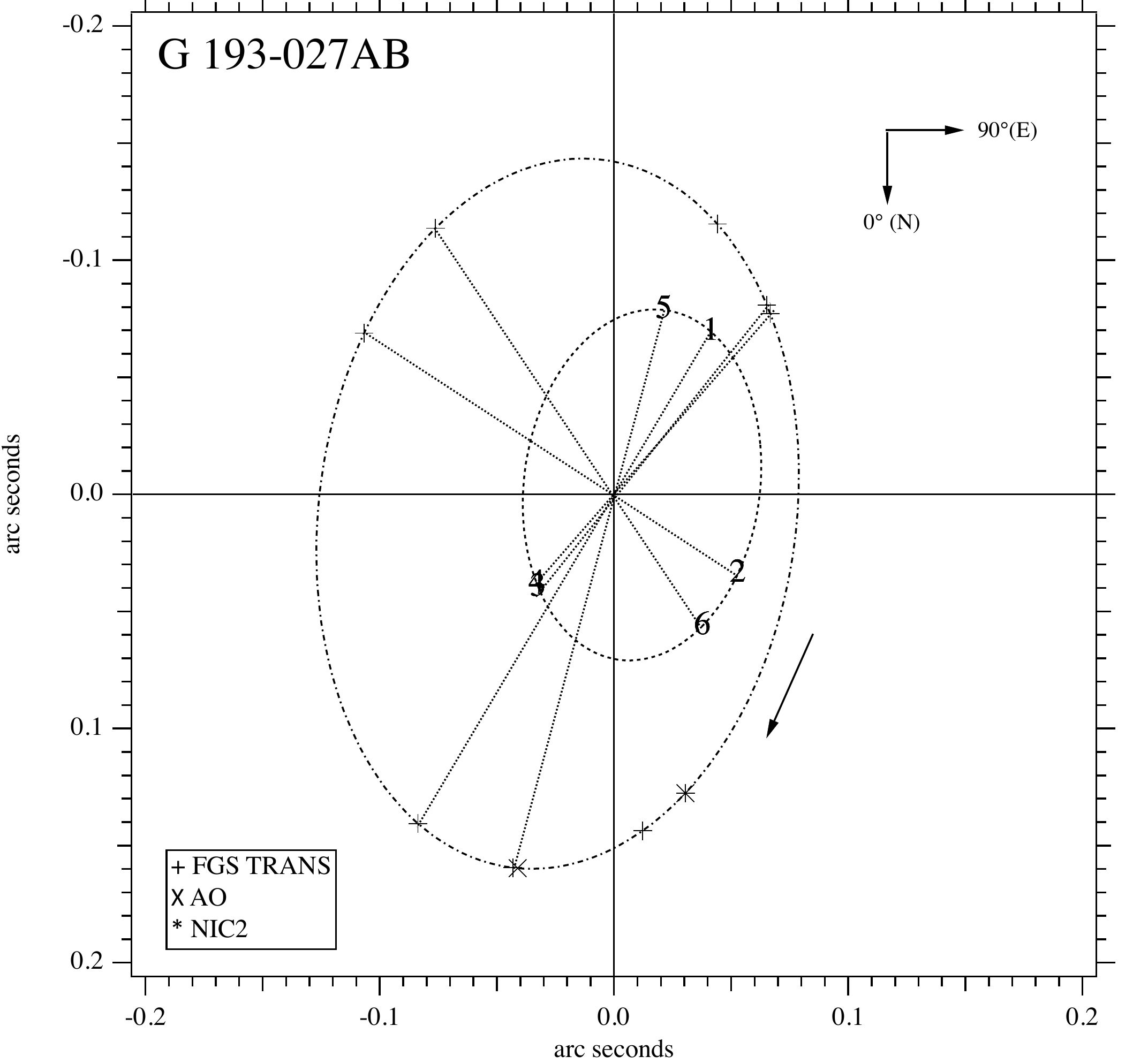}
\caption{G\,193-027\,A (numbers, POS orbit predicted positions),
  component B TRANS (+), AO ($\times$), and NICMOS (*) orbit predicted
  positions. POS, which normally provides astrometry for component A,
  locked onto component B for sets 2, 3, and 4. All observations were
  used to derive the orbital elements listed in
  Table~\ref{tbl-OE}. Component B (TRANS, AO, and NICMOS) and
  component A and B POS astrometric residuals (average absolute value
  for which listed in Table~\ref{tbl-DSR}) are smaller than their
  symbols.  }
\label{G193-027}
\end{figure}

Because of the typical large component separation (100--170 mas), and
small $\Delta V$ = 0.30, the POS observations half the time locked on
component B instead of component A. We modified our model to deal with
this complication by constraining the parallax of component A to equal
the parallax of B and by solving for two POS mode orbits.  In
Figure~\ref{G193-027} we plot observation set numbers on the component
A orbit that should have been established for the six sets of POS
measurements; the POS observations sparsely sample the perturbation of
the primary because they locked on component B for sets 2, 3, and 4.
The relatively large parallax error of 1.4 mas (Table~\ref{tbl-PPM})
reflects the paucity of POS mode observations.

Table~\ref{tbl-OE} contains the \Gon~orbital parameters from a model
not including the \cite{Khr10} parallax prior. Figure~\ref{G193-027}
provides component A and B orbits with observed positions indicated.
Because we had relatively few component A POS observations, we obtain
somewhat poorer mass precision for this system, finding ${\cal M}_{\rm
  A} = 0.126 \pm 0.005{\cal M}_{\sun}$ (4.0\% error) and ${\cal
  M}_{\rm B} = 0.124 \pm 0.005 {\cal M}_{\sun}$ (4.0\%), with the
errors driven almost entirely by the parallax error.  The two
components are of nearly equal mass, consistent with their low $\Delta
V$.  These are the first mass determinations for this system.

\subsection{\Gsfa} \label{G65}

%
%

\begin{figure}
\includegraphics[width=6.5in]{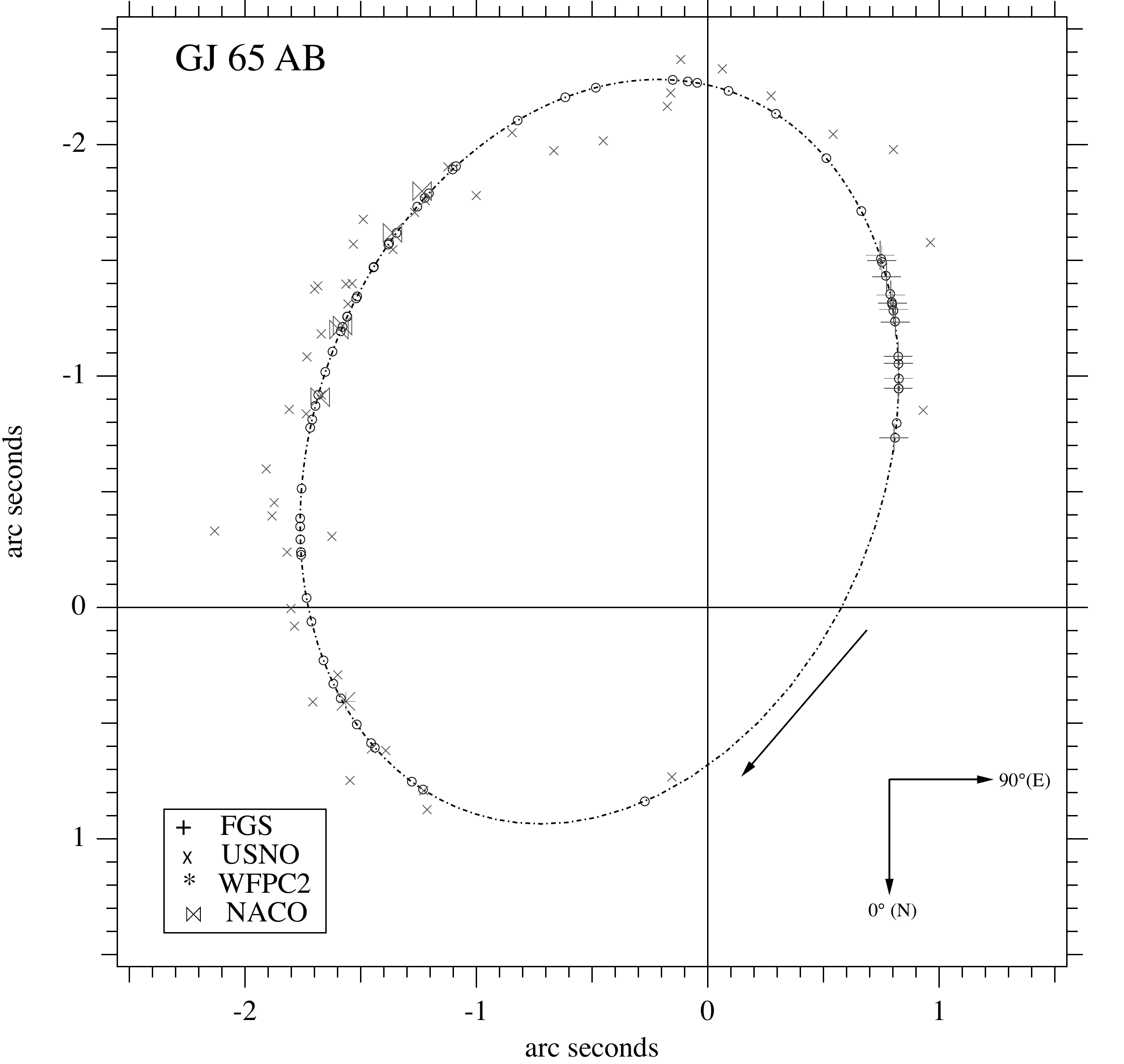}
\caption{GJ\,65  component
  B   orbit predicted (open circles)
 and measured positions (identified in the legend) with sizes proportional to average measurement error.  All A-B position angle and separation observations 
 (sources identified in Table~\ref{tbl-TR}) were used to derive the
  orbital elements listed in Table~\ref{tbl-OE}. 
  No POS mode observations entered into the modeling due to a paucity
  of reference stars.  }
\label{GJ65}
\end{figure}

The 10 \HST TRANS-only observations were made using FGS\,3, covering
only 18 degrees of  orbital position angle during 1.1 yr.  Seven POS mode
observations exist for component A (shown as dots in
Figure~\ref{GJ65}), but the reference frame consisted of a single star
at each epoch, and not always the same single star. Consequently, we
model only the relative orbit. We included a large sample of valuable
visual, photographic, and CCD observations of separation and position
angle \citep{Gey88} for our analysis of \Gsfa. These low-precision
(relative to \HSTns) measurements came primarily from the USNO 61in
reflector and span 48 years.  Even though of lower precision, they are
extremely useful for this system, which has an orbital period of 26.5
yr, the longest among the 15 systems we observed with \HSTns/FGS.
Also useful for extending the orbit sampling; five VLT/NACO \citep{Ker16} and one \HST/WFPC2
\citep{Die12} measure of position angle and separation.

Modeling only relative position angle and separation measurements yields $a$, $P$, $T_0$, $e$,
$i$, $\Omega$, and $\omega_{\rm B}$. Figure~\ref{GJ65} provides
the component AB relative orbit with
observed positions indicated.  The USNO residuals are obviously not
smaller than the circles in the Figure. The average absolute value
residual for the \HST TRANS observations is $\langle |res| \rangle$ =
3.5 mas, considerably smaller than the corresponding 81 mas value for
the USNO and 9.6 mas for the VLT/NACO measurements.  By adopting a mass fraction ($f =
0.494\pm0.004$, Geyer 1988) and absolute parallax ($\pi_{\rm abs} = 373.7\pm2.7$
mas) from the YPC, we obtain component masses of ${\cal M}_{\rm
  A} = 0.120 \pm 0.003{\cal M}_{\sun}$ (2.5\% error) and ${\cal
  M}_{\rm B} = 0.117 \pm 0.003 {\cal M}_{\sun}$ (2.5\%), agreeing with the recent \cite{Ker16}
  values of ${\cal M}_{\rm
  A} = 0.123 \pm 0.004{\cal M}_{\sun}$ (3.5\% error) and ${\cal
  M}_{\rm B} = 0.120 \pm 0.004 {\cal M}_{\sun}$ (3.6\%)

\subsection{\Gfst} \label{GJ473}


The sky within the FGS FOV near \Gfst~contains no usable POS mode
reference stars. Hence, as with \Gsfa~we again model only the relative
orbit with astrometry, using several sources: \HST (TRANS-only) from
FGS\,3 and FGS\,1r, a single \HST FOS measurement \citep{Sch98}, and
optical \citep{Bal94,Bla87,Hor12} and near-infrared speckle
(\cite{Hen92} and Torres et al.~1999) observations.  These
lower-precision (relative to \HSTns) measurements are extremely useful
for this long period (15.8 yr) system.

\begin{figure}
\includegraphics[width=6.5in]{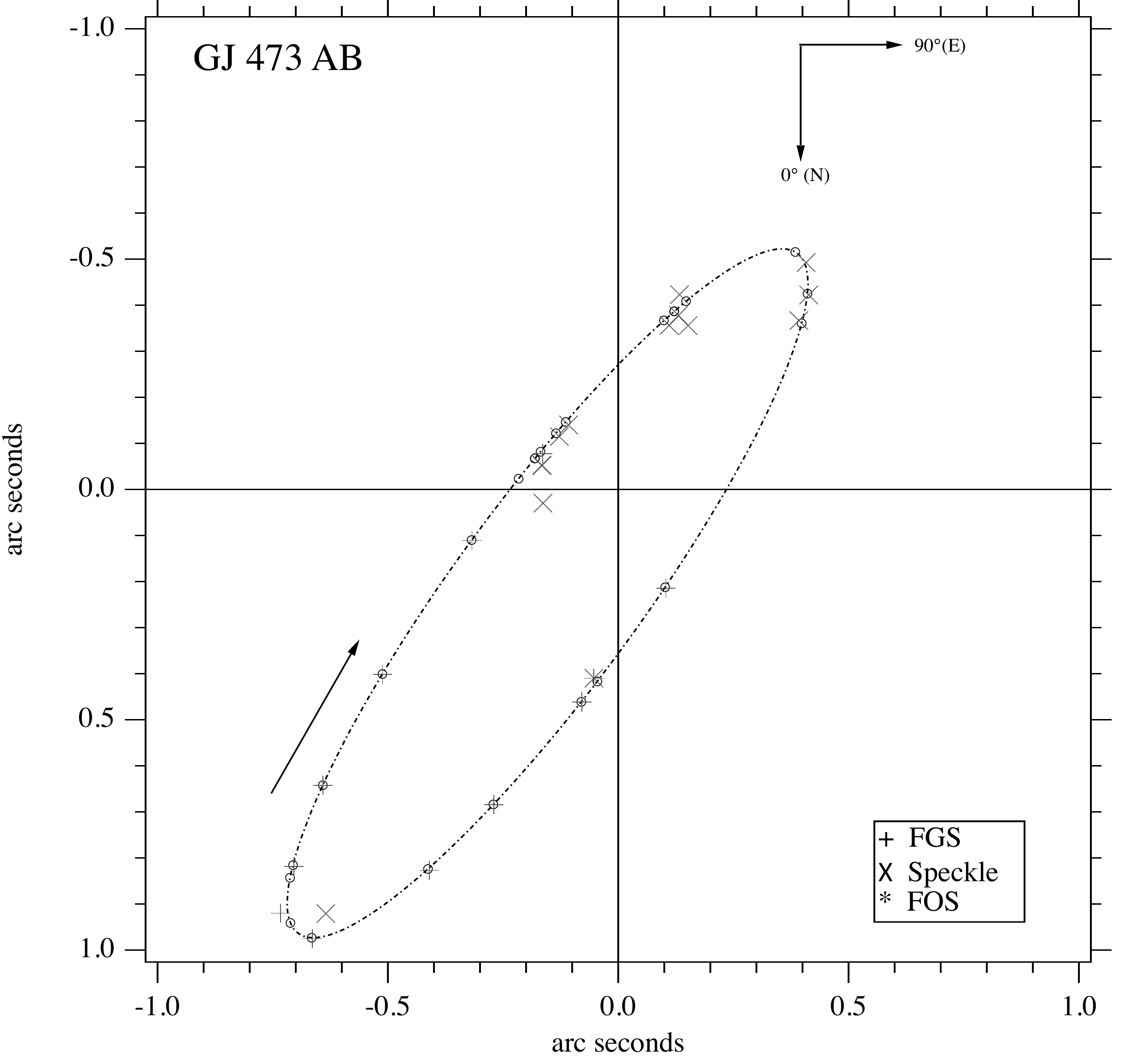}
\caption{GJ\,473\,AB relative orbit. Calculated positions are plotted
  on the orbit as open circles. Measurements from TRANS (+), speckle
  ($\times$), and \HST FOS (*) illustrate the residuals.  All
  observations were used to derive the orbital elements listed in
  Table~\ref{tbl-OE}.  No POS mode observations entered into the
  modeling.  }
\label{GJ473p}
\end{figure}

Figure~\ref{GJ473p} provides the component AB relative orbit with
observed positions indicated.  The speckle residuals are generally not
smaller than the circles in the Figure. The mean absolute value
residual for the \HST TRANS observations is 
2.5 mas, considerably smaller than the corresponding 17 mas value for
the speckle measurements.  By adopting a mass fraction of $f =
0.477\pm0.008$ from Torres et al.~(1999) \nocite{Tor99} and absolute
parallax of $\pi_{\rm abs} = 235.5\pm2.9$ mas from the RECONS
astrometry program at the CTIO/SMARTS 0.9m \citep{Hen06}, we obtain
component masses of ${\cal M}_{\rm A} = 0.124 \pm 0.005{\cal
  M}_{\sun}$ (4.0\% error) and ${\cal M}_{\rm B} = 0.113 \pm 0.005
{\cal M}_{\sun}$ (4.1\%). The relatively large mass errors are driven
almost entirely by the parallax error.  These masses are somewhat
lower than those reported by Torres et al.~(1999), ${\cal M}_{\rm A} =
0.143 \pm 0.011 {\cal M}_{\sun}$ (7.7\% error) and ${\cal M}_{\rm B} =
0.131 \pm 0.010 {\cal M}_{\sun}$ (7.6\%), who used 56 astrometry
observations taken between 1938 and 1998; however, only four
high-quality FGS observations were available then, compared to the 11
FGS observations used here.

\section{Absolute Magnitudes} \label{OGC}

The orbital elements summarized for the 15 systems in
Table~\ref{tbl-OE} are used to derive the component masses given in
Table~\ref{tbl-MMVMK}.  To place these stars on an empirical MLR, we
also require the component absolute magnitudes.  For the MLR at
optical wavelengths, we use $V$ photometry from \citet{Wei84,Wei96}
and the RECONS program at the CTIO/SMARTS 0.9m (see Winters et al. 2015\nocite{Win15}).
For $\Delta V$, we use values in our earlier work \citep{Hen99} as
well as results from the FGS observations presented here to derive
component magnitudes.  The $\Delta m$ measurements from the FGS TRANS
observations using F583W have been transformed to $\Delta V$, as
described in \cite{Hen99}.


At infrared wavelengths, we use 2MASS $K_s$ photometry \citep{Skr06}
for the combined light of all 15 systems in Table~\ref{tbl-MMVMK}.
The $\Delta K$ values have been adopted from infrared speckle
measurements previously reported in \cite{Hen93}, \HSTns/NICMOS
measurements reported in \cite{Die12}, and new Gemini-N observations
using the Near InfraRed Imager (NIRI) with the ALTAIR facility level
adaptive optics system \cite{Die16}.  We have not made any
transformations between the infrared $K$ band filters given that the
$\Delta K$ measurement errors tend to be larger than the slight
adjustments due to differences in the various $K$ filters. 

Using our parallaxes in Table~\ref{tbl-PPM} and the $\Delta V$ and
$\Delta K$ values, we find the component absolute magnitudes presented
in Table~\ref{tbl-MMVMK}.  For these very nearby systems we have
assumed no absorption ($A_V = 0$).  Parallaxes of this precision do
not require correction for Lutz-Kelker-Hanson bias \citep{Lut73,Han79}
in the derived absolute magnitudes.

\section{Additional Low Mass Binary Systems} \label{othersystems}

To boost the number of stars used in the MLRs at both optical and
infrared wavelengths, we augment our sample of 15 binaries with red
dwarfs in 9 additional systems having high-quality mass, $M_V$, and
$M_K$ measurements.  In Table~\ref{tbl-seb}, we list their parallaxes,
$V$ and $K$-band absolute magnitudes, and masses computed by other
investigators.  Ground-based astrometry has been used for all systems,
except the eclipsing binaries GJ 278 CD (YY Gem), GJ 630.1 AB (CM
Dra), GJ 2069 AC (CU Cnc), and GU Boo.  A few systems are particularly
worthy of note:

{\bf GJ\,166C} is a tertiary in the 40 Eri system, in a $\sim$250 yr
orbit orbit with the white dwarf GJ 166B.  Hence, the mass error of
0.029 ${\cal M}_{\sun}$ is the largest for any star in the sample.

{\bf GJ\,2005\,ABC} is an important triple with components near the
end of the stellar main sequence.  Only a small portion of the A--BC
orbit has been observed, so the resulting mass determinations for A
are not of sufficiently high quality to be included in this paper;
here we concentrate on the BC pair.  The three stars have combined $V$
= 15.28 $\pm$ 0.02 from our CTIO/SMARTS 0.9m photometry, and
individual ABC magnitudes of $V =$ 15.35 $\pm$ 0.04, 18.68 $\pm$ 0.06,
and 19.07 $\pm$ 0.07 from Leinert et al.~(2000)\nocite{Lei00}.  In the infrared, $K$
= 8.24 $\pm$ 0.03 for ABC from 2MASS.  Using $\Delta$$K$$_{AB}$ = 1.20
$\pm$ 0.03 and $\Delta$$K$$_{AC}$ = 1.55 $\pm$ 0.09 from \cite{Lei94},
 the components have
individual magnitudes of $K_A$ = 8.73 $\pm$ 0.03, $K_B$ = 9.93 $\pm$
0.04 and $K_C$ = 10.28 $\pm$ 0.04.  The weighted mean parallax of
128.49 $\pm$ 1.50 mas is from the RECONS astrometry program at the
CTIO/SMARTS 1.5m (129.47 $\pm$ 2.48 mas, Costa et al.~2005) and a new
value from the CTIO/SMARTS 0.9m program (127.93 $\pm$ 1.88 mas).  We
adopt the BC orbit and fractional mass (0.48 $\pm$ 0.01) from \cite{Koh12}, with the new
parallax to obtain masses of M$_B$ = 0.079 $\pm$ 0.003 ${\cal
  M}_\odot$ and M$_c$ = 0.073 $\pm$ 0.003 ${\cal M}_\odot$.



{\bf GJ\,2069\,AC} is an eclipsing binary that is part of the
quintuple red dwarf system known as CU Cnc (components A, C, and E)
and CV Cnc (components B and D).  There was no parallax in YPC and the
various parallaxes derived using {\it Hipparcos} data are of poor
quality, with values of 78.05 $\pm$ 5.69 (original reduction), 85.16
$\pm$ 6.42 mas \citep{Pou03}, and 90.37 $\pm$ 8.22 mas \citep{Lee07b}.  The
first parallax value was used by Delfosse et al.~(1999)\nocite{Del99b}, which led
them to conclude that the eclipsing components were subluminous.  Here
we provide a much better value of 64.80 $\pm$ 1.43 mas from the RECONS
astrometry program at the CTIO/SMARTS 0.9m that utilizes 11 years of
astrometry data.  This represents the weighted mean of values
determined for ACE (63.59 $\pm$ 1.92 mas) and BD (66.32 $\pm$ 2.15
mas) --- the system is observed as two sources at the resolution of
the 0.9m.  With this parallax, the components of the eclipsing pair AC
now fall closer to other stars in the MLR, particularly in the $K$-band. 

\section{The MLR} \label{mlr}

\begin{figure}
\includegraphics[width=5.5in]{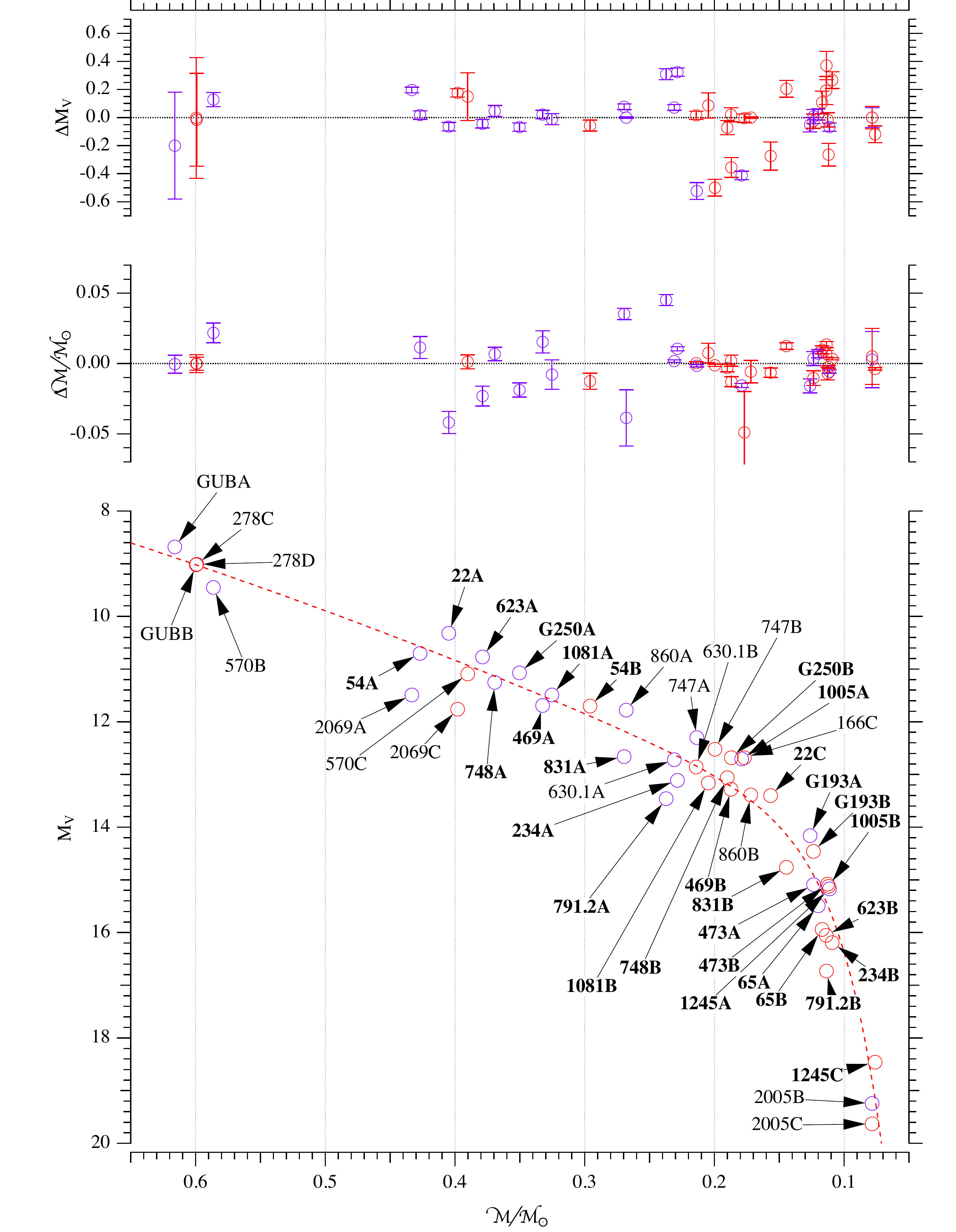}
\caption{ $V$-band MLR .  The dashed line is a double exponential function fit
  (Equation 10) to masses and $M_V$. Primaries are in blue;
  secondaries in red. $M_V$ and ${\cal M}$ residuals show differences
  between observed values and a fit to Equation 10 using GaussFit. We
  list fit coefficients and $ M_V$, ${\cal M}$ residual RMS in
  Table~\ref{tbl-coef}.  Errors for  M$_V$ and mass are plotted on their respective residual points. Points
  are GJ numbers without GJ prefix, except for GU Boo (GUB), G250-029
  (G250), and G193-027 (G193). \textbf{Boldface} denotes new and/or improved
  masses and absolute magnitudes from this work.}
\label{MLRVall}
\end{figure}

\begin{figure}
\includegraphics[width=5.5in]{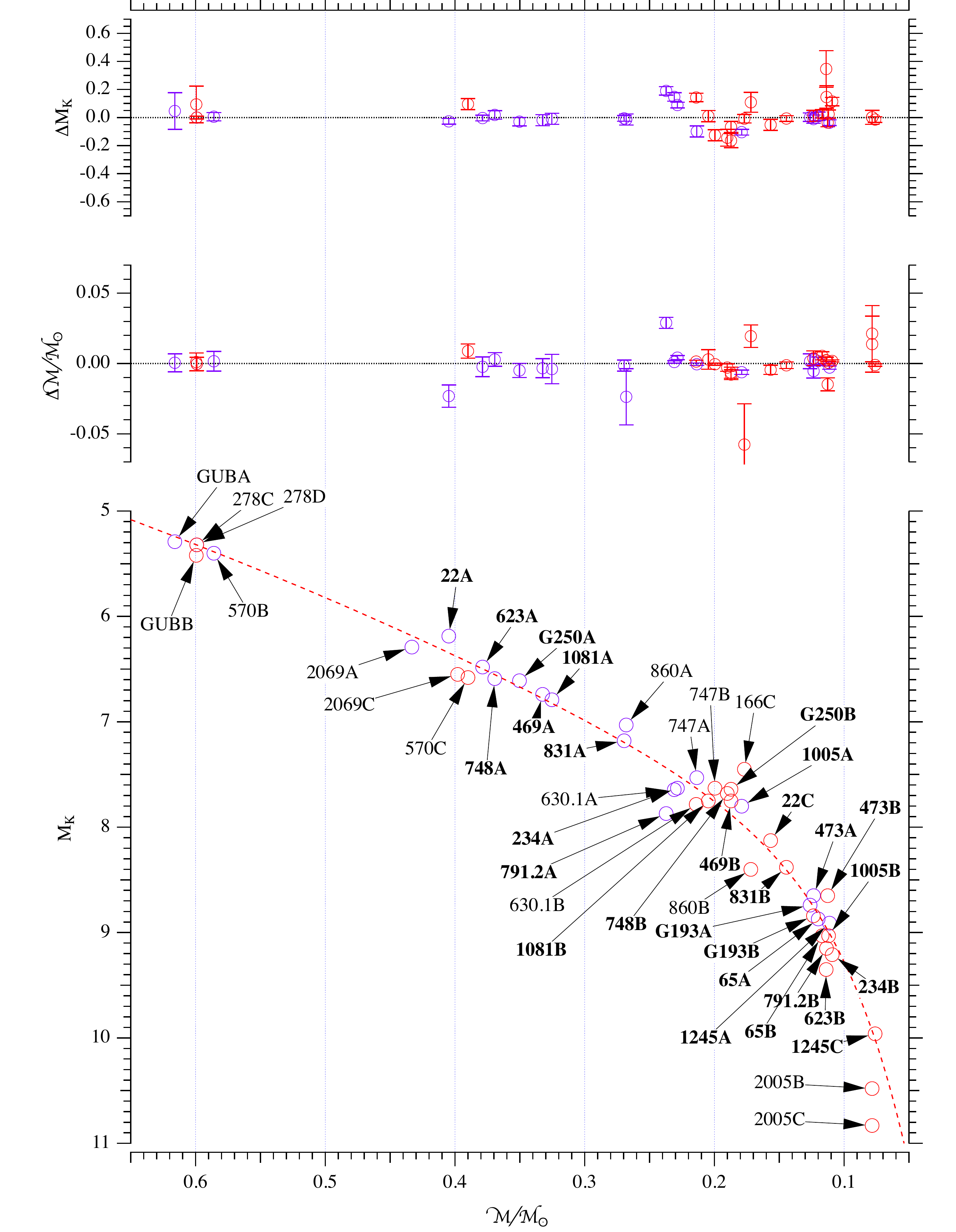}
\caption{ $K$-band MLR .  The dashed line is a double exponential function fit
  (Equation 10) to masses and $M_K$. Primaries are in blue;
  secondaries in red. $M_K$ and ${\cal M}$ residuals show differences
  between observed values and a fit to Equation 10 using GaussFit. We
  list fit coefficients and $ M_K$, ${\cal M}$ residual RMS in
  Table~\ref{tbl-coef}.  Errors for  M$_V$ and mass are plotted on their respective residual points. 
  Note the marked decrease in residual size. Same point labeling as Figure~\ref{MLRVall}. 
  GJ\,54, lacking $K$-band photometry, is not included.}
\label{MLRKall}
\end{figure}

We plot all 47 stars with their masses, ${\cal M}$, and absolute
magnitudes, $M_V$, from Tables~\ref{tbl-MMVMK} and \ref{tbl-seb} on
the $V$- and $K$-band  MLR shown in Figures~\ref{MLRVall} and \ref{MLRKall}.  
 For the
MLR at $K$ only the GJ\,54\,AB system is missing,
lacking individual $M_K$ values. The  object
identifications in bold denote new and/or improved absolute magnitudes
and masses derived for this paper using the \HSTns/FGS and RV data
presented here.  The primaries are plotted in blue; the secondaries in
red.

With the {\it caveat} that we are mixing stars of various ages,
metallicities, and magnetic properties, we fit the $V$ and $K$ mass-absolute
magnitude distribution with a double exponent (with offset) to provide
an empirical MLR for the full range of M dwarf masses from 0.6 ${\cal
  M}_{\sun}$ to the end of the stellar main sequence at 0.08 ${\cal
  M}_{\sun}$.  The form of the fitted equation is:

\beq
M_V = y_0 + A_1 exp\{{{-({\cal M}-x_0)}\over{ \tau_1}} \} +A_2 exp\{{{-({\cal M}-x_0)}\over{ \tau_2}} \}
\eeq

\noindent where the coefficients are  in 
Table~\ref{tbl-coef}.

This function serves as a smoothing tool.  We chose the double
exponent because it yielded a lower $\chi^2$ and smaller RMS residuals
than, for example, a single exponential with offset ($\chi^2$ ten times larger, RMS residuals 4 times larger).  
Because both
absolute magnitudes and masses have errors, we fit Equation 10 to the
points using GaussFit \citep{Jef88}, a modeling tool that fairly
assesses errors in both variable sets.  The top panels of
Figure~\ref{MLRVall} and Figure~\ref{MLRKall} display residuals to the fit for both 
absolute magnitude ($M_V$ and $M_K$) and 
mass.  The RMS values are 0.19 mag in $M_V$ and 0.023 
${\cal M}_{\sun}$ in mass.   The RMS values to the fit are 0.09 
mag in $M_K$ and 0.014 ${\cal M}_{\sun}$ in mass for the MLR at
$K$.  The coefficients for the MLR at $K$ are also given in
Table~\ref{tbl-coef}, along with the RMS values for both MLRs.

%

%

\begin{figure}
\includegraphics[width=6.5in]{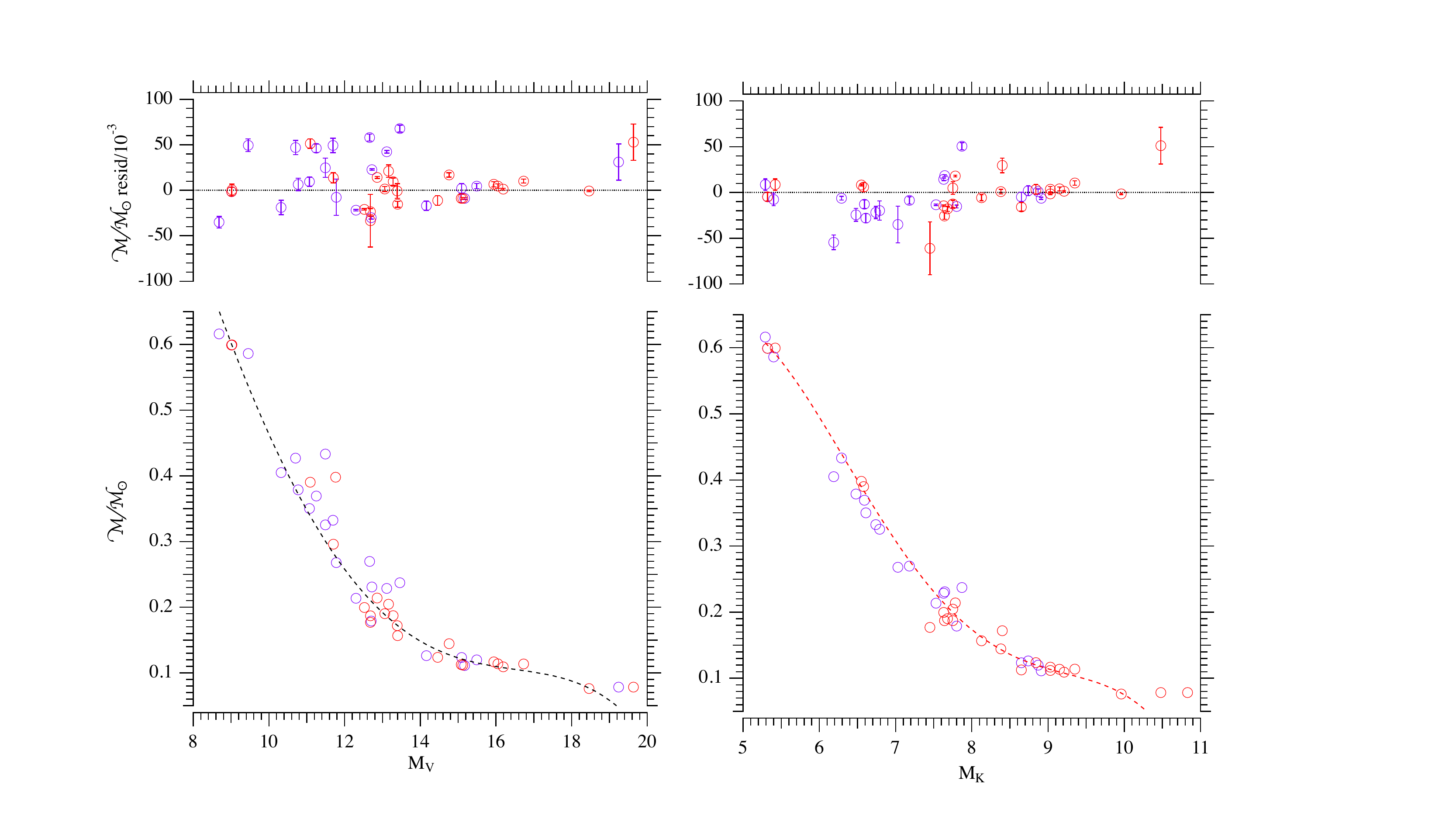}
\caption{ $V$ and $K$-band Luminosity-Mass Relations.  The dashed lines are 5th order polynomial function fits
  (Equation 11) to masses and absolute magnitudes. Primaries are in blue;
  secondaries in red. Residuals are tagged with mass errors from Tables~\ref{tbl-MMVMK} and \ref{tbl-seb}.
  We list fit coefficients  in Table~\ref{LMRT}. The relations lose predictive utility for $M_V >19$ and $M_K>10$.}
\label{LMRall}
\end{figure}

Rather than knowing the mass of an object and needing to derive its
luminosity, astronomers  more typically want
 to estimate a mass from an absolute magnitude.  Therefore, in the
two panels of Figure~\ref{LMRall} we plot mass against $M_V$ and
$M_K$, and provide a method to estimate a mass given an absolute
magnitude in a fundamental band at either optical or near-infrared
wavelengths.  Fit with fifth-order polynomials with magnitude offsets, $x_0$, we obtain
expressions that can be used to estimate masses, ${\cal M}$, as
functions of $M_V$ or $M_K$ (for $M_V\le19, M_K\le10$):

\beq
{\cal M} = C_0 + C_1 (M_V - x_0) + C_2 (M_V - x_0) ^2 + C_3  (M_V - x_0) ^3 + C_4 (M_V - x_0) ^4
\eeq

\noindent where $V$ can be replaced with $K$.  Table~\ref{LMRT}
contains coefficient and offset values for the $M_V$ and $M_K$ relations.  As
examples, a measured $M_V=14$ would yield ${\cal M} = 0.15\pm0.03
{\cal M}_{\odot}$, while $M_K=9$ would yield ${\cal M} = 0.12\pm0.02
{\cal M}_{\odot}$, where the errors are simply the RMS scatter about the 
polynomial fits.


%
%

\begin{figure}
\includegraphics[width=6.5in]{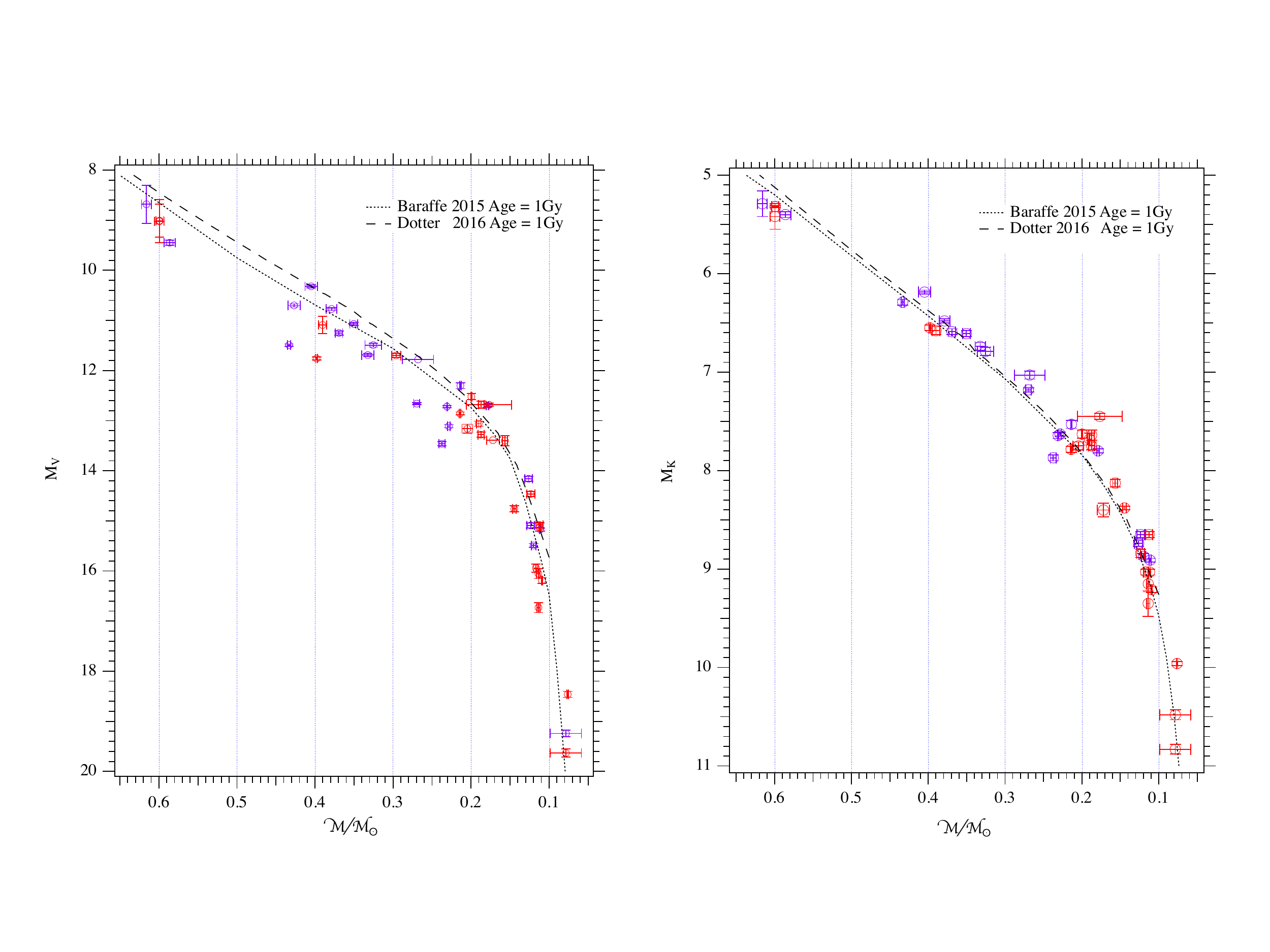}
\caption{$V$ and $K$-band Mass-Luminosity Relations from \cite{Bar15} and \cite{Dot16} 
along with actual measured absolute magnitudes and masses.  Primaries are in blue;
  secondaries in red. Models agree in some mass ranges better than in others with $K$-band models having overall best agreement. 
 }
\label{MLRMod}
\end{figure}

In Figure~\ref{MLRMod}  we compare the $V$ and $K$-band MLRs with
recent models from \cite{Bar15} and \cite{Dot16}.  Note that the models
agree with the $K$-band MLR better than with the $V$-band MLR.  Thus,
the $K$-band fluxes are better modeled better than the $V$-band
fluxes, which show twice the residual offsets to the Equation 10 fits.
These results confirm and extend the
results of \cite{Hen93} and \cite{Del00}.  We shall return to the
larger $V$-band scatter in $\S$9.  These MLRs are the most robust ever
established for M dwarfs, yet could be improved with additional mass
measurements around $\sim$0.5 ${\cal M}_{\sun}$, $\sim$0.3 ${\cal
  M}_{\sun}$, and below $\sim$0.1 ${\cal M}_{\sun}$.

%
%


\section{Discussion} \label{bd}

Why is the scatter in the $V$-band MLR twice that of the corresponding
$K$-band MLR?  To reiterate, age, chemical composition, and magnetic
properties can all affect the luminosities and temperatures of stars,
as well as rotation, mixing length and other properties.  Here we
address the first three properties.

{\bf Age} attributes are particularly complicated because the MLR at
low masses is a mapping that is complicated by both relative age and
absolute age.  Regarding relative age, once formed, M stars descend
slowly to the main sequence along a nearly vertical Hayashi track
\citep{Pal12}, so young stars have higher luminosities than stars on
the main sequence.  In fact, because in the age of the Universe, none
of the low mass stars investigated here have left the main sequence
through evolution, the 24 systems in this sample are a heterogeneous
mix of stars with ages potentially spanning $\sim$10 Gyr, although for a 
roughly uniform star formation rate, only a few percent might lie above the main sequence.
Determining
the age of an M dwarf is notoriously difficult, and usually based on
circumstantial evidence such as x-ray emission, flare rates, etc.,
unless it is associated with a more massive star of known age (e.g.,
Mamajek et al.~2013)\nocite{Mam13} or a particular
cluster or moving group (e.g., Riedel et al.~2011)\nocite{Rie11}.  
As for absolute age, an M star's luminosity is
affected by its composition (metallicity), which is a function of
birth date within the Galaxy.  More metal-rich M dwarfs, generally
formed later in the history of the Galaxy, can appear fainter due to
line blocking.  Absolute ages might be established through
gyrochonology, e.g., \cite{Car14,Mei15}; a more rapidly rotating M
dwarf would presumably be younger than one rotating more slowly,
although a reliable relation for M dwarfs remains elusive.

{\bf Chemical composition (Metallicity)} is, in principle, a more
directly measurable quantity than age.  In the past, accurate M dwarf
metallicities were difficult to quantify.  Relative metallicites could be
inferred from single project efforts \citep{Bea07p,Jen09}
despite their lack of agreement, and such comparisons clearly indicated
that some stars possess fewer metals than others.  For M dwarfs observed at
optical wavelengths, the CaH bands are typically used, although these
features are confounded by the interplay of effects due to both
metallicity and gravity (Jao et al.~2008)\nocite{Jao08}.
There has been considerable work on M dwarf metallicities in recent years 
\citep{Roj12,One12,Ter12,Nev14,Man14,Man15,New15,Lin15}. While there are 
subtle effects that remain 
challenging (including the effect of low log g for young M dwarfs that have
not yet contracted to their main sequence radii), the broad consistency 
between most of this work suggests that accurate metallicities (perhaps at least as accurate 
as $\pm$0.1 dex) are no longer out of reach. 

%


{\bf Magnetic properties} are generally explored via measurements of
chromospheric activity (flaring, x-ray emission, spots) and are
correlated with rapid rotation \citep{Wri11, Ste13}, attributes that
are all typically seen in young M dwarfs.  Photometric studies of M
dwarfs at various timescales have shown that these small stars exhibit
wide ranges in variability.  For periods of minutes to hours, some
stars flare often \citep{Kow13}, while others are quiescent for long periods of
time \citep{Jao11}.  Spots on
the surfaces of M dwarfs reveal the presence of magnetic fields, and
can be used to measure rotation periods of days to months
\citep{Ben98,Irw11}.  Recent work has shown that like the Sun, M dwarfs also
undergo long-period photometric changes lasting several years,
presumably connected to magnetic cycles in the stars \citep{Hos15}. Variations
in magnitude could change a derived absolute magnitude, adding scatter
to the MLR.

The formal errors in Table~\ref{tbl-MMVMK} effectively eliminate mass and
absolute magnitude errors as explanations for the offsets of most of
the stars in the MLRs, although errors in luminosity may be invoked
for GU Boo A and B, as well errors in mass for GJ\,166\,C and
GJ\,860\,A.  Some combination of the three factors outlined above can
explain the remaining scatter we see in Figures~\ref{MLRVall} and
\ref{MLRKall}.  Table~\ref{tbl-Exc} collects the
presently available metallicity, rotation velocity, radius, and x-ray
emission information for the systems considered in this paper.  Here
we attempt to explain a few of the significant departures in the
$V$ and $K$-band  MLR using these data.

The components of GJ\,1005 are elevated above the MLR at $V$.  The
[Fe/H] = $-$0.47 value for this system is among the lowest in the
sample, as are the rotation velocity and x-ray flux.  This system is
likely older than most others in the sample, but absorption that
affects the $V$ band is muted in $K$, so the components fall on the
infrared MLR.


Similar to GJ\,1005, both components of G\,193-027 are elevated above
the MLR at $V$.  The [Fe/H] = $-$0.46 value is virtually identical to
GJ\,1005, although G\,193-027's rotation velocity and x-ray luminosity
fall in the middle of the sample distribution.  We suspect that this
is mildly metal-poor system, again causing it to be elevated in the
MLR at $V$, but not at $K$.

Both components of GJ\,791.2 lie below the MLRs.  The stars exhibit
rapid rotation and relatively large x-ray emission compared to the
other stars in the sample, which suggests relative youth.  This, in
turn, suggests that they are metal-rich and consequently have lower
than expected luminosities.

Both components of GJ\,831 lie below the MLR at $V$, yet fall squarely
on the MLR at $K$.  This is the most metal-rich [Fe/H] = $+$0.3
measurement available for the 24 systems, which may reduce the flux at
$V$ but not at $K$ and lead to the seemingly discordant locations on
the two MLRs.

The GJ\,1245\,ABC triple is in the original {\it Kepler} FOV, and both
the AC and B components have been found to rotate rapidly and exhibit
significant x-ray emission (Lurie et al. 2015\nocite{Lur15}).  X-ray 
emission is an age proxy. 
Youth can be correlated with metallicity which
can effect absolute magnitude.  In
contrast to GJ\,791.2, the puzzle is that neither A nor C strays far
from the MLR at $V$ or $K$.  
We note that at
the very lowest masses in the MLRs presented, there are very few
points with empirically determined masses to constrain the fits.

Two remaining systems worthy of note are GJ\,747 and G\,250-029.  All
four components in these systems are elevated in the MLR at $V$.
There is nothing unusual about the available data for GJ\,747, while
G\,250-029 lacks (Table~\ref{tbl-Exc}) any additional knowledge to
which we might appeal for explanation.

The eclipsing system  GJ\,2069 yielded \citep{Rib03} an extremely high mass precision, 
but the absolute magnitudes would place these components about 0.5 magnitude below the MLR smoothing 
functions in $M_V$.  The system has one of the highest metallicities
in Table~\ref{tbl-Exc}, perhaps partially explaining this deviation. Given that the Figure~\ref{MLRVall} smoothing fits depend on the 
measured mass and luminosity errors, we chose to exclude this system from the $V$-band fitting.


%

%

Turning the question about scatter in the MLRs around, one might ask
why both components of the GJ\,65 system, which have high rotational v\,sin\,i,
perhaps indicative of youth (and presumably higher metallicity), are
not further from other objects with similar masses on any MLR?
Clearly, there remains work to be done to understand the complex
interplay of the effects that set locations on the MLRs.

Ultimately, the predictive utility of an MLR depends on the scatter
about any smoothing function.  In the absence of physically-based
corrections to absolute magnitude, even the lower-scatter $K$-band MLR
(Figure~\ref{MLRKall}) exhibits deviations about the smoothing
function that at some mass levels exceeds 15\%.  For example, at
$M_K$=7.8 the scatter is $\pm$0.035${\cal M}_{\sun}$, or 18\% at the
corresponding mass location of ${\cal M}$ = 0.2${\cal M}_{\sun}$.
This particular mass region, exhibiting the highest scatter, may
signal the onset of full convection (c.f. Houdebine \& Mullan, 2015)\nocite{Hou15}.

\begin{figure}
\includegraphics[width=5in]{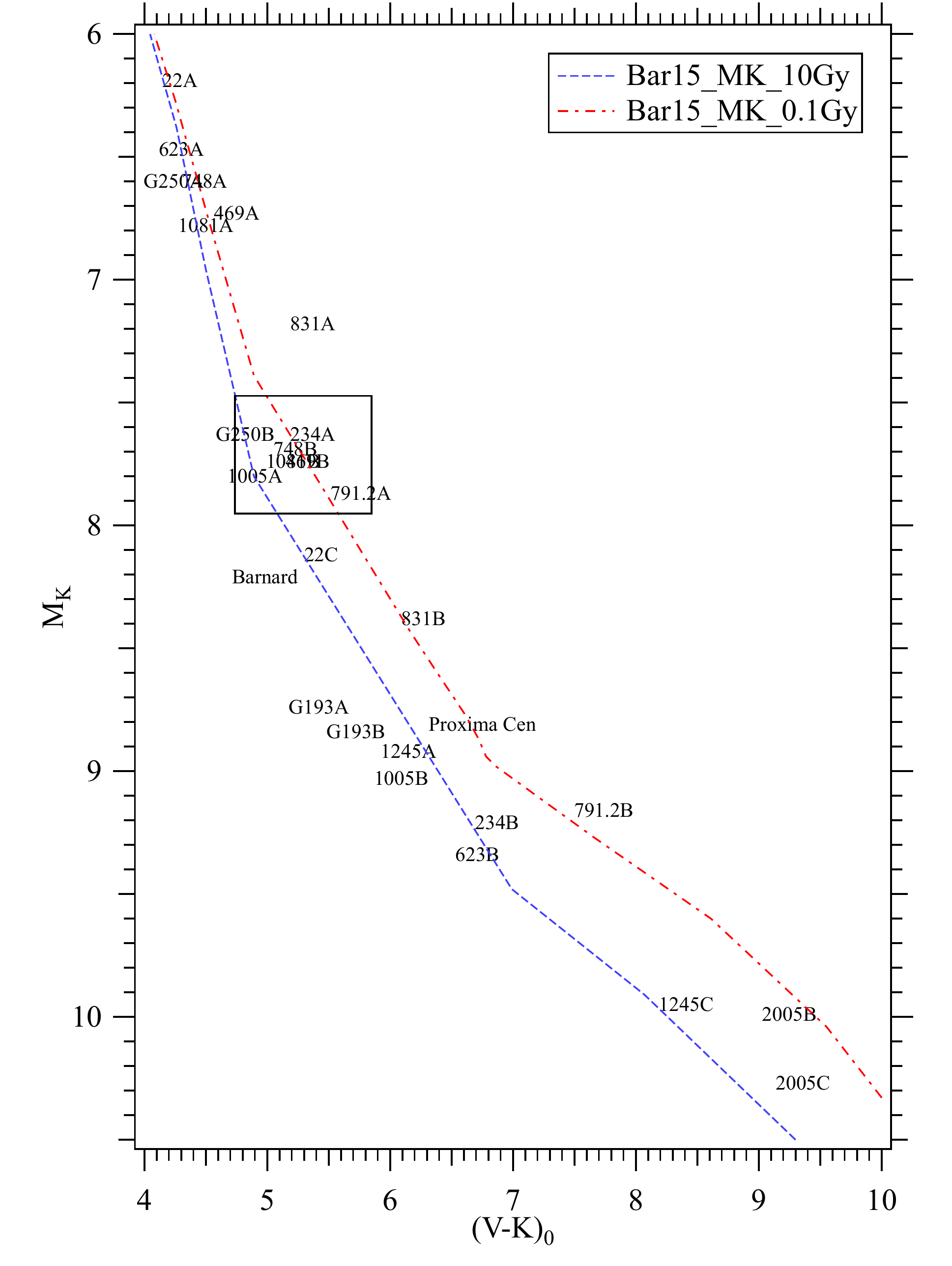}
\caption{M$_K$, (V-K)$_0$ Hertzsprung-Russell diagram for all systems
  with $K$ magnitudes (Tables~\ref{tbl-MMVMK}, \ref{tbl-seb}). We
  derive absolute magnitudes from HST parallaxes
  (Table~\ref{tbl-PPM}). Also plotted are stellar models for 0.1 and
  10 Gy from \cite{Bar15} and the single M dwarfs Proxima Cen and
  Barnard's Star. The box contains a clump consisting of GJ\,234\,A,
  G\,250-029\,B, GJ\,748\,B, GJ\,469\,B, GJ\,1081\,B, GJ\,791.2\,A,
  and GJ\,1005\,A. An age spread can explain some of the scatter. As
  expected most components lie close to the same age track.}
\label{HR}
\end{figure}

Finally, Figure~\ref{HR} presents an M$_K$ vs.~$(V-K)$ HR diagram for
all systems with $\Delta$$K$ measurements (Table~\ref{tbl-MMVMK}) and
masses determined using FGS data.  We derive absolute magnitudes using
our relatively precise \HST parallaxes (Table~\ref{tbl-PPM}).  Also
plotted are stellar models for 0.1 and 10 Gy from \cite{Bar15} and the
single M dwarfs Proxima Centauri and Barnard's Star, using \HST
parallaxes from \cite{Ben99} to derive absolute magnitudes.  As
expected, the presumably co-eval components in systems usually lie
close to the same age track.  We note that the components of GJ\,791.2
lie close to the \cite{Bar15} 0.1 Gyr model, supporting a young age.
Note also that GJ\,1245 A and C lie close to the 10 Gyr model trace,
even though their rapid rotation and x-ray flux (Table~\ref{tbl-Exc})
suggest youth.  Late-type stars like GJ\,1245AC can remain active on a timescale 
of $\sim$8 Gyr (see West et al. 2008\nocite{Wes08}), so it is perhaps not surprising 
that the stars might be quite old but still active. Apparently, this particular bandpass HR diagram is not
an infallible indicator of youth, nor the adopted markers for youth foolproof.



\section{Conclusions} \label{summ}

\begin{enumerate}

\item With \HST Fine Guidance Sensors (FGS) we obtained fringe
  tracking (POS) observations with either FGS\,3 and/or FGS\,1r of 13
  low-mass binary systems, each with 6--16 epochs of observations
  typically spanning 5 years.  These yielded absolute parallaxes with
  an external error better than 1\% and proper motions with average
  errors 0.3 \msy in RA and Dec.

\item Fringe tracking (POS) observations of the primary, including
  photocenter corrections where required, provide a perturbation
  orbit.

\item Fringe scans (TRANS) combined with POS observations provide a
  mass fraction relative to an astrometric reference frame. TRANS
  observations also provide $\Delta V$ measurements \citep{Hen99}.
  For two systems we acquired only TRANS measurements, requiring
  external determinations of the mass fractions.

\item Radial velocities, primarily from the McDonald Observatory 2.1m
  telescope and Cassegrain Echelle spectrograph, were combined with
  astrometry for seven systems, increasing the accuracy of their final
  mass results.

\item We derive 30 component masses in 15 binary red dwarf systems
  with a median precision 1.8\% and an average precision of 2.1\%.
  These are the first mass determinations for 12 stars in 6 systems.
  These masses are augmented with additional red dwarfs with
  high-quality masses in 9 additional systems.

\item We provide system magnitudes at $UBVRIJHK$ and component
  magnitude differences, $\Delta V$ and $\Delta K$, many reported for
  the first time here.  New trigonometric parallaxes from both \HST
  and the RECONS astrometry program at CTIO, as well as literature
  values, are used with the photometry to obtain accurate component
  $M_V$ and $M_K$ values.

\item This study provides high quality data for 47 red dwarfs with
  masses between 0.616${\cal M}_{\sun}$ and 0.076${\cal M}_{\sun}$,
  $M_V$ = 8.68--19.63, and $M_K$ = 5.29--10.27.  Many of the
  secondaries lie below 0.2${\cal M}_{\odot}$, the crucial region
  where age begins to play a significant role in the luminosities of
  stars.  With these data, we establish Mass-Luminosity Relations
  (MLRs) in both the optical $V$ and near-infrared $K$ bands.

\item The $V$ and $K$ MLRs are fit with double exponentials to provide
  empirical conversions of masses to absolute magnitudes, and vice
  versa.  We find that the $K$-band MLR exhibits a scatter in mass
  half that of the corresponding $V$-band relation.

\item The predictive capability of even the smaller-scatter $K$-band
  MLR appears to be limited by age, composition, and magnetic
  differences between low mass stars.  At $M_K$=7.8 the scatter is
  0.035${\cal M}_{\sun}$, or 18\% at the corresponding mass location
  of ${\cal M}$ = 0.2${\cal M}_{\sun}$.

\item Future work will disentangle the effects of age, metallicity,
  and magnetic properties on luminosity, and enable us to construct
  MLRs that consider additional factors that affect low mass stars.

\end{enumerate}

\acknowledgments

Support for this work was provided by NASA through grants GTO
NAG5-1603, GO-6036, 6047, 6566, 6764, 6882, 6883, 6884, 7491, 7493,
7894, 8292, 8728, 8729, 8774, 9234, 9408, 9972, 10104, 10613, 10773,
10929, 11299, and 12629 from the Space Telescope Science Institute,
which is operated by the Association of Universities for Research in
Astronomy, Inc., under NASA contract NAS5-26555.  The RECONS program
has been supported by the National Science Foundation through grants
AST 05-07711, AST 09-08402, and AST 14-12026.

This publication makes use of data products from the Two Micron All
Sky Survey, which is a joint project of the University of
Massachusetts and the Infrared Processing and Analysis
Center/California Institute of Technology, funded by NASA and the NSF.
This research has made use of the SIMBAD and Vizier databases and
Aladin, operated at CDS, Strasbourg, France, the NASA/IPAC
Extragalactic Database (NED) which is operated by JPL, California
Institute of Technology, under contract with the NASA, and NASA's
Astrophysics Data System Abstract Service.

We thank Linda Abramowicz-Reed for her unflagging and expert FGS
instrumental support over the last 25 years.  Cassegrain Echelle (CE)
Spectrograph observing and data reduction assistants included
J.~Crawford, Aubra Anthony, Iskra Strateva, Tim Talley, Amber
Armstrong, Robert Hollingsworth, and Jacob Bean.  We thank Dave Doss,
John Booth, and many other support personnel at McDonald Observatory
for their cheerful assistance over many years, as well as the staff at
CTIO for their continued support of the CTIO/SMARTS 0.9m, where the
RECONS astrometry/photometry program is carried out.  Finally, thanks
to Dr.~Michael Endl for the use of additional unpublished HET radial
velocities for GJ\,623\,A, and to R.~Andrew Sevrinsky for the improved
parallax for GJ\,473\,AB.

Lastly, we thank an anonymous referee for their prompt attention to this paper,
and for the many suggestions that materially improved the final version.


\bibliography{/Active/myMaster}

\clearpage


\begin{center}
\begin{deluxetable}{lcrrlrr}
\tablecaption{Identifications and Coordinates\tablenotemark{a} 
 \label{tbl-ID}}
\tablewidth{0in}
\tablehead{\colhead{ID} &
  \colhead{Giclas}&
\colhead{HIP}&
\colhead{LHS} &
\colhead{Other} &
\colhead{RA} &
\colhead{Dec}}
\startdata
GJ 1005 AB   & 266-076 &1242&1047&L722-22&00 15 27.994&-16 08 01.83\\
GJ 22 AC     &         &&&ADS 433&00 32 29.586&+67 14 03.65\\
GJ 54 AB     &         &5496&1208&L87-59&01 10 22.903& -67 26 41.85\\
GJ 65 AB     & 272-061 &&&L726-8&01 39 01.453&-17 57 02.05\\
GJ 1081 AB   & 096-045 &&&LTT 17822&05 33 19.151&+44 48 57.71\\
GJ 234 AB    & 106-049 &30920&&Ross 614&06 29 23.404&-02 48 50.37\\
G 250-029 AB &         &33142&221&GJ\,3412 &06 54 04.237&+60 52 18.36\\
G 193-027 AB &         &&224&GJ\,3421&07 03 55.655& +52 42 07.60\\
GJ 469 AB    & 012-038 &60910&2565&Wolf 414&12 28 57.486&+08 25 31.26\\
GJ 473 AB    & 012-043 &&333&Wolf 424&12 33 17.383& +09 01 15.77\\
GJ 623 AB    & 202-045 &80346&417&&16 24 09.325&+48 21 10.46\\
GJ 748 AB    & 022-018 &94349&472&Wolf 1062 &19 12 14.676&+02 53 10.71\\
GJ 1245 AC   & 208-044 &&3494&V1581 Cyg&19 53 54.469&+44 24 53.09\\
GJ 791.2 AB  & 144-008 &&3556&HU Del&20 29 48.385&+09 41 20.36\\
GJ 831 AB    & 026-007 &106255&511&Wolf 922&21 31 18.643&-09 47 26.41\\
\enddata
\tablenotetext{a}{HIP = {\it Hipparcos}, LHS = Luyten Half Second. RA, Dec epoch 2000.0.}
\end{deluxetable}
\end{center}


\begin{center}
\begin{deluxetable}{lcrrcrrrcrrr}
\tablewidth{0in}
\tablecaption{System Spectral Types and Photometry \label{tbl-ST}}
\tablehead{  
\colhead{ID}&
\colhead{Sp.T.\tablenotemark{a}} &
\colhead{$U$}&
\colhead{$B$}&
\colhead{ref\tablenotemark{b}}& 
\colhead{$V$}& 
\colhead{$R$}& 
\colhead{$I$}& 
\colhead{ref\tablenotemark{b}}& 
\colhead{$J$\tablenotemark{c}}& 
\colhead{$H$\tablenotemark{c}}& 
\colhead{$K$\tablenotemark{c}}
}
\startdata    
GJ\,1005\,AB    & M3.5V & 14.49 & 13.19 & K07 & 11.48 & 10.27 &  8.70 &  *  & 7.22 & 6.71 & 6.39 \\
GJ\,22\,AC      & M1.5V &  ...  & 11.88 & W96 & 10.28 &  9.24 &  7.97 & W96 & 6.84 & 6.27 & 6.04 \\ 
GJ\,54\,AB      & M2.5V & 12.51 & 11.36 & K07 &  9.82 &  8.70 &  7.32 &  *  & 6.00 & 5.41 & 5.13 \\
GJ\,65\,AB      & M5.0V &  ...  & 13.95 & B90 & 12.06 & 10.40 &  8.34 &  *  & 6.28 & 5.69 & 5.34 \\
GJ\,1081\,AB    & M3.0V & 14.99 & 13.79 & D88 & 12.21 & 11.06 &  9.58 & W84 & 8.20 & 7.59 & 7.34 \\
GJ\,234\,AB     & M4.0V & 13.94 & 12.76 & K07 & 11.14 &  9.78 &  8.04 &  *  & 6.38 & 5.75 & 5.49 \\
G\,250-029\,AB  & M2.5V &  ...  & 12.51 & W96 & 10.95 &  9.86 &  8.48 & W96 & 7.13 & 6.60 & 6.35 \\ 
G\,193-027\,AB  & M5.0V &  ...  & 15.14 & D82 & 13.29 & 11.94 & 10.15 & W96 & 8.54 & 8.09 & 7.78 \\
\enddata
\tablenotetext{a}{Spectral type for combined flux, except for GJ\,65AB, which is for primary only.}
\tablenotetext{b}{References for $UB$ photometry and $VRI$ photometry: 
* = this paper, 
B90 = \cite{Bes90}, 
D82 = \cite{Dah82}, 
D88 = \cite{Dah88}, 
K07 = \cite{Kil07}, 
K10 = \cite{Koe10}, 
W84 = \cite{Wei84}, 
W96 = \cite{Wei96}} 
\tablenotetext{c}{from 2MASS}
\end{deluxetable}
\end{center}

\begin{deluxetable}{llrr}
\tablecaption{ Radial Velocities\tablenotemark{a} 
 \label{tbl-RVs}}
\tablewidth{0in}
\tablehead{\colhead{ID} &
  \colhead{mJD\tablenotemark{b}}&
\colhead{Primary (\kms)}&
\colhead{Secondary (\kms)} 
}
\startdata
\\
GJ\,22 AC   &49962.9    &  $-$5.86$\pm$0.22 &                 \\
            &50405.7    &  $-$5.52     0.38 &                 \\
            &50723.8    &  $-$6.12     0.26 &                 \\
            &51136.8    &  $-$5.74     0.24 &                 \\
            &52809.9    &  $-$2.52     0.58 &                 \\
            &53936.4    &  $-$3.93     0.30 &                 \\
            &54305.9    &  $-$4.48     0.29 &                 \\
\enddata
\tablenotetext{a}{All velocity units in \kms.}
\tablenotetext{b}{mJD=JD-2400000.0}
\tablenotetext{c}{Velocities from H$\alpha$ line only}
\end{deluxetable}


\begin{deluxetable}{lllccccccc}
\tablecaption{Results from Radial Velocities\tablenotemark{a} \label{tbl-RV}}
\tablewidth{6.5in}
\tablehead{\colhead{ID}
 & \colhead{$V$}
 & \colhead{$\Delta V$}
 &  \colhead{ K$_{\rm A}$ }
 &  \colhead{ K$_{\rm B}$}
 &  \colhead{ $\gamma$}
 & \colhead{\#A}
 & \colhead{A$\langle |res| \rangle$}
 & \colhead{\#B}
 & \colhead{B$\langle |res| \rangle$}
 }
\scriptsize
\startdata
\Gtt&10.05&3.08&1.92&&-4.45&8&0.21&&\\
&&&0.02&&0.05&&&&\\
\Gttf&11.99&3.08&2.28&&16.18&6&0.23&&\\
&&&0.01&&0.04&&&&\\
\Gfs&12.06&1.59&3.82&6.77&-6.62&11&0.17&7&0.56\\
&&&0.03&0.06&0.02&&&&\\
\Gst&10.28&5.28&2.20&&-26.79&14&0.01&(HET)&\\
&&&0.02&&0.01&14&0.10&(CE)&\\  
\Gsf&11.12&1.81&5.07&10.03&-39.61&16&0.10&12&0.17\\
&&&0.04&0.09&0.02&&&&\\
\Gsn&13.37&3.27&4.09&8.54&-34.24&12&0.70&11&0.66\\
&&&0.08&0.13&0.07&&&&\\
\Get&12.06&2.10&5.17&9.64&-61.37&13&0.39&13&0.96\\
&&&0.06&0.10&0.11&&&&\\
\enddata

\tablenotetext{a}{All velocity units in \kms. Errors for RV amplitudes K$_{\rm A}$,
  K$_{\rm B}$, and system velocity, $\gamma$, are just below each entry.  \#A, \#B are
  numbers of unique epochs of RV measurements. A$\langle |res|
  \rangle$ and B$\langle |res| \rangle$ are average absolute value RV
  residuals in \kms.}

\end{deluxetable}

\clearpage


\begin{deluxetable}{lccccccc}
\tablecaption{GJ\,831\,AB Reference Star Characteristics and Final Parallaxes \label{tbl-pis}}
\tablewidth{6in}
\tablehead{\colhead{ID}
 &  \colhead{ $V$ }
 &  \colhead{$V-K$}
 &  \colhead{$J-K$}
 & \colhead{Sp.T}
 & \colhead{$M_V$}
 & \colhead{$\pi_{est}$ (mas)}
 & \colhead{$\pi_{abs}$ (mas)}
 }
\scriptsize
\startdata
ref-2  & 13.41 & 1.38$\pm$0.10 & 0.34$\pm$0.03 & G0V  & 4.4 & 1.6$\pm$0.5 & 1.7$\pm$0.2 \\
ref-3  & 13.65 & 1.73     0.11 & 0.47     0.04 & G7V  & 5.4 & 2.2     0.7 & 2.2     0.2 \\
ref-4  & 13.02 & 1.62     0.10 & 0.42     0.04 & G7V  & 5.4 & 3.0     1.0 & 3.0     0.3 \\
ref-5  & 14.35 & 3.01     0.10 & 0.87     0.05 & M1V  & 9.2 & 9.0     3.0 & 8.4     0.9 \\
ref-10 & 14.67 & 1.58     0.11 & 0.45     0.04 & G5IV & 2.5 & 0.4     0.2 & 1.4     0.2 \\
\enddata
\end{deluxetable}


\begin{center}
\begin{deluxetable}{lclcccccc}
\tablecaption{Astrometry Data Spans and Residuals\tablenotemark{a} \label{tbl-DSR}}
\tablewidth{6.5in}
\tablehead{\colhead{ID} 
& \colhead{\#ref}
& \colhead{$\langle V \rangle$ref}
& \colhead{POS}
& \colhead{POS\tablenotemark{b}}
& \colhead{POS}
& \colhead{TRANS}
& \colhead{TRANS}
& \colhead{TRANS}
\\
\colhead{}&\colhead{} & \colhead{} &\colhead{span (yr)}& \colhead{\#obs}& \colhead{$\langle |res| \rangle$ (mas)}& \colhead{span (yr)}& \colhead{\#obs}& \colhead{$\langle |res| \rangle$ (mas)}
}
\scriptsize
\startdata
\Lst                   & 2 & 14.5  & 4.36 & 12 & 1.2 &  4.36 & 17 &  1.4 \\
\Gtt                   & 4 & 14.4  & 6.06 &  6 & 0.8 &  9.22 &  9 &  3.3 \\
\Gtt  &   &       &      &    &     & 15.04 & 23\tablenotemark{c} & 10.1 \\
\Gff                   & 3 & 13.22 & 4.56 &  6 & 0.9 &  9.15 &  9\tablenotemark{c} &  2.2 \\
\Lsts                  &   &       &      &    &     &  1.47 & 13 &  3.5 \\
\Lsts &   &       &      &    &     & 63.0  & 51\tablenotemark{d} & 81.4 \\
\Goz                   & 6 & 14.05 & 4.00 &  5 & 0.7 & 10.47 &  11 &  2.1 \\
\Gttf                  & 5 & 14.21 & 5.5  &  7 & 0.8 & 13.90 & 11 &  2.4 \\
\enddata

\tablenotetext{a}{\#ref = number of astrometric reference stars;
  $\langle V \rangle$ref = average $V$ magnitude of the reference
  stars; POS span = length of POS campaign; \#POS = number of epochs
  of POS observation; POS$\langle |res| \rangle$ = average absolute
  value of POS astrometric residuals; TRANS span = length of TRANS
  campaign; \#TRANS = number of epochs of TRANS observation;
  TRANS$\langle |res| \rangle$ = average absolute value of TRANS
  astrometric residuals, unless noted with other observations.}

\tablenotetext{b}{For some orbit plots the number of POS plotted appears less than listed 
here because of overplotting and plot resolution.}
\tablenotetext{c}{includes measurements in addition to \HST TRANS}
\tablenotetext{d}{Ground-based astrometry only.}
\end{deluxetable}
\end{center}


\begin{center}
\begin{deluxetable}{lrrrrr}
\tablecaption{Parallaxes and Proper Motions \label{tbl-PPM}}
\tablewidth{0in}
\tablehead{
\colhead{ID} & \colhead{$\pi_{abs}$}& \colhead{$\pi_{abs}$}& \colhead{$\pi_{abs}$}& \colhead{$\mu_{RA}$} & \colhead{$\mu_{DEC}$} \\
&\colhead{YPC / other}&\colhead{$HIP$\tablenotemark{a} }&\colhead{\HST}&\colhead{\HST}&\colhead{\HST}\\
\colhead{}&\colhead{mas}&\colhead{mas}&\colhead{mas}&\colhead{\msy}&\colhead{\msy}
}
\startdata
\Lst\tablenotemark{b} & 182.1 $\pm$ 6.8                  & 200.5 $\pm$ 9.4 & 166.6 $\pm$ 0.3 &    642.0 $\pm$ 0.3 & $-$616.3 $\pm$ 0.4 \\
\Gtt                  &  ...                             &  99.4       2.2 &  99.2       0.6 &   1749.7       0.3 & $-$254.3       0.2 \\
\Gff                  &  ...                             & 122.0       2.4 & 126.9       0.4 &    386.2       0.2 &    579.7       0.1 \\
\Gsfa                 & 373.7       2.7 &  ...            &  ...            &    ...             &    ...             \\
\Goz                  &  65.2       1.8                  &  ...            &  65.2       0.4 &     54.9       0.2 & $-$365.2       0.2 \\
\Gttf                 & 244.2       3.7                  & 242.3       3.1 & 241.0       0.4 &    700.0       0.4 & $-$675.0       0.3 \\
\Gtf                  &  84.0       4.8                  &  95.4       2.4 &  95.6       0.3 &    518.0       0.5 & $-$995.4       0.2 \\
\Gon\tablenotemark{b} & 108.8       1.9\tablenotemark{c} &  ...            & 112.6       1.4 &    677.6       0.6 & $-$914.5       0.4 \\
\enddata
\tablenotetext{a}{from \cite{Lee07a}}
\tablenotetext{b}{no parallax prior in modeling}
\tablenotetext{c}{relative parallax from \cite{Khr10}}
\tablenotetext{d}{absolute parallax from RECONS astrometry program at CTIO/SMARTS 0.9m (see Henry et al.~2006\nocite{Hen06})}
\end{deluxetable}
\end{center}


\begin{deluxetable}{l l r r r r r r r}
\tablecaption{Component Position Angles, Separations, Residuals and Sources\tablenotemark{a}\label{tbl-TR}}
\tablewidth{0in}
\tablehead{\colhead{ID} 
&  \colhead{ set}
&  \colhead{ mJD }
&  \colhead{ $\rho$}
& \colhead{$\Delta \rho$}
&  \colhead{ $\theta$}
& \colhead{$\Delta \theta$}
& \colhead{$\rho \Delta \theta$}
& \colhead{Sources\tablenotemark{a}}\\
&&&(mas)&(mas)&(deg)&(deg)&(mas)&
}
\startdata
\Gozz&&&&&&&&\\
&1&49208.5915&397.7&0.0&123.9&-0.2&-1.4&FGS\,3\\
&2&49494.5823&286.6&2.7&159.8&0.5&2.4&FGS\,3\\
&3&49579.3203&242.1&1.3&178.3&0.5&1.9&FGS\,3\\
&4&49640.6823&213.8&-0.8&198.3&-1.2&-4.3&FGS\,3\\
&5&49687.0690&194.1&1.4&213.4&0.9&3.1&FGS\,3\\
&6&49893.4353&193.6&2.4&303.2&0.5&1.8&FGS\,3\\
&7&49930.6908&198.3&0.8&318.3&0.1&0.5&FGS\,3\\
\enddata

\tablenotetext{a}{Non-FGS measurements: 
B04=\cite{Beuz04}, adaptive optics;
B87=\cite{Bla87}, optical speckle;
B94=\cite{Bal94}, optical speckle;
FOC=\cite{Bar96}, \HSTns/FOC;
FOS=\cite{Sch98}, \HSTns/FOS; 
H12=\cite{Hor12}, optical speckle; 
H92=\cite{Hen92}, infrared speckle; 
M07=\cite{Mar07}, aperture masking; 
M91=\cite{McC91}, infrared speckle; 
NIC=\cite{Gol04}, \HSTns/NICMOS; 
VIS/CCD= \cite{Gey88}; visual, photographic, and CCD measurements;
W03=\cite{Woi03}, infrared speckle;
WFPC2=\cite{Sch00}, \cite{Die12} \HSTns/WFPC2;  
NACO=\cite{Ker16}  VLT/NACO
}


\end{deluxetable}


\begin{landscape}
\begin{deluxetable}{l l l r r r r r r r r}
\tabletypesize{\tiny}
\tablecaption{Orbital Elements\tablenotemark{a} \label{tbl-OE}}
\tablewidth{0in}
\tablehead{\colhead{ID} 
& \colhead{ $a$ ($\arcsec$)}
& \colhead{ $\alpha$$_A$ ($\arcsec$)}
& \colhead{ $inc$ (\arcdeg) }
& \colhead{ $\Omega$ (\arcdeg)}
& \colhead{P (days)}
& \colhead{T$_0$}
& \colhead{ecc}
& \colhead{$\omega_{\rm A}$ (\arcdeg) }
& \colhead{$\omega_{\rm B}$ (\arcdeg) }
& \colhead{f}
}
\scriptsize
\startdata
\Lst&0.3037&0.1169&146.1&62.8&1666.1&49850.4&0.364&166.6&-13.4&0.385\\
&0.0005&0.0004&0.2&0.4&2.5&0.8&0.001&0.5&0.5&0.002\\
\Gtt&0.5106&0.1425&43.7&178.3&5694.2&51817.2&0.163&284.5&104.5&0.279\\
&0.0007&0.0011&0.2&0.2&14.9&5&0.002&0.5&0.5&0.002\\
\Gff&0.1264&0.0517&126.6&92.2&418.5&53944.6&0.174&46.8&226.8&0.410\\
&0.0007&0.0003&0.4&0.5&0.2&1.6&0.004&1.4&1.4&0.003\\
\Lsts&2.0458&&128.3&147.6&9612&41358.4&0.617&&104.6&0.494\\
&0.0066&&0.3&0.5&12&12&0.004&&0.5&0.003\\
\enddata
\tablenotetext{a}{a=semimajor axis of relative orbit,
  $\alpha$=semimajor axis of perturbation of component A, $\omega_{\rm
    A}$= longitude of periastron passage of component A, f = mass
  fraction = ${\cal M}_{\rm sec} $/(${\cal M}_{\rm pri} $+${\cal
    M}_{\rm sec} $).}
\end{deluxetable}
\end{landscape}

\clearpage


\begin{landscape}
\begin{deluxetable}{lcccccccccccc}
\tabletypesize{\tiny}
\tablecaption{Masses, $M_V$, and $M_K$ for Red Dwarfs observed with \HSTns/FGS\tablenotemark{a} \label{tbl-MMVMK}}
\tablewidth{0in}
\tablehead{
   \colhead{ID}
&  \colhead{${\cal M}_{\rm pri} $}
&  \colhead{${\cal M}_{\rm sec} $}
&  \colhead{V\tablenotemark{b}}
&  \colhead{$\Delta V$}
&  \colhead{ref\tablenotemark{c}}
&  \colhead{M$_V\rm pri$}
&  \colhead{M$_V\rm sec$}
&  \colhead{K\tablenotemark{d}}
&  \colhead{$\Delta K$}
&  \colhead{ref\tablenotemark{c}}
&  \colhead{M$_K\rm pri$}
&  \colhead{M$_K\rm sec$}
}
\startdata     
GJ\,1005\,AB   & 0.179$\pm$0.002 &  0.112$\pm$0.001 &  11.48$\pm$0.01  & 2.39$\pm$0.05 & * &  12.70$\pm$0.01 &  15.12$\pm$0.09 &  6.39$\pm$0.02 &  1.23$\pm$0.02 & * &  7.80$\pm$0.02 &  9.03$\pm$0.03 \\
GJ\,22\,AC     & 0.405     0.008 &  0.157     0.003 &  10.28     0.03  & 3.08     0.10 & H99 &  10.32     0.03 &  13.40     0.10 &  6.04     0.02 &  1.93     0.03 &  *  &  6.19     0.02 &  8.12     0.04 \\
GJ\,54\,AB     & 0.432     0.008 &  0.301     0.006 &   9.82     0.02  & 1.04     0.05 &  *  &  10.70     0.03 &  11.70     0.04 &  5.13     0.02 &       ...      & ... &       ...      &       ...      \\ 
GJ\,65\,AB     & 0.120     0.003 &  0.117     0.003 &  12.08     0.02  & 0.45     0.08 & H99 &  15.49     0.04 &  15.94     0.05 &  5.33     0.02 &  0.16     0.02 & * &  8.87     0.03 &  9.03     0.03 \\ 
GJ\,1081\,AB   & 0.325     0.010 &  0.205     0.007 &  12.21     0.03  & 1.67     0.10 & H99 &  11.49     0.04 &  13.16     0.09 &  7.34     0.04 &  0.96     0.01 &  *  &  6.79     0.04 &  7.75     0.04 \\
GJ\,234\,AB    & 0.223     0.002 &  0.109     0.001 &  11.14     0.03  & 3.08     0.05 & H99 &  13.11     0.03 &  16.19     0.06 &  5.49     0.02 &  1.58     0.02 & * &  7.63     0.02 &  9.21     0.03 \\ 
G\,250-029\,AB & 0.350     0.005 &  0.187     0.004 &  10.95     0.03  & 1.61     0.08 & H99 &  11.07     0.03 &  12.68     0.07 &  6.35     0.02 &  1.03     0.07 &  *  &  6.61     0.03 &  7.64     0.05 \\
G\,193-027\,AB & 0.126     0.005 &  0.124     0.005 &  13.29     0.03  & 0.30     0.05 &  *  &  14.16     0.05 &  14.46     0.05 &  7.78     0.02 &  0.10     0.02 &  *  &  8.74     0.03 &  8.84     0.04 \\
\enddata
\tablenotetext{a}{Masses (${\cal M}$) in solar masses (${\cal M}_\odot$). Components are Primary = pri, Secondary = sec.}
\tablenotetext{b}{$V$ values from references in Table 2.}
\tablenotetext{c}{References for $\Delta V$ and $\Delta K$:
* = this paper,     
B96 = \cite{Bar96}, 
B00 = \cite{Ben00b}, 
H93 = \cite{Hen93}, 
H99 = \cite{Hen99}} 
\tablenotetext{d}{from 2MASS}
\end{deluxetable}
\end{landscape}

\begin{center}
\begin{deluxetable}{lcrcrrcrc}
\tablecaption{ Other Mass Determinations  \label{tbl-seb}}
\tablewidth{0in}
\tablehead{
\colhead{ID} 
& \colhead{SpT}
& \colhead{$\pi_{abs}$ [mas]}
&  \colhead{ref\tablenotemark{a}}
&  \colhead{M$_V$}
&  \colhead{M$_K$}
&  \colhead{ref\tablenotemark{a}}
&  \colhead{${\cal M}_\odot$}
&  \colhead{ref\tablenotemark{a}}
}
\startdata
GJ\,166\,C    & M4.0V & 200.7$\pm$0.2 & H12       & 12.68$\pm$0.03 &  7.45$\pm$0.03 & H12 & 0.177$\pm$0.029 & H99 \\
GJ\,278\,C    & M1~~V &  66.9     0.6 & T02       &  9.01     0.33 &  5.32     0.01 & T10 & 0.599     0.005 & T02 \\
GJ\,278\,D    &  ...  &               &           &  9.01     0.33 &  5.32     0.01 &     & 0.599     0.005 &     \\
GJ\,570\,B    & M1.5V & 169.8     0.9 & F99       &  9.45     0.05 &  5.40     0.03 & D00 & 0.586     0.007 & F99 \\
GJ\,570\,C    &  ...  &               &           & 11.09     0.17 &  6.58     0.04 & F99 & 0.390     0.005 &     \\
GJ\,630.1\,A  & M5~~V &  69.2     2.5 & YPC       & 12.72     0.02 &  7.64     0.03 & M09 & 0.231     0.001 & M09 \\
GJ\,630.1\,B  &  ...  &               &           & 12.86     0.03 &  7.78     0.03 &     & 0.214     0.001 &     \\
GJ\,747\,A    & M3~~V & 120.2     0.2 & S00       & 12.30     0.06 &  7.53     0.04 & D00 & 0.214     0.001 & S00 \\
\enddata
\tablenotetext{a}{Sources: 
*   = this paper, 
C05 = \cite{Cos05}, 
D00 = \cite{Del00}, 
D15 = \cite{Die16}, 
F99 = \cite{For99},
H12 = \cite{Hol12}, 
H99 = \cite{Hen99}, 
L00 = \cite{Lei00}, 
L05 = \cite{Lop05}, 
M09 = \cite{Mor09}, 
P97 = \cite{Per97}, 
R03 = \cite{Rib03}, 
S00 = \cite{Seg00}, 
S08 = \cite{Sei08}, 
T02 = \cite{Tor02}, 
T06 = \cite{Tam06}, 
YPC = \cite{WvA95}}
\end{deluxetable}
\end{center}


\begin{center}
\begin{deluxetable}{l c c }
\tablecaption{ MLR Fits and Residuals\tablenotemark{a} \label{tbl-coef}}
\tablewidth{0in}
\tablehead{\colhead{Param}
& \colhead{$M_V$ }
& \colhead{$M_K$}
}
\startdata
y$_0$&-2.59&-11.41\\
A$_1$&4.77&1.64\\
$\tau_1$&0.03&0.05\\
A$_2$&16.98&19.81\\
$\tau_2$&1.38&3.10\\
x$_0$&0.076&0.076\\
RMS\tablenotemark{b} $M$ &0.19 &0.09\\
RMS\tablenotemark{b} ${\cal M}_{\sun}$ & 0.023 & 0.014\\
$\chi^2$ & 1331 & 384 \\
DOF\tablenotemark{c} & 42 & 40\\
\enddata
\tablenotetext{a}{Parameters in the Equation 10 double
  exponential.}
  \tablenotetext{b}{RMS of absolute magnitude, $M$, and mass, ${\cal M}$, residuals}
  \tablenotetext{c}{DOF = degrees of freedom for each fit}
\end{deluxetable}
\end{center}


\begin{center}
\begin{deluxetable}{l c c }
\tablecaption{ Polynomial Fits for Mass as a Function of Absolute Magnitude\tablenotemark{a}   \label{LMRT}}
\tablewidth{0in}
\tablehead{\colhead{Param} 
& \colhead{$M_V$}
& \colhead{$M_K$}
}
\startdata
C0&0.19226$\pm$0.000424&0.2311$\pm$0.0004\\
C1&--0.050737 0.000582&-0.1352 0.0007\\
C2&0.010137  0.00021&0.0400 0.0005\\
C3&-0.00075399 4.57e-05 &0.0038 0.0002\\
C4&--1.9858e-05 1.09e-05 &-0.0032 0.0001\\
$x_0$ & 13.0 & 7.5  \\
\enddata
\tablenotetext{a}{Parameters in the Equation 11 polynomial.}
\end{deluxetable}
\end{center}


\begin{center}
\begin{deluxetable}{l c c c c l}
\tablecaption{ Factors Potentially Influencing Deviations from MLR \label{tbl-Exc}}
\tablewidth{0in}
\tablehead{\colhead{ID} 
&  \colhead{[Fe/H]}
&  \colhead{V\,sin\,i }
&  \colhead{R/R$_\sun$}
& \colhead{log(L$_X$/L$_{bol}$)}
&  \colhead{Refs\tablenotemark{a}}\\
& & [\kms] &&&
}
\startdata
GJ\,1005\,AB    &$-$0.47&$<$3&0.23&$-$4.97&G14,S13,F13\\
GJ\,22\,AC      &$-$0.28&&0.47&&NC15,R12,P01\\
GJ\,54\,AB      &&&0.37&$-$4.78&G14,S13\\
GJ\,65\,A       &$-$0.42&32&0.165&&K12,G05,J09,K16\\
GJ\,65\,B       &&30&0.159&&G05,J09,K16\\
GJ\,1081\,AB    &&&0.4&&G14\\
GJ\,234\,AB     &0.18&6&0.26&$-$3.65&NC15,G05,P01,S13\\
G\,250-029\,AB  &&&0.35&&G14\\
\enddata
\tablenotetext{a}{F13 = \cite{Fei13}, G05 = \cite{Gle05}, I11 =
  \cite{Irw11}, J09 = \cite{Jen09}, NB14 = \cite{Nev14}, NC15 =
  \cite{New15}, P01 = \cite{Pas01}, R12 = \cite{Roj12}, E13 =
  \cite{Eir13}, S13 = \cite{Ste13}, W11 = \cite{Wri11}, T15 =
  \cite{Ter15}, G14 = \cite{Gai14}, K12 = \cite{Kol12}, K16 = \cite{Ker16}}
\tablenotetext{b}{Assumes same [Fe/H] as GJ\,570\,A, a K5V star. }
\end{deluxetable}
\end{center}


\clearpage


\clearpage


\clearpage


\clearpage


\clearpage


\clearpage


\clearpage


\clearpage


\clearpage


\clearpage


\clearpage


\clearpage


\clearpage


\clearpage


\clearpage


\end{document}